\documentclass[%
 aip,
 amsmath,amssymb,
 reprint,%
]{revtex4-1}
\usepackage{hyperref}
\usepackage{graphicx}
\usepackage{dcolumn}
\usepackage{bm}
\usepackage{natbib} 
\usepackage{amsthm} 
\usepackage{subfigure}
\usepackage[utf8]{inputenc}
\usepackage[T1]{fontenc}
\usepackage{mathptmx}
\usepackage{etoolbox}
\makeatletter
\def\@email#1#2{%
 \endgroup
 \patchcmd{\titleblock@produce}
  {\frontmatter@RRAPformat}
  {\frontmatter@RRAPformat{\produce@RRAP{*#1\href{mailto:#2}{#2}}}\frontmatter@RRAPformat}
  {}{}
}%
\makeatother
\begin{document}

\preprint{AIP/123-QED}

\title{Multiple-Nanowire Superconducting Quantum Interference Devices: Critical Currents, Symmetries, and Vorticity Stability Regions}
\author{Cliff Sun}
 \affiliation{Department of Physics, University of Illinois, Urbana-Champaign}
\author{Alexey Bezryadin}%
\affiliation{Department of Physics, University of Illinois, Urbana-Champaign}%

\date{\today}

\begin{abstract}
An ordinary superconducting quantum interference device (SQUID) contains two weak links connected in parallel. We model a multiple-wire SQUID (MW-SQUID), generalized in two ways. First, the number of weak links, which are provided by parallel superconducting nanowires, is larger than two. Second, the current-phase relationship of each nanowire is assumed linear, which is typical for a homogeneous superconducting thin wire. For such MW-SQUIDs, our model predicts that the critical current ($I_c$) is a multi-valued function of the magnetic field. We also calculate vorticity stability regions (VSR), i.e., regions in the current-magnetic field plane in which, for a given distribution of vortices, the currents in all wires are below their critical values, so the vortices do not move between the cells. The VSRs have rhombic shapes in the case of two-wire SQUIDS and have more complicated shapes in the case of many nanowires. We present a classification of such VSRs and determine conditions under which VSR is disjoint, leading to 100\% supercurrent modulation and quantum phase transitions. According to the model, the maximum critical current curves obey $IB$ symmetry, while each VSR obeys $IBV$ symmetry.  The model predicts conditions at which MW-SQUID exhibits a perfect diode effect in which the critical current of one polarity is zero while it is not zero for the opposite polarity of the bias current. We also provide a classification of the stability regions produced by (1) completely symmetric, (2) phase disordered, (3) position disordered, (4) critical current disordered, and (5) completely disordered multi-wire SQUIDs.
\end{abstract}

\maketitle

\section{Introduction}

One of the current promising fields of research involves studying superconducting weak links, or superconducting nanowires, which, if connected in parallel, form an array of loops. Each superconducting loop is characterized by its vorticity, which is the winding number of the phase of the condensate wave function divided by $2\pi$. If thermal and quantum fluctuations are negligible, the vorticity cannot change unless the current in at least one wire exceeds the critical current parameter of that wire. The vorticity is an integer because the wave function must be single-valued, and, therefore, the phase of the superconducting order parameter can only change by an integer multiple of $2\pi$ on a closed contour coinciding with each superconducting loop.  
    
    Superconducting nanowires possess several unique characteristics, including a linear current-phase-relationship \cite{tinkham_book, murphy-2017, murphy}, thermal \cite{PhysRev.156.396, langer-1967, PhysRevB.5.864} and quantum \cite{PhysRevLett.61.2137,bezryadin-2000, PhysRevLett.87.217003, quantum_suppression, ARUTYUNOV20081, sahu-2009, bezryadin-2012, arutyunov-2012, mooij-2006, shaikhaidarov-2022, PhysRevLett.97.017001, PhysRevLett.107.137004} phase slips, high kinetic inductance \cite{belkin-2011, murphy, friedrich-2019}, and the ability to trap and hold fluxons in closed superconducting loops and act as superconducting memory elements \cite{eduard_memory}. The ability of superconducting nanowire closed loops to trap vortices in various metastable configurations allows them to maintain multiple stable states, making nanowire SQUIDs particularly promising for use in superconducting memory devices \cite{murphy-2017, Zhao_Toomey_Butters_McCaughan_Dane_Nam_Berggren_2018, chen-2020, friedrich-2019}. Note that this is a significant advantage compared to ordinary memory devices based on Josephson junctions \cite{manheimer-2015, 6377273, guarcello-2017, goldobin-2013, niedzielski-2015}, in which case additional macroscopic inductors are needed in the circuit to achieve multiple metastable states and the corresponding memory functionality.
    Previous research has demonstrated precise control over these metastable states by selectively pushing vortices into the device, thereby “setting” its state \cite{murphy}. Note that whenever we discuss vortices in the device we mean vortices located between the nanowires, and we assume there are no vortices inside the nanowire and no vortices inside the superconducting electrodes. Due to their high non-linear inductance, superconducting nanowires are also attractive for applications in quantum computing, especially in developing kinetic-inductance-based qubits \cite{manuch, kineticon}. Beyond computing, superconducting nanowires have been studied as single-photon detectors because of their property to switch to the normal state if a single photon strikes the nanowire of the detector \cite{single_photon_detector}. Future research might show that photons can program vorticity states of superconducting nanowire arrays.

Previous research has focused on a dual nanowire configuration\cite{hopkins_SQUID}, realized as a nanowire superconducting quantum interference device (nanowire SQUID) containing only two parallel superconducting nanowires, connected in parallel and linking two macroscopic superconducting thin-film electrodes. The critical current has a sawtooth shape and at sufficiently low temperatures shows up as a multi-valued function of the magnetic field \cite{gurt-yo, hopkins_SQUID, ultra_low_noise_SQUID}. For two-wire devices, the region in the $I$-$B$ plane, within which the number of vortices in the superconducting loop remains fixed (VSR region), is a parallelogram. Such was observed experimentally and within linear CPR models \cite{murphy-2017}. In addition, multiple stable states were observed, which showcased the metastable nature of superconducting nanowire loops. These states have been used for the information storage in superconducting memory elements . 


\begin{figure}[t]
    \centering
    \includegraphics[width=0.9\linewidth]{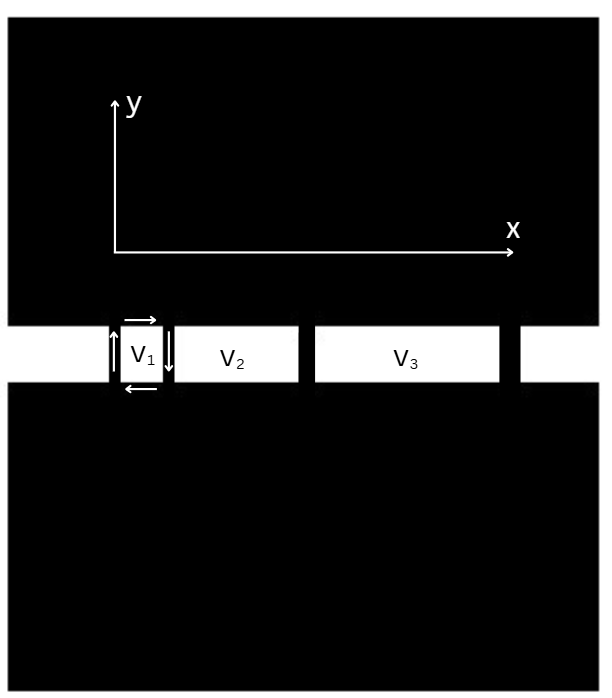}
    \caption{A generic illustration of a random nanowire array. Two bulk superconducting films (black) are connected by four (in this example) parallel nanowires. In the presented model, each nanowire is characterized by three parameters, namely its position, its critical current, and its critical phase. Each closed loop is characterized by its vorticity number, three of which are shown in the picture. If the vorticity equals, for example, one, then it means that one fluxon is trapped in the loop, i.e., the phase increases by 2$\pi$ on a closed contour going around the loop. The x-axis is anti-parallel with the nanowires.}
    \label{fig:randomarray}
\end{figure}
\vspace{-10pt}
The behavior of SQUIDs with more than two nanowires remains largely unexplored. To address this gap, we developed a critical current model of an n-nanowire system, referred to as multiple wire SQUID or MW-SQUID. The nanowires are assumed to be connected in parallel, linking two large electrodes. The model assumes that each nanowire has a linear CPR, as was established in previous investigations. Note that the linear CPR assumption is important, since previous work on arrays of wires modeled with sinusoidal CPR generated different results\cite{pekker-2005}. Our present model predicts the shape of the critical current versus magnetic field curves, $I_c(B)$, and the shapes of vorticity stability regions (VSR). The borders of the ensemble if VSRs represent the multi-valued critical current function. We use the critical phase ($\phi_c$) model, where each nanowire may only support a phase difference up to $\phi_c$ between the two ends of the nanowire. 
    
Our model shows that the presence of fluxons (vortices) in the loops formed by nanowires has a strong impact on the critical current and VSRs. The shapes of the stable vorticity regions are predicted to be rhombic for the two-nanowire devices and become much more complex in the case of multiple nanowires. A classification of the stable vorticity regions and their symmetries and topology is developed and presented here. Moreover, we derive the conditions when the VSR regions become disjoint, thus predicting a sequence of quantum transitions. 
We also investigate the response of VSRs to disorder, classifying superconducting nanowire configurations into the following categories: (1) Phase disordered arrays are such in which all nanowires are equidistant and have the same critical currents, but have different critical phases, (2) Critical current disordered arrays are such that all nanowires are equidistant \& have the same critical phases, but different critical currents, (3) Position-disordered arrays are such that all nanowires have the same critical phases and critical currents, but the nanowires are not equidistant, \& (4) random arrays of nanowires are such that nanowires have different critical phases, they are located at random positions, and have different critical currents. We elucidate the effects of various disorders on the shape of VSR. It is predicted that position disorder is the only form of disorder that can change the period of the device's critical current. 

To control the vorticity states, it is important to understand hidden symmetries. We find that MW-SQUIDs exhibit symmetry (invariance) if three quantities are inverted simultaneously, namely the bias current $I$, the magnetic field $B$, and persistent currents of the vortices trapped on the superconducting loops. If all three inversions are applied (i.e., $IBV$-inversion is performed), then the VSR remains invariant, and the Meissner phase equation, describing relative phase shifts between the wires, remains invariant. Thus, such device symmetry can be called $IBV$-symmetry. Since the positive/negative critical current is defined by maximizing/minimizing the total current over the entire set of possible vorticity states, the two critical current curves show a simplified $IB$ symmetry, i.e., the critical current is invariant if the current and the magnetic field are both inverted simultaneously. The $IBV$ symmetry is present even in SQUIDS with a large number of parallel nanowires, all of which are different and randomly positioned. Moreover, this symmetry still holds even when the wires are not parallel to the $y$ axis. The only condition required for $IB$ symmetry to hold is that all wires have to have an antisymmetric current-phase dependence. In this paper, we assume that there are no trapped vortices in the superconducting electrodes. On the other hand, if such vortices would be trapped inside the electrodes of the SQUID then they would cause a violation of the $IB$ symmetry.

\section{Modeling of multiple nanowire superconducting quantum interference device: MW-SQUID}

The device under consideration is a parallel array of superconducting nanowires connected to two macroscopic thin film electrodes. It is analogous to an ordinary SQUID, but the traditionally used Josephson junctions are replaced in our considered system with nanowires. One key assumption here is that the nanowires having a linear current-phase relationship. Another important generalization in our model is that the number of nanowires in such a nanowire device can be larger than two. In this model, the phase difference on each wire is related to the phase difference on other wires, due to the fact the the whole device is superconducting and thus phase coherent. The relationship between the phase difference 
values is determined by only two factors, specifically (1) by the Meissner currents in the thin film electrodes and (2) by the number of fluxons trapped between nanowires (on the superconducting loops formed by the wires). Such an approach is analogous to the one proposed previously, in the experiments and modeling of DNA-templated nanowire SQUIDs \cite{hopkins_SQUID}. 

A schematic of the devices is given in Fig.\ref{fig:randomarray}. There, the electrodes are shown as black rectangles, and the nanowires are shown as vertical black bars. For this model device we assume that the inversion of the y-coordinates is a symmetry of the device and the inversion of z-coordinates is a symmetry. Therefor, the device is assumed symmetrical with respect to a rotation about the x-axis by 180 degrees.

Since the wires are connected in parallel, the total supercurrent can be calculated by summing all the supercurrents flowing through individual nanowires. In other words, $I_{\text{tot}} = \sum_{i=1}^{n}I_i$, which the integer $i$ is just the number of the nanowire, which runs from one to the maximum $n$, which is the number assigned to the rightmost nanowire (assume $n>1$).

The model developed here is a generalization of the two-wire SQUID model proposed in the works related to two-nanowire  SQUIDs \cite{murphy-2017, murphy}. The SQUID with two nanowires showed periodic oscillations of its resistance as well as its critical current. It was derived that if two coplanar thin film electrodes are linked with two coplanar parallel nanowires, and the distance between the wires is $d$ and the width of the electrodes is $W$ then the period of oscillations of the critical current and/or the device resistance (in case if the temperature is elevated and thermal phase slips are present) is $\Delta B(z)=(\pi^2/8G)(\phi_0/dW)$, where G=0.916 is the Catalan constant and $\phi_0=2.068$x$10^{-15}$Wb is flux quantum. A detailed derivation can be found in \cite{hopkins_SQUID}. The formula is exact if the electrodes are wide, i.e., if $W>>d$. For example, if the width of the electrodes is 20$\mu$m and the distance between the wires is 5$\mu$m, then the period is $\Delta B = 0.0000276T = 0.276G$. 

The periodic oscillation is analogous to the well-known Little-Parks oscillations \cite{little_parks}. The origin of the periodicity is that the electrodes impose a rigid relationship between the phase bias of all the wires. (The phase bias term stands for the phase difference of the condensate wave function taken between the ends of the wire.) This means that if the phase bias of one wire is given and the magnetic field is given then the phases bias of all other wires can be calculated. If $B=0$, then the phase difference on the first wire equals the phase difference on the second wire, assuming there are no vortices between the wires. The terms ''phase difference'' and ''phase bias'' are used interchangeably. If $B$ is not zero then the Meissner currents generate phase gradients along the edges of the electrodes. These gradients will cause a difference in the phase bias of the nanowires. It is assumed that the electrodes are much stronger superconductors compared to the nanowires, so that phase gradients in the electrodes depend only on the applied $B$ but not on the currents in the nanowires. These phase gradients define how the phase difference on nanowires changes with the position of the wire. The period of the SQUID, $\Delta$, is defined such that if the applied field is $B=\Delta B$, then the shift imposed by the Meissner currents in the electrodes on the SQUID loop equals $2\pi$. Note also that the phase gradients are linear in $B$. In such case, if the phase difference on the wire $i=1$ is $\phi_1$ and the phase difference on the wire $i=2$ is $\phi_2$ then $\phi_2=\phi_1+2\pi B/\Delta B=\phi_1+2\pi dB/[(\pi^2/8G)(\phi_0/W)]=\phi_1+2\pi zB/\Delta B_1$. 

We have introduced a new normalization constant $\Delta B_1=(\pi^2/8G)(\phi_0/W)$, which is the period of oscillations evaluated assuming that the distance between the wire equals unity ($d=1$). Suppose the coordinate of the first wire is $X_1$ and the coordinate of the second wire is $X_2$. Then the equation becomes: $\phi_2=\phi_1+2\pi (X_2-X_1)B/\Delta B_1$. This is the main equation governing nanowire SQUID. We will refer to it as the Meissner phase equation.

If there are many wires then for each pair of wires, $i$ and $i+1$, the equation has the same format, namely the phase bias on the wires are related to each other as $\phi_{i+1}=\phi_i+2\pi (X_{i+1}-X_i)B/\Delta B_1$. Suppose the total number of wires in the device is $n$. Then we can introduce normalized coordinates $x_i=(X_i-X_1)/(X_n-X_1)$, so that the first wire is at the normalized position $x_1=0$ and the last wire is at $x_n=1$. Then the Meissner phase evolution equation becomes $\phi_{i+1}=\phi_i+2\pi (X_n-X_1)(x_{i+1}-x_i)B/\Delta B_1$. Next step, introduce a normalized magnetic field as $b=(X_n-X_1)B/\Delta B_1$. Then $\phi_{i+1}=\phi_1+2\pi x_ib$. Here and below, we use normalized coordinates and the normalized magnetic field, as defined above.

So far, we have assumed that there are no persistent supercurrents in the superconducting loops formed by the parallel nanowires and the superconducting electrodes. If vortices (can also be called fluxons, fluxoids) are present in the space between the wires, then they create persistent supercurrents on the superconducting loop, and therefore a phase bias on the wires. It is assumed that the persistent supercurrents are weak, so they do not produce additional phase shifts in the electrodes. Then, for two neighboring wires, we can write the following expression: 

\begin{equation}
    \phi_{i+1}=\phi_i+2\pi b (x_{i+1}-x_i)-2\pi v_{i,i+1}
\label{e1}    
\end{equation}
where $v_{i,j}$ is the number of vortices trapped between wires $i$ and $j$. Note that these are coreless vortices, which are manifested by persistent supercurrents in the corresponding loops. In general, for a superconducting loop formed by nanowires $i$ and $j$, the phase change on a closed contour coinciding with the loop is always an integer multiple of $2\pi$. This requirement is needed to ensure that the wave function of the superconducting condensate is a single-valued function of the coordinates. Therefore, $v_{i,j}$ is an integer.

Suppose the number of vortices trapped in the space between wires $i=1$ and $i$ is $v_{1, i}$ (always an integer), then the relationship between the phase bias (or the phase difference) on the corresponding wires is

\begin{equation}
    \phi_i = \phi_1 + 2\pi bx_i - 2\pi v_{1,i}
    \label{e2}
\end{equation}

Note that $v_{1,i}=v_{1,2}+v_{2,3}+v_{3,4}+...v_{i-1,i}$ since the number of vortices is an additive quantity. The supercurrent $I$ flowing in each nanowire obeys a linear CPR \cite{murphy-2017, murphy}, as given by:

\begin{equation}\label{nanowire_cpr}
    I_i = I_{c,i}\frac{\phi_i}{\phi_{c,i}}
\end{equation}
where $i$ is the wire order number, $\phi_i$ is the phase difference of the superconducting order parameter evaluated between the two ends of the wire (phase bias), and $\phi_{c, i}$ is the critical phase of the wire $i$. In our model, the normalized supercurrent for wire $i$ is defined as $j_{i} = I_i/\langle I_{c, i} \rangle = j_{c,i}\phi_i/\phi_{c, i}$ where $\langle I_{c, i} \rangle$ is the average of the critical currents of all wires in the device, and $j_{c, i}=I_{c, i}/\langle I_{c,i} \rangle$ is the normalized critical current for the i-th wire. Such normalization implies that the absolute maximum of the supercurrent equals the number of wires, $n$, in the MW-SQUID. For example, in Fig.\ref{fig:2-wire-stability-regions} the maximum current equals $j_{max}=2$ because there are only two wires in the device. Note that in the general case, this maximum is attainable only if all the wires have the same critical phase, or if one can tune the magnetic field such that all the wires can achieve the critical phase at the same time. If there are many wires and they are all different then it might not always be possible, generally speaking, to bring all of the wires to the critical phase state simultaneously. 

The critical phase constant $\phi_{c,i}$ represents the maximum phase difference of the superconducting order parameter that the $i$-th nanowire can hold between its ends and remain superconducting. The critical phase criterion says that once $|\phi_i| \geq |\phi_{c,i}|$ in the nanowire number $i$, then this nanowire cannot sustain superconductivity and either undergoes a phase slip by $2\pi$ (which is equivalent to a vortex crossing the wire) to reduce the phase difference or transitions to the normal state. If a phase slippage occurs, heat is generated, and the entire device typically switches to the normal (resistive) state since its temperature jumps up. So, effectively, the condition  $|\phi_i| \geq |\phi_{c,i}|$ for any $i$ defines a critical current of the entire device. Due to phase coherence, the phase difference defined on one wires determines the phase difference on the other wires, and this dependence is determined by the perpendicular magnetic field, $B$, and the spacing between the wires. Consequently, the critical current is a function of $B$. Moreover, the phase difference on wires depends on the number of vortices trapped between the wires. Thus, the device exhibits a multivalued critical current function, where different branches corresponds to different distributions of vortices in the cells. The critical current which is the largest between different branches will be denoted as $I_{c,+}(B)$. The lowest critical current, which is usually negative, is denoted as $I_{c,-}(B)$.

\section{\label{sec:2-wire} Two-Nanowire SQUID}

We proceed with the analysis for two superconducting nanowires in parallel. Such a device is similar to an ordinary SQUID. Yet, nanowires, unlike traditional superconductor-insulator-superconductor Josephson junctions (JJ), exhibit an approximately linear current-phase relationship (CPR) \cite{murphy-2017,murphy,tinkham_book}. Here, for simplicity, we assume that the CPR is exactly linear. For a JJ, the phase difference at which the current is the maximum is $\pi/2$. In contrast, for a superconducting nanowire, its critical phase is larger than $\pi/2$ ($\phi_c>\pi/2$) and can be very large if the wire is very long. In fact, the critical phase of a nanowire is proportional to its length. Because of that, the number of vortices in the loop formed by the nanowires is not always zero, like in the SQUID loop, which has no extra inductors. 

The maximum number of vortices can be calculated. As an example, we do this for $B=0$ and assume that the wires are identical ($\phi_{c,1}=\phi_{c,2}=\phi_c$). If $B=0$, then the phase bias of both wires is the same, i.e., $\phi_1=\phi_2$. The phase change in the electrodes is zero since $B=0$ and the Meissner current is zero. Therefore $\phi_1-\phi_2=2\pi v_{1,2}$, by the definition of vorticity $v_{1,2}$. Assume for this example that the bias current is zero, $I=0$, then $\phi_2=-\phi_1$. Then $2\phi_1=2\pi v_{1,2}$. If we now take into account the critical phase condition $\phi_1<\phi_c$, then the conclusion is $v_{1,2}<\phi_c/\pi$. Since $v_{1,2}$ is an integer, we conclude that the SQUID nanowire loop can hold at least one vortex only if $\phi_{c,1}>\pi$. The same condition applies to the second wire, $\phi_{c,2}>\pi$, since in this example it is assumed that the wires are identical. Therefore, a symmetrical SQUID can trap at least one vortex in the loop if the wires have a critical phase of larger than $\pi$. It can trap at least two vortices if the critical phase is larger than $2\pi$, etc. In the general case, the maximum number of vortices that can occupy the loop of the symmetrical nanowire SQUID is the integer part $\phi_c/\pi$. 

If the SQUID is asymmetrical, a similar analysis can be performed. If there is a vortex, then the sum of the phase biases on both wires equals $2\pi$, and the current in one wire is opposite to the other if the net bias current is zero. As a result, one obtains two conditions which restric the number of vortices on the loop of the considered 2-SQUID, specifically $v{1,2}<(\phi_{c,1}/2\pi)(1+1/\beta)$ and $v{1,2}<(\phi_{c,2}/2\pi)(1+\beta)$, where the asymmetry parameter is defined as $\beta=\phi_{c,1}I_{c,2}/(\phi_{c,2}I_{c,1})$.

If the bias current is above zero, $I>0$, then it adds the same phase bias to each wire (assuming $B=0$), i.e., $j=I/I_c=\phi_1/\phi_c+\phi_2/\phi_c$. Note that here we use a normalized bias current $j=I/I_c$, which is consistent with the definition above since here we assume that both wires have the same critical current $I_c$ and the same critical phase $\phi_c$. At the same time, the presence of $v_{1,2}$ vortices requires that $\phi_1-\phi_2=2\pi v_{1,2}$. In addition to this, we still have the conditions $\phi_1<\phi_c$ and $\phi_2<\phi_c$. Combining these inequalities with the two equations above, one gets the maximum vorticity as the integer part of $(1-j/2)(\phi_c/\pi)$, namely $v_{max}$=int$[(1-j/2)(\phi_c/\pi)]$. For nanowires of critical phases $\phi_{c,1}$ and $\phi_{c,2}$, this equation can be generalized to $v_{max}=\text{int}[<(1-j/2)(\phi_{c,i}/\pi)>]$. Here the triangle brackets represent the mean value. This formula shows that one can reduce the number of metastable states by increasing the bias current. This fact can be used to create a desired vorticity state. The algorithm is as follows: Apply the bias current such that there is only one stable vorticity state. If the device is superconducting, then the desired $v_{1,2}$ value is achieved. If the device is resistive, reduce the bias current to zero and repeat the procedure.

\begin{figure}[t]
    \centering
    \includegraphics[width=\linewidth]{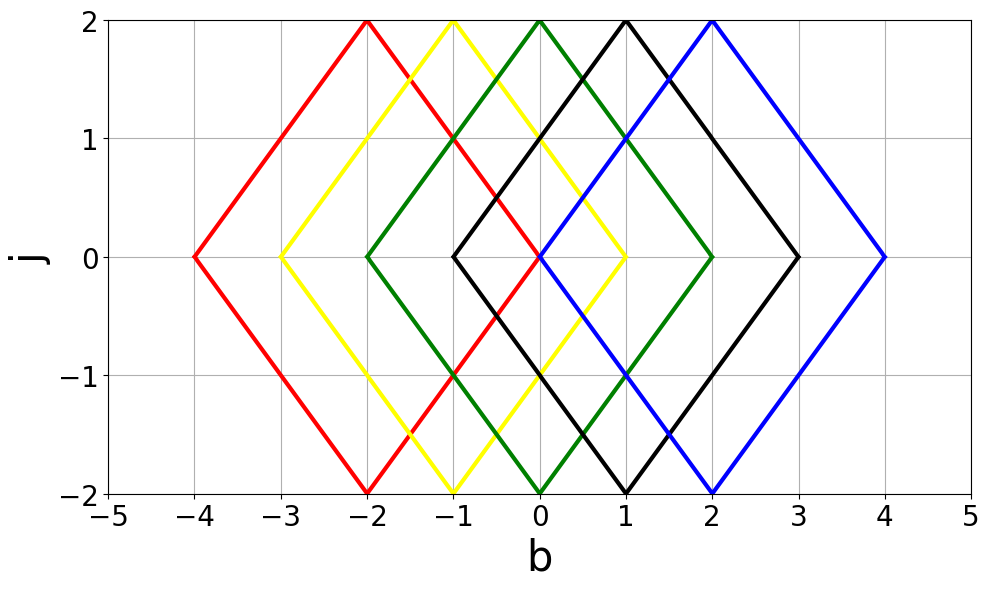}
    \caption{Two nanowires with critical phases ($\phi_{c,i}$) set to \(2\pi\) and critical currents ($I_{c,i}$) set to 1. Vorticity values, which are always integers, range from -2 to 2 (left to right). The produced stability regions exhibit exclusively diamond-shaped patterns for this system. The modulation of the maximum critical current is only 25 percent in this case.}
    \label{fig:2-wire-stability-regions}
\end{figure}

Let us now use the linear model outlined above to calculate the VSRs (vorticity stability regions) for 2-SQUID (a SQUID with only two nanowires). The VSR is such a domain in the I-B (or j-b) plane that the current in each wire is less than the critical current as long as $I$ and $B$ remain within this domain. If the current is less then critical then the number of vortices cannot change, i.e., the vorticity is stable. This analysis assumes that thermal and quantum fluctuations are negligible. When both wires are identical, the VSR is a rhombus ("diamond"), as shown in Fig.\ref{fig:2-wire-stability-regions}. If there are no vortices in the loop (i.e., $v_{1,2}=0$), then the maximum and the minimum critical current take place at $B=0$. The sequence of VSR corresponding to different (always integer) $v_{1,2}$ values is a periodic sequence of diamonds. The diamond corresponds to one vortex in the loop ($v_{1,2}=1$), has its maximum at $b=1$, and the peak of the $v_{1,2}=m$ VSR is at $b=m$, where $m$ is an integer. Note also that the physical meaning of the VSR is that its boundaries give the critical current of the devices, for a fixed vorticity value.

\begin{figure*}[t]
    \centering
    \subfigure{%
        \includegraphics[width=0.32\linewidth]{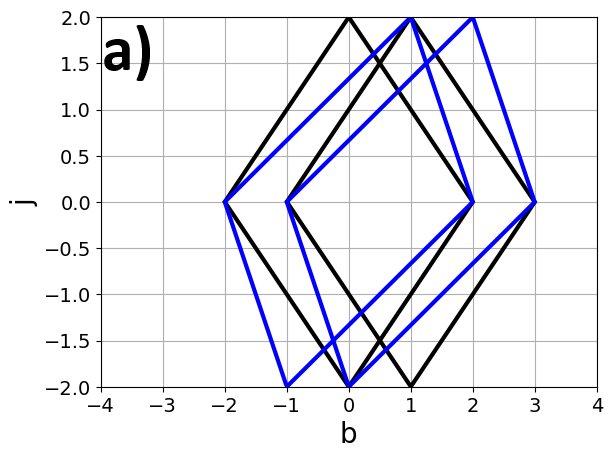}
    }
    \hfill
    \subfigure{%
        \includegraphics[width=0.32\linewidth]{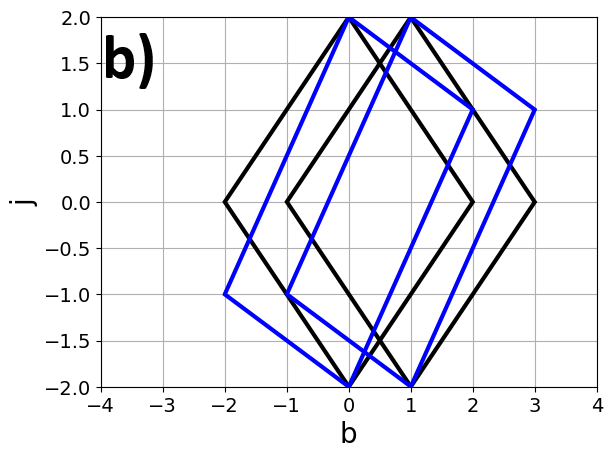}
    }
    \hfill
    \subfigure{%
        \includegraphics[width=0.32\linewidth]{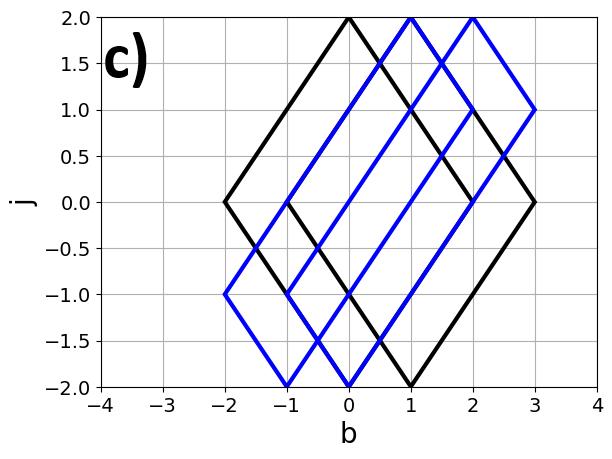}
    }
  \caption{Vorticity stability regions (VSR) for devices with two parallel nanowires, i.e., 2-SQUIDs. (a) Critical phase disorder for two nanowires with identical critical currents, namely $I_{c,1}=I_{c,2}=1$. The symmetrical devices (black line) corresponds to $\phi_{c,1} = 2\pi$ and $\phi_{c,2}=2\pi$. The distorted (blue) diamond corresponds to a device with different wires, namely $\phi_{c,1} = \pi$ and $\phi_{c,2}=3\pi$. Two vorticity states, $v_{1,2}=0$ and $v_{1,2}=1$, are shown as examples. (b) Critical current disorder for two nanowires with identical critical phases, namely $\phi_{c,1} = \phi_{c,2}=2\pi$. The black diamond is the symmetrical case where the critical current is set to 1 for both wires. The asymmetrical case (blue) corresponds to $I_{c,1} = 0.5$  and $I_{c,2}=1.5$, with $v_{1,2}=0$ and $v_{1,2}=1$ displayed from left to right. (c) Complete disorder for two nanowires. The symmetric case (black) corresponds to $\phi_{c,1} = \phi_{c,2} = 2\pi$ and $I_{c,1} = I_{c,2} = 1$. The asymmetric case (blue) is calculated for a 2-SQUID that has critical phases $\phi_{c,1} = \pi$, $\phi_{c,2} = 3\pi$, and critical currents $I_{c,1} = 0.5$, $I_{c,2} = 1.5$. The vorticity states 0 and 1 are displayed from left to right.}

    \label{fig:2-wire-disorder}
\end{figure*}

Let us consider how the VSRs change if the wires are not identical. If the wires have different critical phase values but identical critical currents, then the maximum of the VSR shifts from zero field, while the side corners of each VSR do not shift from their original positions, as shown in Fig.\ref{fig:2-wire-disorder}a. The shift of the top corner of the VSR can be calculated as follows. The top vertex of the VSR, which represents the maximum critical current, occurs at such a magnetic field that the phase bias equals its critical value for both wires. Suppose the phase bias on the first wire is critical, i.e., $\phi_1=\phi_{c,1}$. The phase bias on the second wire then is given by the Meissner phase equation, which becomes $\phi_2 = \phi_1 + 2\pi b-2\pi v_{1,2}$, as follows form Eq.\ref{e1}. There, we take into account that if there are only two wires, then the normalized coordinates for them are $x_1=0$ and $x_2=1$, by definition. Therefore, the phase bias of the second wire is critical under the condition $2\pi b=\phi_{c,2} - \phi_{c,1}+2\pi v_{1,2}$. Thus, the requirement that both wires have the phase bias equal to their critical values translates into the following condition $b = (\phi_{c,2} - \phi_{c,1})/2\pi + v_{1,2}$. Since $v_{1,2}$ can be any integer number, an infinite sequence of critical current peaks occurs.

We also investigate the effect of the critical current disorder. For this purpose, we keep the critical phases the same and assign different critical currents to the wires forming the SQUID. The model predicts that the side vertices of the VSR diamonds shift in the vertical direction (along the bias current axis, $I$ or $j$), and they shift in the opposite directions with respect to each other. However, the maximum of the VSR does not shift in this case as shown in Fig.\ref{fig:2-wire-disorder}b. In symmetrical devices, we always find that the maximum supercurrent through the device is positive and the minimum critical current has a negative value.

Interestingly, our model predicts that if $I_{c,1}\neq I_{c,2}$, the maximum and the minimum total current $I$ values can have the same sign, assuming the vorticity is not allowed to change. For example, in Fig.\ref{fig:2-wire-disorder}b, for $v_{1,2} = 0$ and $b=1.5$, we find $I_{c,+}^0\approx 1.25$ and $I_{c,-}^0\approx 0.25$, i.e., both the maximum and the minimum are positive. The subscript "0" signifies that the critical current corresponds to zero vorticity. This tells us that zero vorticity is only stable at a finite bias current. Similarly, there exist some range of magnetic fields at which the maximum current and the minimum current are both negative. For example, in Fig.\ref{fig:2-wire-disorder}b, for $v_{1,2} = 0$ and $b=-1.5$, both $I_{c,+}^0$ and $I_{c,-}^0$ are negative. Note, if the vorticity is allowed to adjust, i.e., undergo phase slips, then the maximized-over-the-vorticity critical current, $I_{c,+}$, is always positive and the minimized-over-the-vorticity critical current $I_{c,-}$ is always negative. In the example of Fig.\ref{fig:2-wire-disorder}b, assuming $b=1.5$ we find that $I_{c,+}=I_{c,+}^0\approx1.25$ and  $I_{c,-}=I_{c,-}^1\approx-1.25$, i.e., to reach the maximum absolute value of the negative current the system has to allow a single phase slip and adjust the vorticity from zero to one.

When applying both critical phase disorder and critical current disorder to a 2-SQUID, we observe a combination of the maximum of the VSR shifting horizontally and the side vertices of the VSR shifting vertically. The shape of the VSR in this case remains a parallelogram, as shown in Fig.\ref{fig:2-wire-disorder}c. The top vertex of the VSR shifts if and only if there exists a difference between the critical phases. Therefore, if an experiment shows that the critical current maximum, which is the nearest to zero field, occurs away from zero field then one concludes that the critical phase values are different for the two wires. Note that according to $IB$ symmetry (to be discussed below), the top vertex and the bottom vertex shift in opposite directions along the B-axis.  Therefore such shifts give rise to the superconducting diode effect\cite{Song2023,diode_with_memory}. 



 Consider VSR for a SQUID with two identical nanowires. The top vertex of the VSR always occurs at a magnetic field of $b = v_{1,2}$ because under this condition, the phase bias imposed on the wires by the vortex is equal and opposite to the phase bias imposed by the Meissner currents. Due to this compensation effect (i.e., the Little-Parks effect), the total current in the wires is zero (assuming zero bias current), and the free energy of the system is at its minimum. Thus, if $b=v_{1,2}$ then the device can accept the maximum possible bias current, i.e., the VSR is maximally extended along the bias current axis.
 
 It is instructive to analyze different symmetries of nanowire SQUIDs. Suppose we perform a vortex anti-vortex transformation (V-inversion), that is, reversing the polarity of all vortices, i.e., changing $v_{1,2}$ to $-v_{1,2}$. Then the phase bias on each wire, induced by the vortex, will change its polarity. Now let us perform a B-inversion transformation, i.e., change $B$ to $-B$. Then the phase bias on each wire, induced by the Meissner current, changes its sign. Finally, perform I-inversion, i.e., change $I$ to $-I$. Then the phase induced on each wire by all possible causes will change sign. Yet, the 
 CPR for each wire is assumed to be antisymmetric. Therefore, the maximum magnitude of the current that can be injected into the device is an invariant of these three transformations if they are performed together. Thus, the VSRs are said to obey $IBV$ symmetry. Note that the borders of a VSR represent the maximum possible value of the supercurrent flowing through the device. If the task is to find the critical current envelope, i.e., the maximum critical current, $I_{c,+}$, for a given magnetic field, then a maximization over all possible vorticity values is done. In this case, it is not necessary to perform V-inversion. Therefore, while VSR, defined for a fixed vorticity, obeys $IBV$-symmetry, the $I_{c,\pm}(B)$ functions obey the $IB$ symmetry.


For identical, equidistant nanowires, the $I_{c,\pm}(B)$ curves are analogous to ordinary SQUIDS, which are analogous to optical diffraction grating patterns having two slits, with maximums at integer values of the normalized magnetic field $b$ and minima occurring at half integer $b$ values. Note that in analogy to the diffraction grating, the $b$ value effectively controls the phase increment from one wire to the next one. 


The minimum value of the absolute value of the critical current is $0$ if all critical phases are $\pi/2$. (The critical phase cannot be less than $\pi/2$.) If the critical phases of the nanowire are above $\pi/2$, the VSRs will take up more space on the I-B plane and overlap. Therefore, the absolute value of the critical current will be above zero at any magnetic field, and the 100 percent modulation of the critical current does not happen, like in regular symmetric SQUIDs. On the other hand, overlapping VSRs allow for programmable superconducting memory devices, which require more than one metastable state.

\section{\label{sec:3-wire}Symmetrical Three-Nanowire SQUID (3-SQUID): Stability Regions}
\begin{figure*}[htbp]
    \centering
    \subfigure{%
        \includegraphics[width=0.32\linewidth]{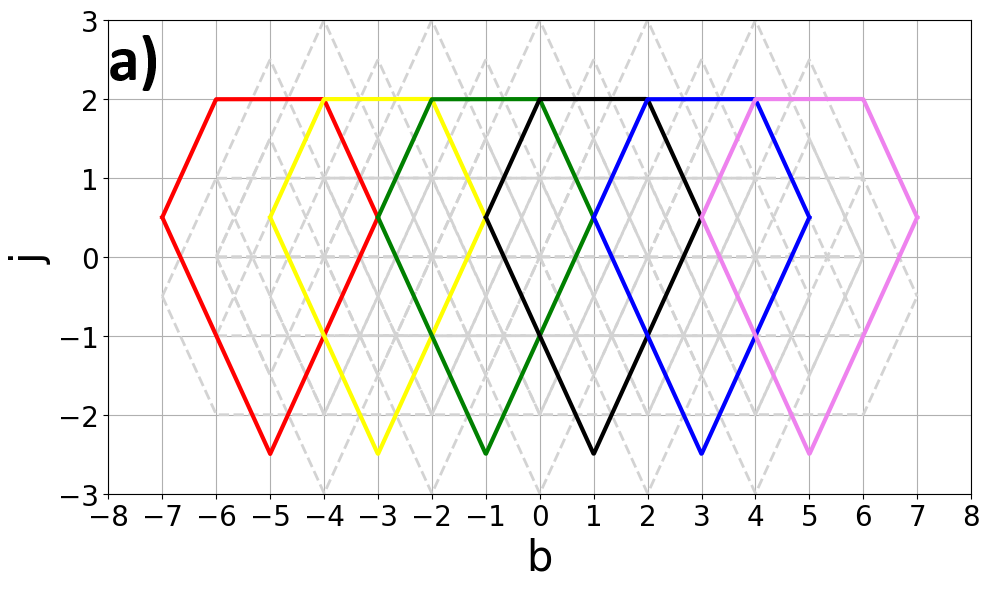}
    }
    \hfill
    \subfigure{%
        \includegraphics[width=0.32\linewidth]{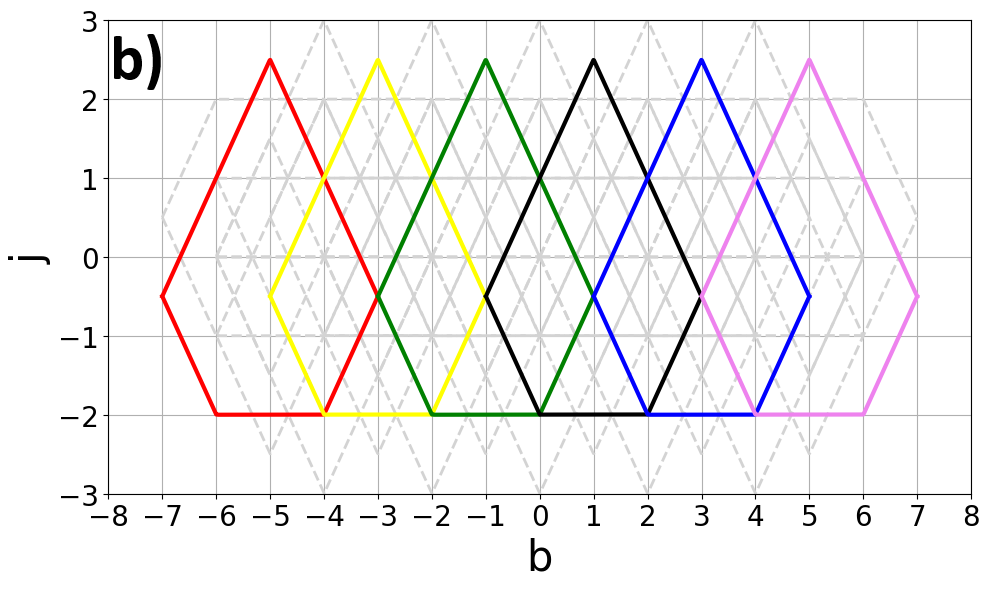}
    }
    \hfill
    \subfigure{%
        \includegraphics[width=0.32\linewidth]{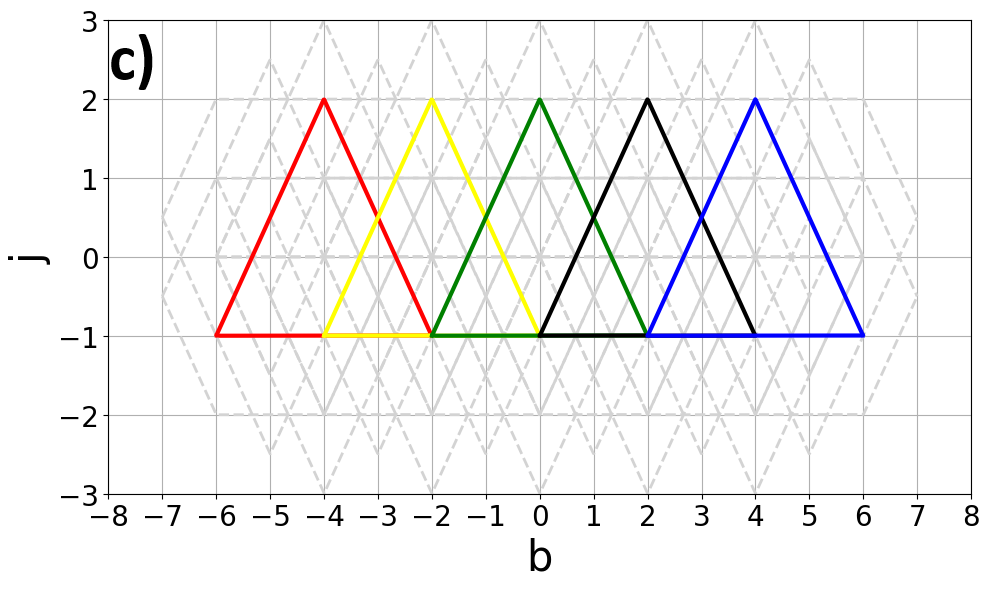}
    }
    \hfill
    \subfigure{%
        \includegraphics[width=0.32\linewidth]{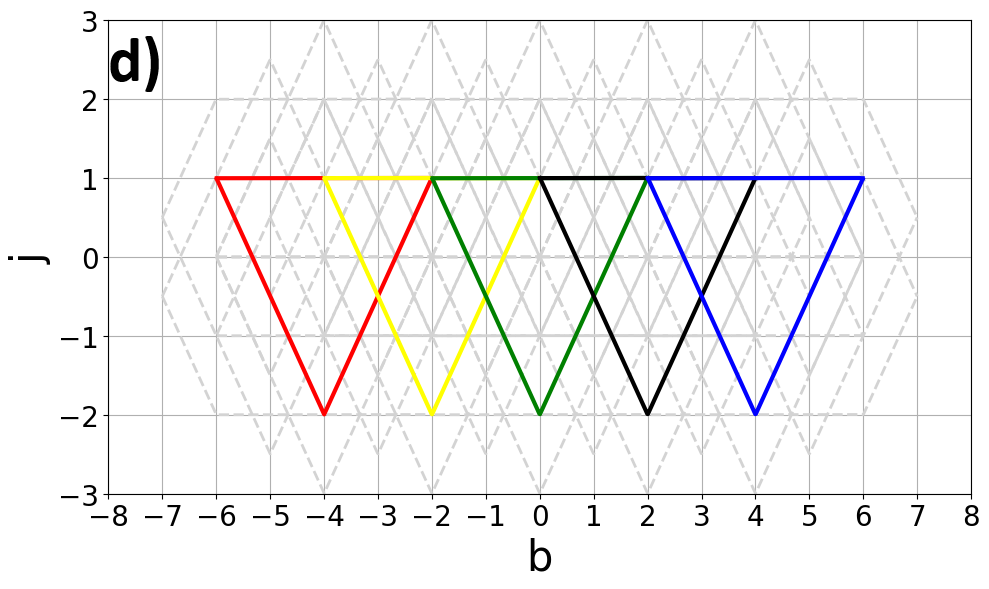}
    }
    \hfill
    \subfigure{%
        \includegraphics[width=0.32\linewidth]{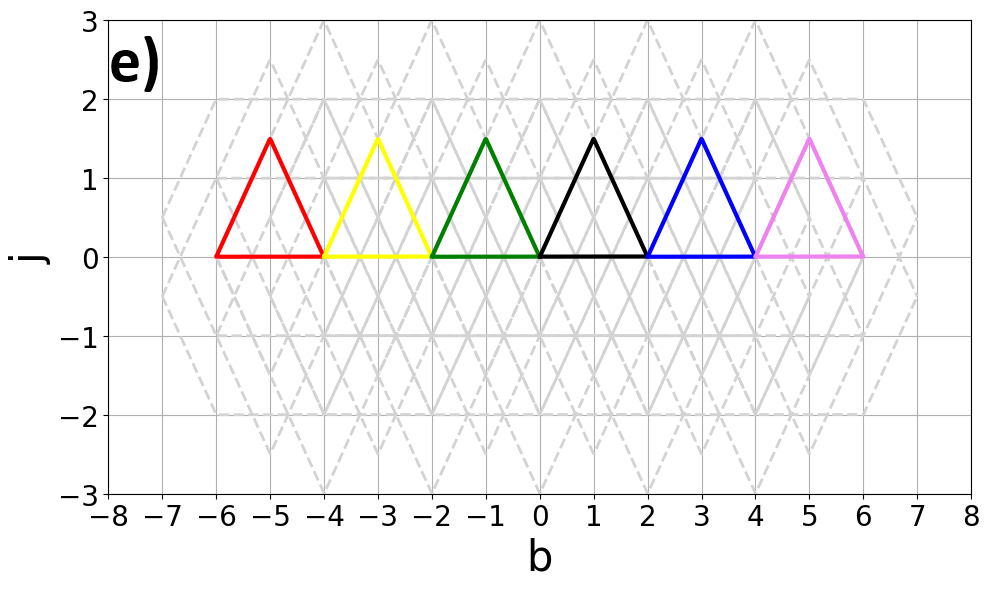}
    }
    \hfill
    \subfigure{%
        \includegraphics[width=0.32\linewidth]{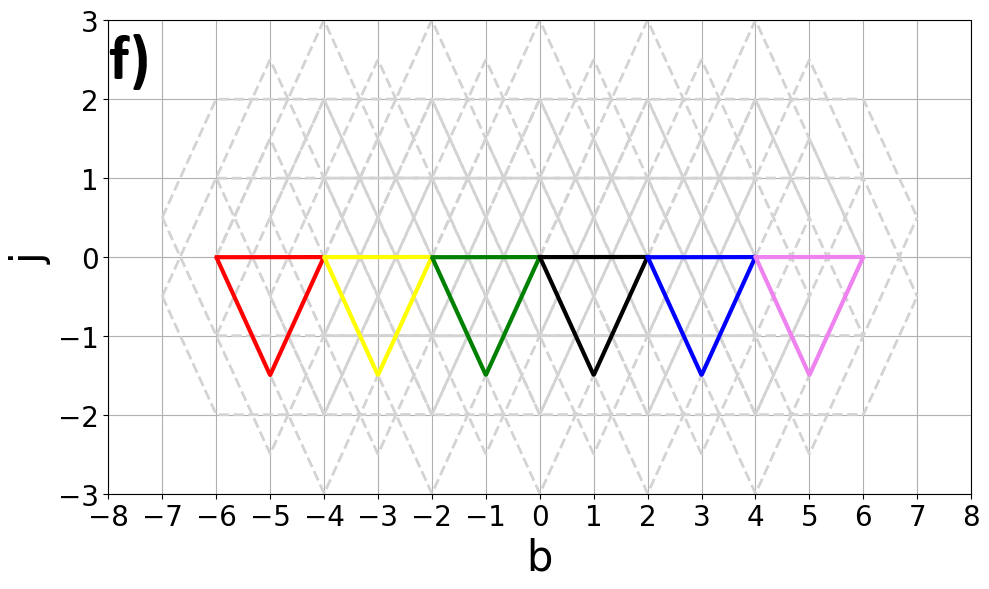}
    }
    \subfigure{%
        \includegraphics[width=0.32\linewidth]{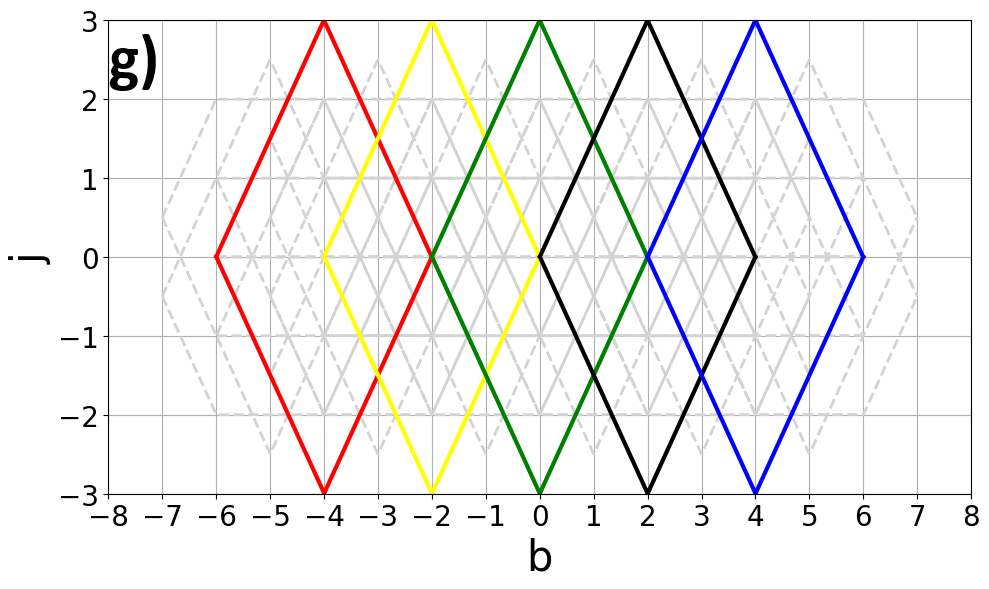}
    }
\end{figure*}
\begin{figure*} [t!]
  \caption{Vorticity stability regions (VSR) for a device containing three equidistant nanowires, the coordinates being 0, 0.5, and 1. The x-axis is the normalized magnetic field, and the y-axis is the normalized total bias current flowing through the device. The critical phases are set to \(2\pi\) and the critical currents are set to 1 for all three wires. Note that the gray dashed lines in all figures indicate the boundaries of various possible stability regions. (a) Flat-top diamond VSR. Such correspond to the following vortex configurations: $[-3, -2]$, $[-2, -1]$, $[-1, 0]$, $[0, 1]$, $[1, 2]$, and $[2, 3]$ from left to right. In this case, the left loop has one fewer vortex than the right loop. (b) Flat-bottom diamond VSR. Such occur with the following vorticity configurations $[-2, -3]$, $[-1, -2]$, $[0, -1]$, $[1, 0]$, $[2, 1]$, and $[3, 2]$, et cetera, corresponding to the stability regions, shown from the left to the right. (c) Large flat-bottom triangular VSR. Such correspond to the following vortex configurations: $[-1, -3]$, $[0, -2]$, $[1, -1]$, $[2, 0]$, and $[3, 1]$, from the left to the right. (d) Large flat-top triangular VSR. Such correspond to the following vortex configurations: $[-3, -1]$, $[-2, 0]$, $[-1, 1]$, $[0, 2]$, and $[1, 3]$, appearing from left to right. (e) Small flat-bottom triangular VSR. Such correspond to the following vortex configurations: $[-1, -4]$, $[0, -3]$, $[1, -2]$, $[2, -1]$, $[3, 0]$, and $[4, 1]$, from left to right. Such configurations represent a perfect superconducting diode such that the negative critical current is zero while the positive critical current is peaked. (f) Small flat-top triangular VSR. Such correspond to the following vortex configurations: $[-4, -1]$, $[-3, 0]$, $[-2, 1]$, $[-1, 2]$, $[0, 3]$, and $[1, 4]$ from left to right. The positive critical current is zero, while the negative critical current reaches the absolute value of 1.5 at its peak. (g) Diamond-shaped VSR. Such correspond to the following vortex configurations: $[-2, -2]$, $[-1, -1]$, $[0, 0]$, $[1, 1]$, and $[2, 2]$, from left to right. These VSRs occur if each loop contains the same number of vortices. The maxima occur if the normalized magnetic flux is equal to the vorticity value in each cell.}
  \label{fig:3-wire-stability-regions}
\end{figure*}
\begin{figure*}[htbp]
    \centering
    \includegraphics[width=0.32\linewidth]{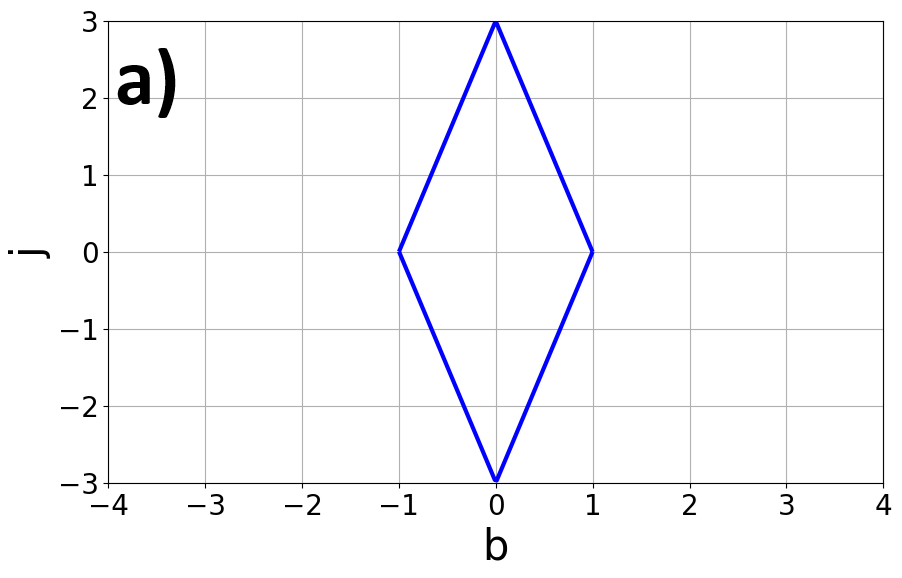}
    \includegraphics[width=0.32\linewidth]{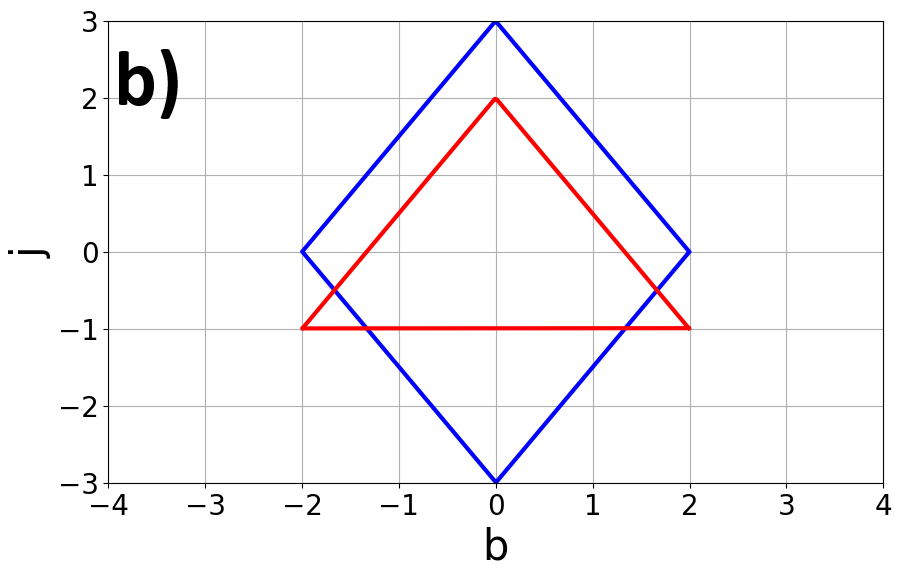}
    \includegraphics[width=0.32\linewidth]{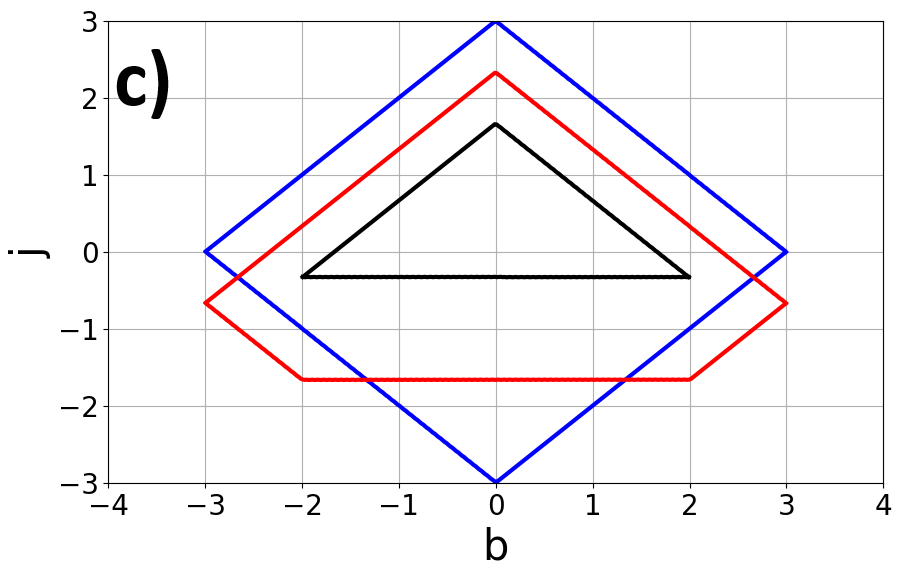}
    \caption{Increase in metastable states as the critical phase homogeneously increases for three identical, equidistant nanowires. We define $\big|v_{2,3}-v_{1,2}\big|$ as the absolute difference in the number of fluxons between the first superconducting loop (cell) and the second cell. (a) Critical phases are set to $\pi$, the only VSR present is the diamond shape (blue curve) with $\Delta v_{(2,3) - (1,2)} = 0$. (b) Critical phases are set to $2\pi$, the diamond VSR (blue curve) with $\Delta v_{(2,3) - (1,2)} = 0$ and triangular VSR (red curve) with $\Delta v_{(2,3) - (1,2)} = 2$ manifest. (b) Critical phases are set to $3\pi$, the diamond VSR with $\Delta v_{(2,3) - (1,2)} = 0$ (blue curve), the flat-top diamond VSR (red curve) with $\Delta v_{(2,3) - (1,2)} = 2$, and the triangular VSR (black curve) with $\Delta v_{(2,3) - (1,2)} = 4$ manifests. Increasing the critical phase from $2\pi$ to $3\pi$ transformed the triangular shape of $\Delta v_i = 2$ to a flat-bottom diamond. The area for each VSR increases on the $B-I$ plane. }
    \label{fig:3-wire-c_p-v_n}
\end{figure*}
We proceed with the analysis of MW-SQUIDs containing three parallel nanowires. Such a device contains two superconducting loops (two "cells"). The position of the second nanowire can now affect the symmetry of the device as well as the relative sizing of the first and second superconducting cells. This new degree of freedom drastically changes the behavior of the 3-wire MW-SQUID, as compared to that of the well-understood 2-nanowire SQUID. This section focuses on classifying the stability regions of the 3-nanowire SQUID. 
\begin{figure}[b]
    \centering
    \includegraphics[width=0.9\linewidth]{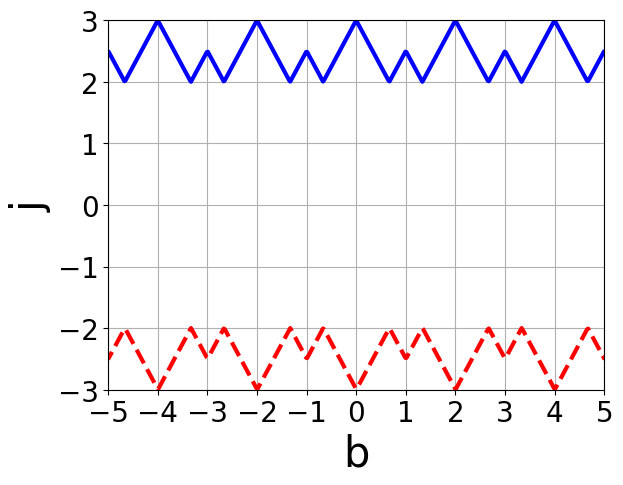}
    \caption{Critical current curve for three equidistant, identical superconducting nanowires. All critical phases are set to $2\pi$. The critical current curves (blue and red) showcase a pattern that mimics the three slit optical diffraction pattern.}
    \label{fig:3-wire-diffraction}
\end{figure}

In this section, we assume that the nanowires are identical and equidistant. Thus, the device is symmetrical with respect to a 180-degree rotation around the y-axis. As always, the device is also symmetrical with respect to the 180-degree rotation around the x-axis. Since the nanowires are equidistant therefore their normalized coordinates are $x_1=0$, $x_2=0.5$, and $x_3=1$. For 3-nanowire SQUIDs, our linear CPR model predicts that the stability regions are no longer all rhombic in shape. Possible VSRs produced by the three identical equidistant nanowires are limited to the following list: rhombus shapes (diamonds) (Fig.\ref{fig:3-wire-stability-regions}g), flat top and flat bottom diamonds, big triangles and smaller triangles, as shown in Fig.\ref{fig:3-wire-stability-regions}(a-g). 
 If either the current or the magnetic field is shifted outside the stability region, then the current becomes larger than the critical current in at least one of the wires. In this case, a phase slip must happen, which can, with some probability, switch the device to the normal state. So the borders of each VSR represent a critical current of the device.

For the 3-SQUID considered, which has three identical and equidistant nanowires, the shape of each VSR is uniquely determined by the difference between the number of fluxons in the 2nd cell and the 1st cell. As always, we define $v_{1,2}$ (always an integer) to be the number of fluxoids in the first cell and $v_{2,3}$ the number of fluxoids in the second cell.  

In Fig.\ref{fig:3-wire-stability-regions}g, one can see that if $\big|v_{2,3}-v_{1,2}\big| = 0$, then the VSR has a diamond (rhombic) shape. We note in passing that even for devices with more than three nanowires, the diamond-shaped shape vorticity stability region occurs if and only if all cells have the same number of trapped vortices. Now we continue our discussion of the VSRs produced by the 3-wire-SQUID (Fig.\ref{fig:3-wire-stability-regions}) where $\phi_{c,i} = 2\pi$. The model predicts that if $\big|v_{2,3}-v_{1,2}\big| = 1$ then the VSR shape is a flat-top diamond (which has the largest number of vertices, specifically five vertices), as shown in Fig.\ref{fig:3-wire-stability-regions}(a,b). If $\big|v_{2,3}-v_{1,2}\big| = 2$, then the VSR is a large triangle, as shown in Fig.\ref{fig:3-wire-stability-regions}(c,d). When $\big|v_{2,3}-v_{1,2}\big| = 3$, then the VSR shape is a small triangle, as shown in Fig.\ref{fig:3-wire-stability-regions}(e,f). VSRs with $\big|v_{2,3}-v_{1,2}\big| > 3$ do not occur since for such VSR any combination of $I$ and $B$ values would produce currents which exceed the critical current in at least one of the three nanowires. However, increasing the magnitude of all critical phases allows higher vorticity differences to manifest on the I-B plane, as shown in Fig. \ref{fig:3-wire-c_p-v_n}. In general, for 3 nanowires of the same critical phases, the shapes of the generated VSRs are regular diamond shapes, flat-top/flat-bottom diamonds of various sizes (depending on the magnitude of the critical phases and the vorticity differences between the loops), and triangular shapes of various sizes (depending on the magnitude of the critical phases and the vorticity differences between the loops).
Based on these observations, one concludes that adding the same number of fluxons to each cell does not change $\big|v_{2,3}-v_{1,2}\big|$ and therefore does not change the VSR shape. Suppose we add $k=\text{integer}$ vortices to each cell. Then the VSR shifts to the right by $2k$. 

Diamond VSRs are produced when the number of fluxons trapped in each cell is the same for all cells. In this case, the current generated in the middle wire by the vortex located in the left cell cancels the current generated by the vortex present in the right cell. Now, to reduce the total current in the left wire and the right wire to zero (assuming zero bias current), the normalized magnetic flux $b$ must be equal to the total number of vortices in the entire device. Therefore, the top vertex of each diamond VSR occurs at $b=v_{1,3}$ where
$v_{1,3} = v_{1,2} + v_{2,3}$ is the total number of fluxons in the device. This is shown in Fig.\ref{fig:3-wire-stability-regions}g. 

Our model predicts a perfect superconducting diode in the SQUIDS containing three nanowires. This means that the critical current on one polarity is large, while the critical current of the opposite polarity is zero. It is illustrated in Fig.\ref{fig:3-wire-stability-regions}(e,f) where $\phi_{c,i} = 2\pi$, where the small triangular VSRs ($\big|v_{2,3} - v_{1,2}\big| = 2$) exhibit a flat side coinciding with the $I=0$ axis. In this case, its minimum ($I_{c,-}$) is at zero current, and its maximum ($I_{c,+}$) normalized current reaches 1.5 at its peak. Such a perfect superconducting diode (rectifier) might have applications in superconducting memory as a useful Boolean device. A significant diode effect is also present in the large triangular VSR, shown in Fig.\ref{fig:3-wire-stability-regions}(c,d), where the magnitude of its maximum is twice as large as its minimum.

\section{\label{sec:3-wire}Three-Nanowire SQUID: Critical current versus magnetic field}

When plotting all possible VSRs for 3 identical, equidistant superconducting nanowires, the maximum envelope curve of all VSRs is defined as the positive critical current ($I_{c,+}(b)$) and the minimum envelope curve of all VSRs is the negative critical current ($I_{c,-}(b)$). Both of these curves can be found in Fig.\ref{fig:3-wire-diffraction}, where the blue curve is $I_{c,+}(b)$ and the red dotted curve is $I_{c,-}(b)$. (All the critical phases are set to $2\pi$.) The minimum of the diffraction curve is $I_{c,+} \approx 2$ and the maximum is $I_{c,+} \approx 3$, which is a $33\%$ modulation depth. For a 3-wire SQUID, the principal maxima of the critical current occur when the Meissner phase difference between neighbor wires is $2\pi * \text{integer}$, which happens if $b=2*\text{integer}$. There are also secondary maxima in the critical current function, which occur if the Meissner phase difference between the first wire and the last wire is $2\pi*\text{integer}$, which is the case if $b=2*\text{integer}+1$. 

Note that in this respect the critical current curves are similar to the intensity pattern of an optical diffraction grating (ODG) with three slits. With an ODG, principle maxima also occur when the phase shift of the light waves propagating through neighbor slits is an integer multiple of $2\pi$. Thus, if ODG has three slits, then, for the first principle maximum, for example, the phase shift between the first slit and the last (third) slit is $4\pi$. In this case, a secondary maximum occurs if the phase shift between the first and the last slit is $2\pi$. 

In the case of a 3-wire SQUID, the maximum peaks are produced by the diamond stability regions, and the smaller peaks are produced by the flat top/bottom diamond stability regions. The periodicity of this curve is the result of the periodicity of the ensemble of all types of VSRs, which is the result of the Little-Parks effect.

\begin{figure}[t]
    \centering
    \includegraphics[width=0.49\linewidth]{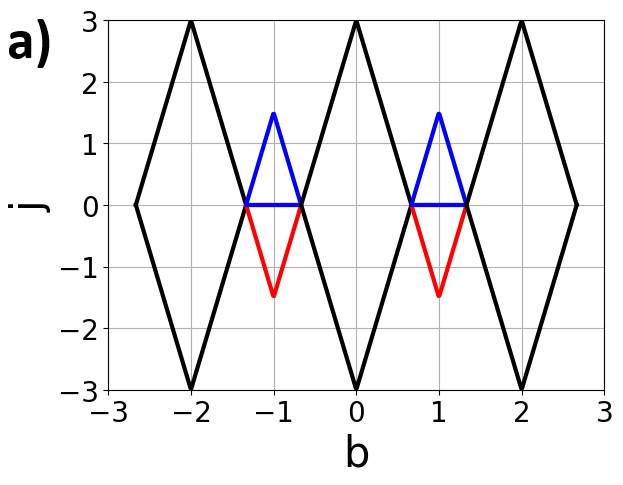}
    \includegraphics[width=0.49\linewidth]{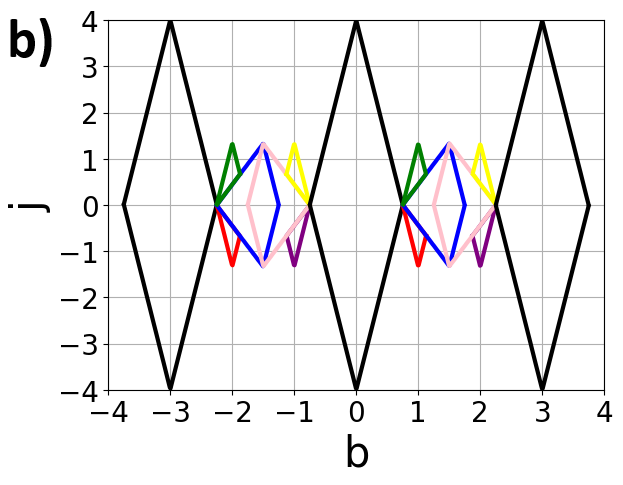}
    \caption{Examples of vorticity stability regions for a three-wire SQUID (a) and a four-wire SQUID (b). Both examples correspond to symmetrical devices in which all wires are identical. (a) Three equidistant, identical nanowires, all of critical phases set to $\frac{2}{3}\pi$, critical currents set to $1$. The black diamond shapes represent vorticity states of type $(m,m)$. Here $m$ is an integer. The period of the pattern is $\Delta b=2$. The blue shapes represent $(m+1,m)$, and the red shape represent $(m,m+1)$. (b) VSR regions of an ideal SQUID containing four equidistant, identical nanowires, all of the critical phases $\frac{3}{4}\pi$. The critical currents of each wire are set to $1$. The black diamond shapes represent $(m,m,m)$ vorticity, the red shapes represents $(m,m,m+1)$, the blue shapes represent $(m,m+1,m)$ vorticity, the green shapes represent $(m+1,m,m)$, the purple shapes represent $(m-1,m,m)$, the pink shapes represent $(m,m-1,m)$, and the yellow shapes represent $(m,m,m-1)$ vorticity. The period of the sequence is $\Delta b=3$. }
    \label{fig:disjoint-VSR-equidistant}
\end{figure}

For 3 identical, equidistant superconducting nanowires, the minimum of the critical current pattern hits zero when all the critical phases are $\frac{2}{3}\pi$ as shown in Fig. \ref{fig:disjoint-VSR-equidistant}a. Thus, if the critical phase is less than  $\frac{2}{3}\pi$, then a 100\% supercurrent modulation can be achieved. If $\max(\phi_{c,i}) < \frac{2}{3}\pi$, then the VSRs start becoming disjoint, and quantum transitions will occur at zero temperature. This means that if the magnetic field is increased, the system will transition between superconducting and normal states (the one having zero critical current), even if the temperature is zero. On the other hand, programmable superconducting memory devices can be developed when there is an overlap between VSRs. Such a situation is achieved if the critical phase of the wires is larger than $\frac{2}{3}\pi$. In this case, the state of the system is history dependent, which is the essence of the memory effect.
\section{\label{sec:3-wire-disorder}Three-Nanowire SQUID: Effect of Disorder}

In real systems, it is impossible to make the nanowires identical. Therefore, in this section, we study how the conclusions change if some variation of the parameters is introduced. We applied (1) critical phase disorder, (2) critical current disorder, (3) position disorder, and (4) critical phase + critical current + position disorder to a 3-SQUID. We analyzed how disorder affects all different types of VSRs for $\big|v_{1,2} - v_{2,3}\big| = 0,1,2,3$. Note that the shape of the VSR is defined by the absolute value of the difference of the number of vortices present in the first loop and the second loop (cell) of the considered 3-wire SQUID, as was concluded in the previous sections.

\begin{figure*}[t]
    \centering
    \subfigure{%
        \includegraphics[width=0.24\linewidth]{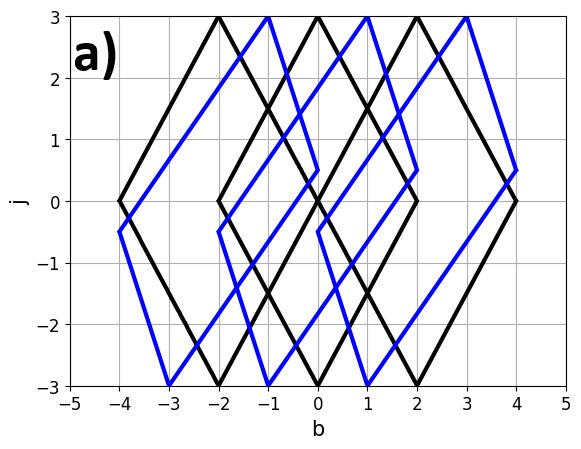}
    }
    \hfill
    \subfigure{%
        \includegraphics[width=0.24\linewidth]{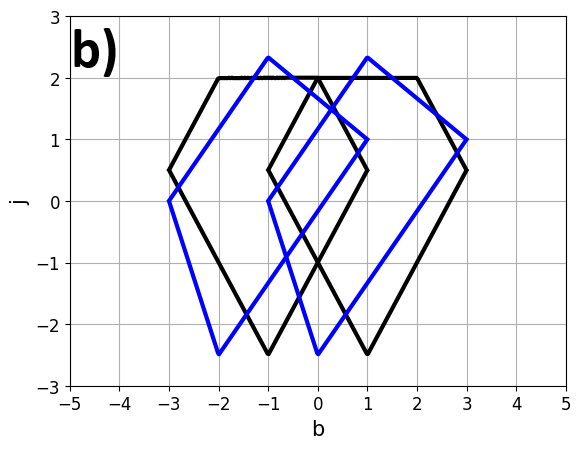}
    }
    \hfill
    \subfigure{%
        \includegraphics[width=0.24\linewidth]{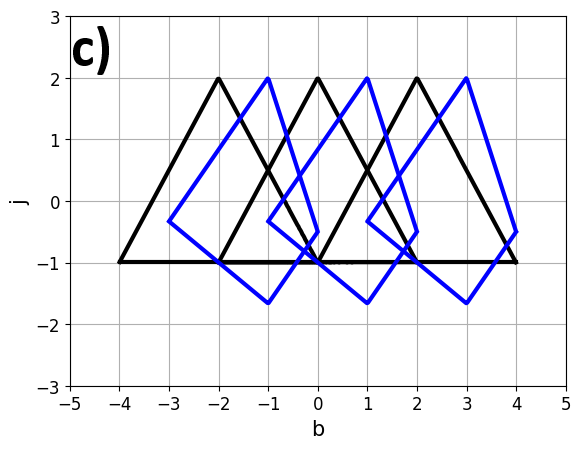}
    }
    \hfill
    \subfigure{%
        \includegraphics[width=0.24\linewidth]{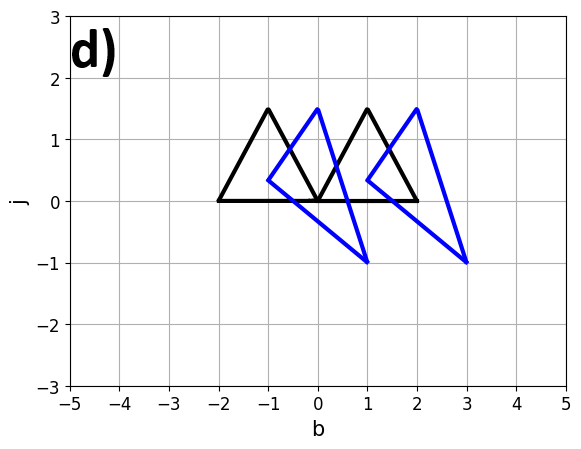}
    }
    \caption{Phase disorder for three equidistant nanowires with critical currents set to 1. In the figure, a symmetric device (black curve) corresponds to \(\phi_c = [2\pi, 2\pi, 2\pi]\). A disordered device (blue curve) gives \(\phi_c = [\pi, 2\pi, 3\pi]\). We apply the phase disorder analysis to the following stability region shapes:
        (a) Diamond-shaped regions with vortex configurations $[-1, -1][0,0][1, 1]$
        (b) Flat-top diamond regions with vortex configurations $[-1,0], [0,1]$
        (c) Big triangular regions with vortex configurations $[0,-2], [1,-1], [2,0]$
        (d) Small triangular regions with vortex configurations $[1, -2][2, -1]$ A difference of $2\pi$ between the critical phases of the 1st and 3rd nanowire results in a shift of $\Delta B = 1$ in the peaks of each VSR, if there is one. 
        }
    \label{fig:3-wire-phase_disorder}
\end{figure*}

Firstly, we applied a phase disorder to a system of 3 identical equidistant nanowires. We analyzed how the phase disorder affected diamond VSRs (Fig.\ref{fig:3-wire-phase_disorder}a), flat-top diamond VSRs (Fig.\ref{fig:3-wire-phase_disorder}b), triangular VSRs (Fig.\ref{fig:3-wire-phase_disorder}c), and small triangular VSRs (Fig.\ref{fig:3-wire-phase_disorder}d). The mean critical phase was fixed to $2\pi$. A phase disorder in the form of $[\pi, 2\pi, 3\pi]$ was applied, while keeping the critical currents all the same, $j_c=1$. The wires are assumed to be equidistant. The periodicity of each type of VSR is still preserved along the b-axis. The period is unchanged. Yet the shape of each VSR becomes distorted. We find that, for any vorticity difference, the top vertex shifts horizontally by 1 and the bottom vertex shifts horizontally by -1. While the horizontal shift is the same for all vorticity differences, the vertical shift is not. In the case of $0$ and $1$ vorticity difference (Fig.\ref{fig:3-wire-phase_disorder}(a,b)), the left vertex shifts vertically by $-0.5$ and the right vertex shifts vertically by $0.5$. For $\big|v_{1,2} - v_{2,3}\big| = 2$ vorticity difference (Fig.\ref{fig:3-wire-phase_disorder}c), the left vertex shifts by $0.75$ and the right vertex shifts by $0.5$. For the $\big|v_{1,2} - v_{2,3}\big| = 3$ vorticity difference (Fig.\ref{fig:3-wire-phase_disorder}d), the left vertex shifts by $0.25$ and the right vertex shifts by $-1$. This is unlike the 2-nanowire system (Fig.\ref{fig:2-wire-disorder}), where applying a phase disorder only shifted the top and bottom vertices of each VSR.

\begin{figure}[b]
    \centering
    \includegraphics[width=0.48\linewidth]{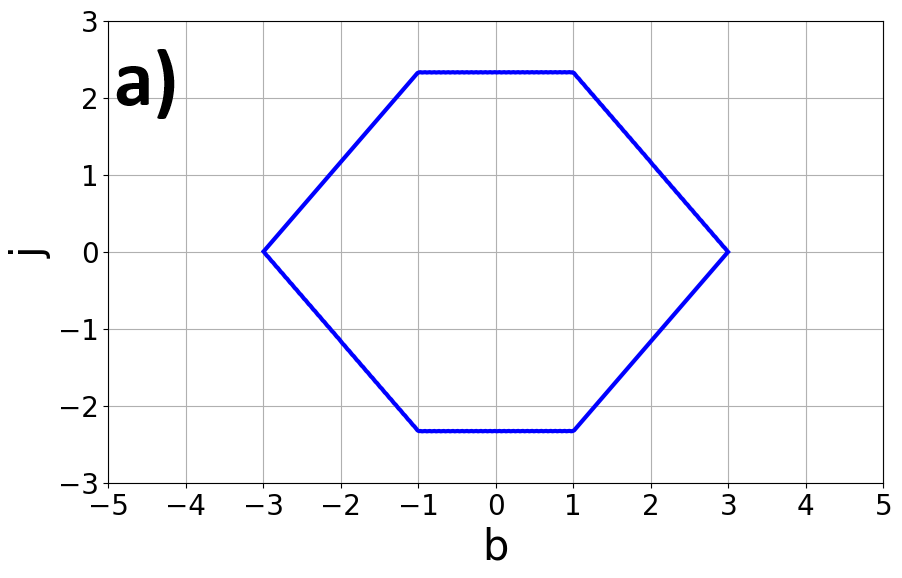}
    \includegraphics[width=0.48\linewidth]{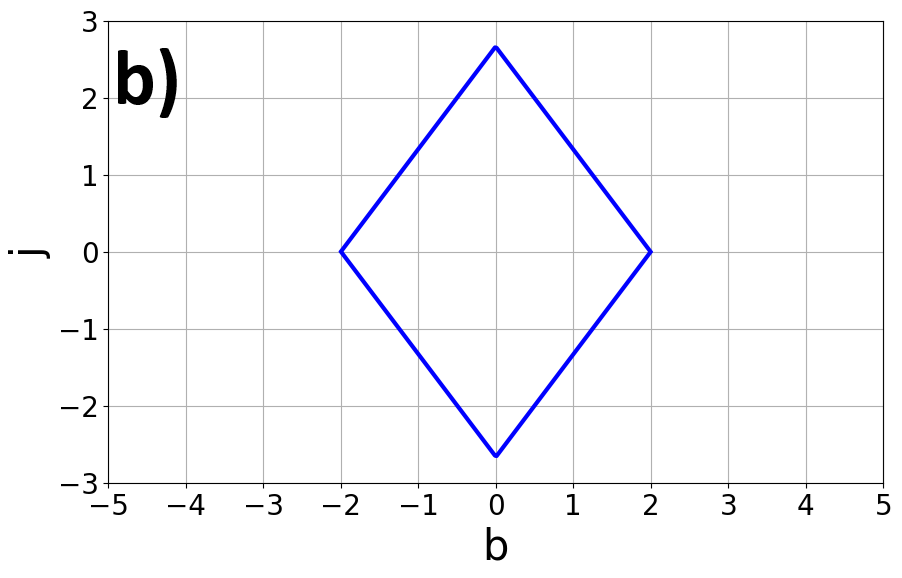}
    \caption{Flat top regime for the zero vorticity state for three identical, equidistant superconducting nanowires. In (a), the critical phases are $[3\pi, 2\pi, 3\pi]$. The flat top regime manifests because the central nanowire switches first. In (b), the critical phases are $[2\pi, 3\pi, 2\pi]$. The flat top regime does not manifest because the central nanowire does not switch first.}
    \label{fig:3-wire-flat-region}
\end{figure}

One of the interesting phenomena that occurs in systems with phase disorder is the flat-top VSR (different from the flat-top diamond VSR), the physical meaning of which is that the critical current does not change with magnetic field, in some finite range of $b$ values. One can obtain such a result using a simple calculation, as follows. Consider a device in which the critical phase of the middle wire is lower, for example $\phi_c=[3\pi,2\pi,3\pi]$, and assume that there are no vortices, $v_{1,2}=v_{2,3}=0$. The current is $j=\phi_1/3\pi+\phi_2/2\pi+\phi_3/3\pi$. The Meissner phase imposed by the electrodes leads to the following dependence of the phase bias of the wires, $\phi_2 = \phi_1 + 2\pi b(x_2 - x_1) $ and $\phi_3 = \phi_1 + 2\pi b(x_3 - x_1) $. Assume the coordinates of the wires are $x=[0,0.5,1]$. Then $\phi_2 = \phi_1 + \pi b$ and $\phi_3 = \phi_1 + 2\pi b$. The total normalized current is $j=\phi_1/3\pi+(\phi_1 + \pi b)/2\pi+(\phi_1 + 2\pi b)/3\pi$. Now we can express the total current through the phase difference on the second wire, which has the lowest critical value. Take into account $\phi_1 = \phi_2 - \pi b$. The current then is $j=(\phi_2 - \pi b)/3\pi+\phi_2/2\pi+(\phi_2  + \pi b)/3\pi$. The magnetic field dependence cancels out as $j=\phi_2/3\pi+\phi_2/2\pi+\phi_2 /3\pi=7\phi_2/6\pi$. If the bias current is increased, then $\phi_2$ will increase until it reaches its critical value, i.e., until $\phi_2=\phi_{c,2}$. At this moment, the current will be the critical current of the device $j_c(b)=7\phi_{c,2}/6\pi$. Notice that it does not depend on the magnetic field. This is in agreement with our numerical model as shown in Fig.\ref{fig:3-wire-flat-region}a. This flat top has a finite range. The derivation above assumes that the superconductor-normal transition happens in the middle wire, which is true as long as $b$ is sufficiently low so that $\phi_3 = \phi_2  + \pi b<\phi_{c,3}$. This inequality must hold for any value of $\phi_2$ up to its critical value, i.e., $\phi_{c,2}  + \pi b<\phi_{c,3}$. Only if the latter condition is satisfied, the middle wire switches first as $j$ is ramped up. This condition breaks down if the magnetic field exceeds $b_v=(\phi_{c,3}-\phi_{c,2})/\pi$. The field $b_v$ defines the position of the right vertex of the VSR. If $b>b_v$, then the right wire reaches the critical phase faster than the middle wire, due to the contribution from the Meissner phase. Thus, the right wire would undergo the switching event first, as the bias current is increased. As always, we assume that if one wire switches to the normal state, then the entire device switches to the normal state. The vertex location calculated above is also in agreement with the calculation as shown in Fig.\ref{fig:3-wire-flat-region}a. 

This proof can also be generalized to devices with critical phases of the form $[\phi_c, \phi_{c,2}, \phi_c]$. If $\phi_{c,2} < \phi_c$, then the flat top region does occur, as shown in Fig.\ref{fig:3-wire-flat-region}a. If $\phi_{c,2} > \phi_c$, then the flat top region does not occur, as shown in Fig.\ref{fig:3-wire-flat-region}b, for a device with critical phases of $[2\pi, 3\pi, 2\pi]$. 
\begin{figure*}[t]
    \centering
    \subfigure{%
        \includegraphics[width=0.24\linewidth]{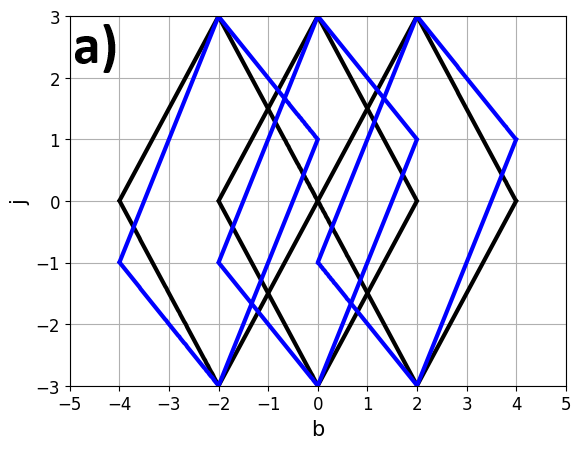}
    }
    \hfill
    \subfigure{%
        \includegraphics[width=0.24\linewidth]{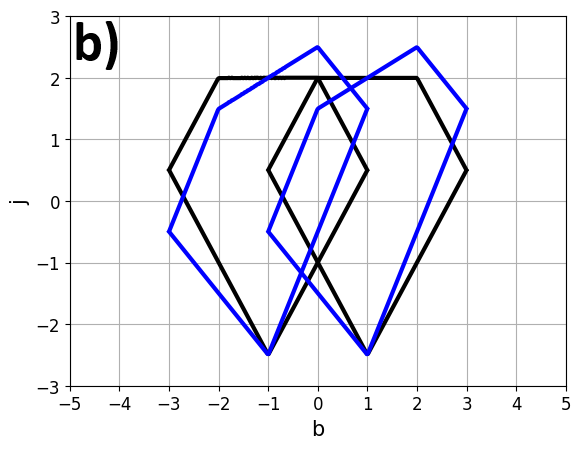}
    }
    \hfill
    \subfigure{%
        \includegraphics[width=0.24\linewidth]{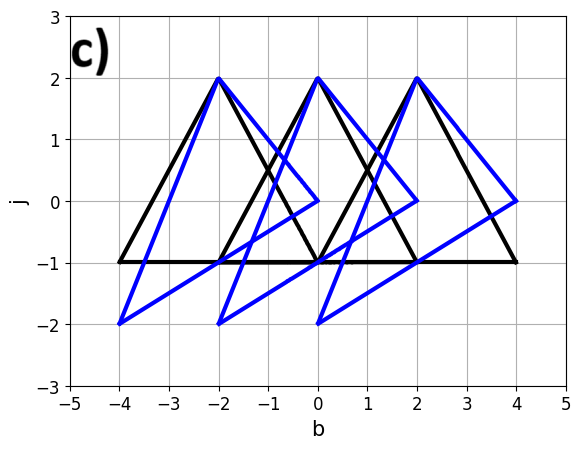}
    }
    \hfill
    \subfigure{%
        \includegraphics[width=0.24\linewidth]{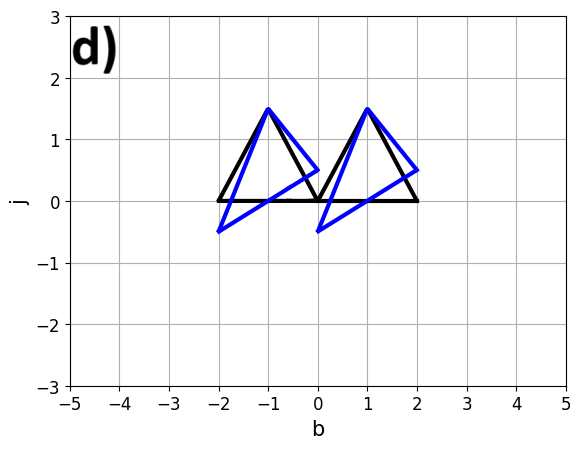}
    }
    \caption{Critical current disorder for three equidistant nanowires (positions: [0, 0.5, 1]) with all critical phases set to 2$\pi$. A symmetric array (blue curve) corresponds to $j_{c}=[1, 1, 1]$.  An asymmetric device (black curve) corresponds to $j_{c}= [0.5, 1, 1.5]$. We apply the critical current disorder analysis to the following stability region shapes: 
        (a) Diamond-shaped regions with vortex configurations $[-1, -1][0,0][1, 1]$
        (b) Flat-top diamond regions with vortex configurations $[-1,0], [0,1]$
        (c) Big triangular regions with vortex configurations $[0,-2], [1,-1], [2,0]$
        (d) Small triangular regions with vortex configurations $[1, -2][2, -1]$
    }
    \label{fig:3-wire-current_disorder}
\end{figure*}
The pattern of VSR is periodic due to the Little-Parks effect. Assume that there is one vortex in each cell, that is, $v_{1,2}=v_{2,3}=1$. As well, assume that the wires are equidistant, that is $x_1= 0$, $x_2 =0.5$, and $x_3 = 1$. Then the phase shifts generated by these vortices can be compensated by shifting the magnetic field by 2. Indeed, if $b=2$ we have $\phi_2 = \phi_1 +2\pi b(x_2-x_1)- 2\pi v_{1,2} = \phi_1 + 2\pi - 2\pi v_{1,2}$ and $\phi_3 = \phi_1 + 2\pi b(x_3-x_1)- 2\pi (v_{1,2}+v_{2,3}) = \phi_1 + 4\pi - 2\pi (v_{1,2}+v_{2,3})$. Now remember that we consider the case when each vorticity equals one, then we get $\phi_2 = \phi_1 + 2\pi-2\pi=\phi_1$. We obtain $\phi_3=\phi_1$ through a similar analysis. This is the essence of the Little-Parks effect: If each cell has an integer number of vortices, say $m$, then we can choose an integer normalized magnetic field $b=2m$ such that the phase shift between the wires caused by the vortices exactly compensate the phase shift generated by the Meissner currents \cite{little_parks}. This is the reason why VSRs are a periodic sequence of identical shapes, with the period being $\Delta b=2$. It is 2, not 1, because there are two cells. If the number of cells were $k$, then the period would be $\Delta b=k$. The derivation above assumes all cells are identical.

 Next, as an example, we calculate analytically the position of the top vertex of the VSR in a 3-wire SQUID with phase disorder. We also assume here that the critical currents of all wires are equal. Assume also that there are no vortices in each cell. For this calculation, we number the wires such that $\phi_{c,1}\leq \phi_{c,2}\leq \phi_{c,3}$. If the bias current of the device ($j$) is ramped up from zero, at $b=0$, then the wire-1 switches to the normal state first. As it happens, the entire device switches to the normal state. (There might be a situation when a phase slip occurs and the vorticity changes so that the device state changes to a different VSR.) The reason why the wire-1 switches first is that at zero field, the phase bias is the same for all wires, and wire-1 has the lowest critical phase, according to the numbering scheme adopted in this section. If the magnetic field is sufficiently small, then the wire with the lowest critical phase would still switch first. Yet, there will be some critical magnetic field value above which some other wire will start switching to the normal state first as the bias current of the devices ($j$) is increased from zero up. The critical field at which two wires reach their critical phase simultaneously defines the vertex of the VSR. Next, we calculate this critical field analytically, as an example. 
Generally speaking, there is a phase correlation between the wires imposed by the electrodes. If the magnetic field is zero, then, as discussed above, all wires have the same phase difference since the phase of the superconducting condensate wavefunctions is the same within each electrode. If the magnetic field is not zero, then the phase correlation between the phase biases for the j-th and i-th wires is: $\phi_j = \phi_{i} + 2\pi b(x_j - x_i) - 2\pi v_{i,j}$. Let us simplify the discussion further in this example and assume zero vorticity in both loops ($v_{1,2}=v_{2,3}=0$). According to this, if the first wire is in the critical state then $\phi_1=\phi_{c,1}$, $\phi_2 = \phi_{c,1} + 2\pi b(x_2 - x_1)$, and $\phi_3 = \phi_{c,1} + 2\pi b(x_3 - x_1)$. Such phase bias state is only possible if $\phi_2<\phi_{c,2}$ and $\phi_3<\phi_{c,3}$. At some critical value of the magnetic field, these conditions break down. The second wire (wire-2) will be switching simultaneously with the wire-1 if $b$ is such that $\phi_{c,2} = \phi_{c,1} + 2\pi b(x_2 - x_1)$, i.e., if $b=b_{1,2} \equiv (\phi_{c,2} - \phi_{c,1})/(2\pi(x_2 - x_1))$. Here $b_{1,2}$ is the field at which wire-1 and wire-2 switch simultaneously (provided that wire-3 is not switched). Same way, we can calculate the field at which wire-1 and wire-3 switch simultaneously (provided that wire-2 is not switched). The critical field for wires-1 and wire-3 switching simultaneously (as the bias current is ramped up from zero) is $b_{1,3}=(\phi_{c,3} - \phi_{c,1})/(2\pi(x_2 - x_1))$. This condition defines the magnetic field at which, if wire-1 reaches the critical phase, then the phase bias of wire-3 is also at its critical value. Then the VSR vertex location field $b_v$ is the minimum of the two critical fields defined above, namely $b_v=\min(b_{1,2},b_{1,3})$.
This analytical calculation is in agreement with our numerical computation (see, for example, Fig.\ref{fig:3-wire-phase_disorder}a). From this formula, it can be concluded that if $\phi_{c,1} = \phi_{c,2} \leq \phi_{c,3}$, then the top vertex is at $b=0$. Conversely, if $\phi_{c,3} = \phi_{c,2} > \phi_{c,1}$, then the top vertex is shifted from $b=0$.

Secondly, we applied a critical current disorder to the system of 3 identical superconducting nanowires. We analyzed how the critical current disorder affected diamond VSRs (Fig.\ref{fig:3-wire-current_disorder}a), flat-top diamond VSRs (Fig.\ref{fig:3-wire-current_disorder}b), triangular VSRs (Fig.\ref{fig:3-wire-current_disorder}c), and small triangular VSRs (Fig.\ref{fig:3-wire-current_disorder}d). The mean normalized critical current is fixed to 1 by definition. Physically speaking, changing the critical current of a nanowire is equivalent to altering its width. A critical current disorder in the form of $[0.5, 1, 1.5]$ was applied, while keeping all critical phases fixed to $2\pi$, and wires were kept equidistant. Similar to phase disorder, the system with the critical current disorder remains periodic, the period being equal to 2, see Fig.\ref{fig:3-wire-current_disorder}(a-d). For all vorticity differences, the left vertex shifts vertically by $-1$, and the right vertex shifts vertically by $1$. As predicted from our derived formula from the above paragraph, the top and bottom vertices remained fixed for every VSR (if there is a vertex) because all critical phases are identical. As well, in VSRs with a flat-top or flat-bottom, the central point of this flat region remained fixed, see Fig.\ref{fig:3-wire-current_disorder}(c,d). The horizontal positions of all vertices are determined by the critical phase and not by the critical current.
\begin{figure*}[t]
    \centering
    \subfigure{%
        \includegraphics[width=0.24\linewidth]{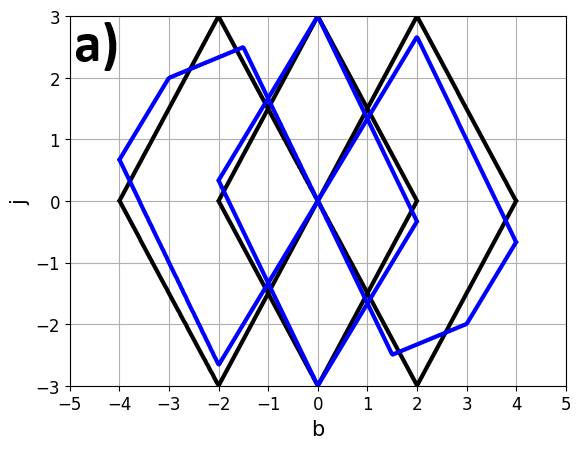}
    }
    \hfill
    \subfigure{%
        \includegraphics[width=0.24\linewidth]{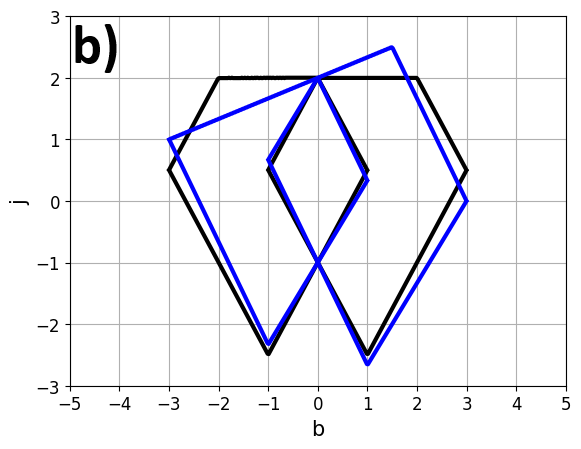}
    }
    \hfill
    \subfigure{%
        \includegraphics[width=0.24\linewidth]{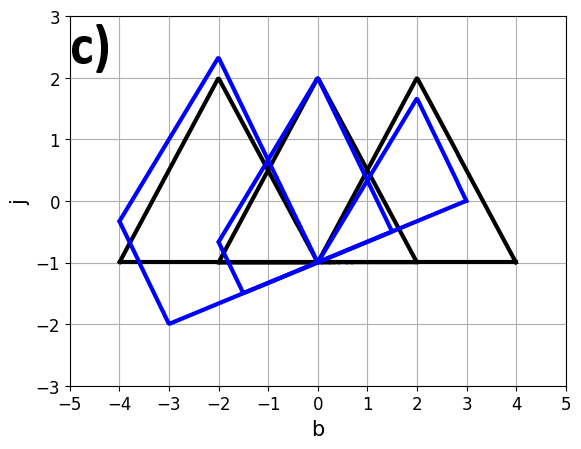}
    }
    \hfill
    \subfigure{%
        \includegraphics[width=0.24\linewidth]{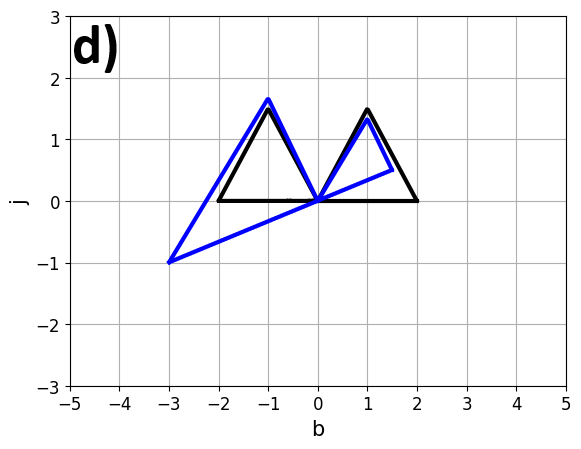}
    }
    \caption{Position disorder for three nanowires with periodicity breaking. All critical phases are 2$\pi$, and all critical currents are 1. A symmetric device (black curve) corresponds to the wire geometry of $[0, 0.5, 1]$. An asymmetric device (blue curve) corresponds to the wire geometry of $[0, 1/3, 1]$. We apply the position disorder analysis to the following stability region shapes: 
        (a) Diamond-shaped regions with vortex configurations $[-1, -1][0,0][1, 1]$
        (b) Flat-top diamond regions with vortex configurations $[-1,0], [0,1]$
        (c) Big triangular regions with vortex configurations $[0,-2], [1,-1], [2,0]$
        (d) Small triangular regions with vortex configurations $[1, -2][2, -1]$
    }
    \label{fig:3_wire_position_disorder}
\end{figure*}

\begin{figure*}[t]
    \centering
    \subfigure{%
        \includegraphics[width=0.24\linewidth]{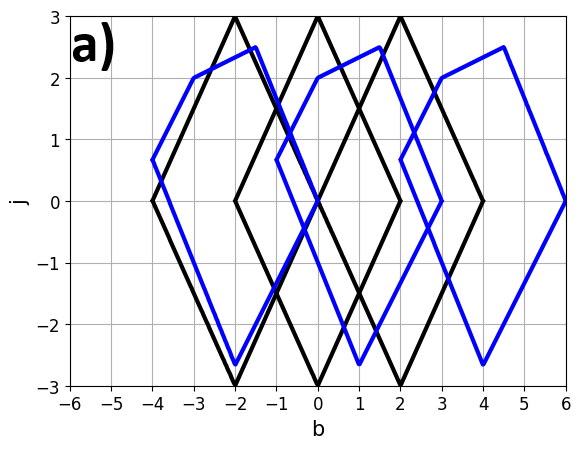}
    }
    \hfill
    \subfigure{%
        \includegraphics[width=0.24\linewidth]{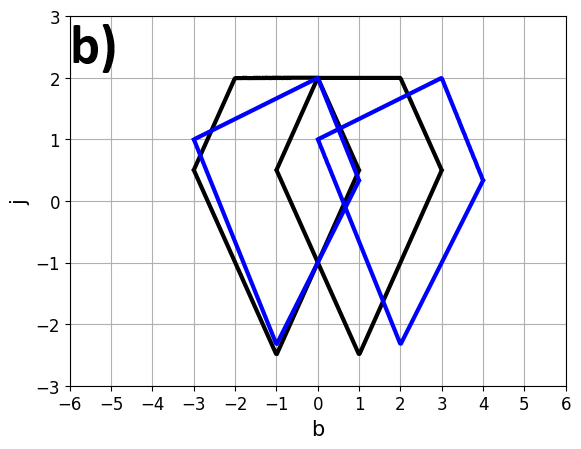}
    }
    \hfill
    \subfigure{%
        \includegraphics[width=0.24\linewidth]{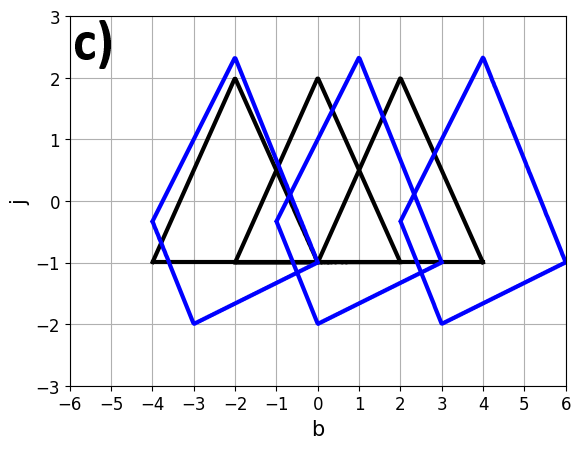}
    }
    \hfill
    \subfigure{%
        \includegraphics[width=0.24\linewidth]{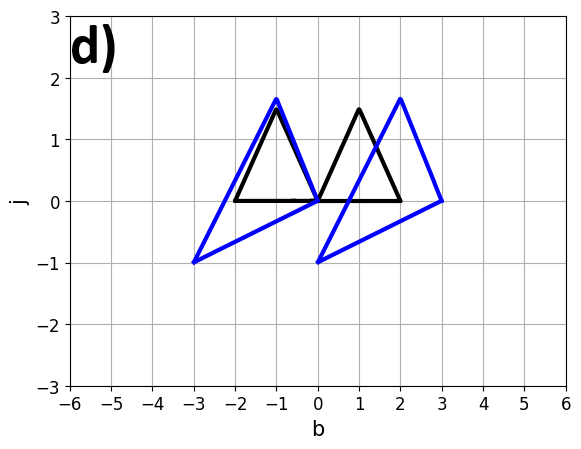}
    }
    \caption{Position disorder for three nanowires with periodicity restored. All critical phases are 2$\pi$, and all critical currents are 1. A symmetric device (black curve) corresponds to the wire geometry of $[0, 0.5, 1]$. The period of its VSRs is 2. An asymmetric device (blue curve) corresponds to the wire geometry of $[0, 1/3, 1]$. The new period of its VSRs is 3. We show the change in periodicity of each VSR when applying a position disorder:
        (a) Diamond-shaped regions with vortex configurations $[-1, -1],[0,0],[1, 1]$ (black). Distorted flat-top diamond regions with vortex configurations $[-1,-1], [0,1], [1,3]$ (blue).
        (b) Flat-top diamond regions with vortex configurations $[-1,0], [0,1]$ (black). Trapezoidal regions with vortex configurations $[-1,0], [0,2]$ (blue).
        (c) Big triangular regions with vortex configurations $[0,-2], [1,-1], [2,0]$ (black). Distorted trapezoidal regions with vortex configurations $[0,-2], [1,0], [2,2]$ (blue).
        (d) Small triangular regions with vortex configurations $[1, -2][2, -1]$ (black). Distorted triangular regions with vortex configurations $[1,-2], [2,0]$ (blue). 
    }
    \label{fig:3_wire_position_disorder_periodic}
\end{figure*}

Thirdly, we applied a position disorder to the system of three identical superconducting nanowires. We analyzed how the position disorder affects the ideal case shapes, namely the diamond VSRs (Fig.\ref{fig:3_wire_position_disorder}a), flat-top diamond VSRs (Fig.\ref{fig:3_wire_position_disorder}b), triangular VSRs (Fig.\ref{fig:3_wire_position_disorder}c), and small triangular VSRs (Fig.\ref{fig:3_wire_position_disorder}d). A position disorder of the form $[0, 1/3, 1]$ was used (blue curve), while fixing all critical phases to $2\pi$ and critical currents to be $1$. 
Unlike the phase disorder and the current disorder, here we find a qualitatively new phenomenon, namely that the periodicity of the shapes appears broken. This means that VSRs of each type (corresponding to a certain value of $\big|v_{1,2} - v_{2,3}\big|$) no longer have the same shape. However, the periodicity of the VSRs can be restored by adding a fluxon arrangement of $[1,2]$ to any given vorticity state. This means adding $1$ fluxon to the first cell and $2$ fluxons to the second cell. This can be seen in Fig.\ref{fig:3_wire_position_disorder_periodic}(a-d). 
This change in the periodicity can be explained by taking into account that shifting the second nanowire increases the size of one of the superconducting loops (cells) and decreases the size of the other. In this case, the Little-Parks effect has to be generalized. We now calculate the new period of a three-nanowire SQUID with different-sized superconducting cells. The positions of these 3 wires are $[0, x_2, 1]$. Then the size of the left cell is $x_2$ and the size of the right cell is $1-x_2$. Suppose the ratio of the cell sizes is $(1-x_2)/x_2=k/m$ is an irreducible ratio and $k$ and $m$ are integers. The generalized Little-Parks effect for a 3-wire SQUID can be formulated as follows. The simple condition when all three wires have the same phase bias occurs if $b=0$ and the vorticity of each loop is zero. Next, we need to find the lowest non-zero magnetic field such that by introducing vortices into the cell, we can again obtain a state in which the phase bias on all the wires is the same. 

\begin{figure*}[t]
    \centering
    \subfigure{%
        \includegraphics[width=0.24\linewidth]{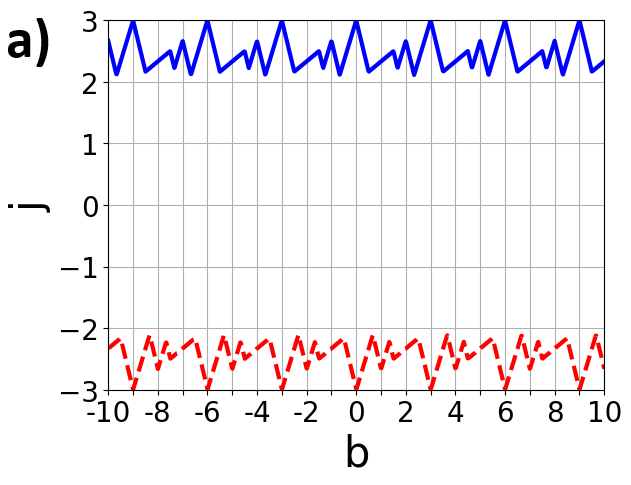}
    }
    \hfill
    \subfigure{%
        \includegraphics[width=0.24\linewidth]{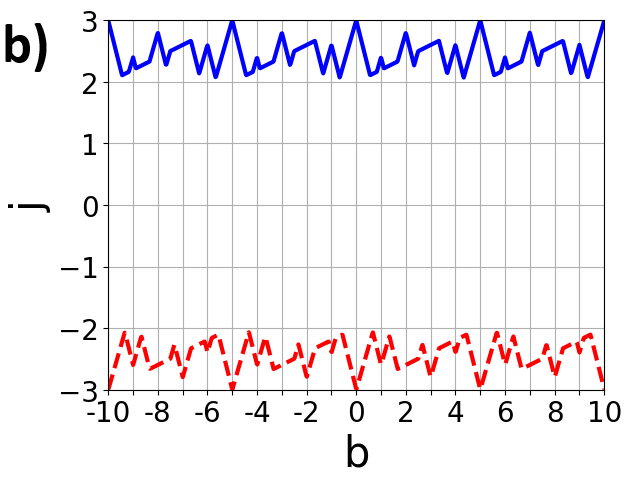}
    }
    \hfill
    \subfigure{%
        \includegraphics[width=0.24\linewidth]{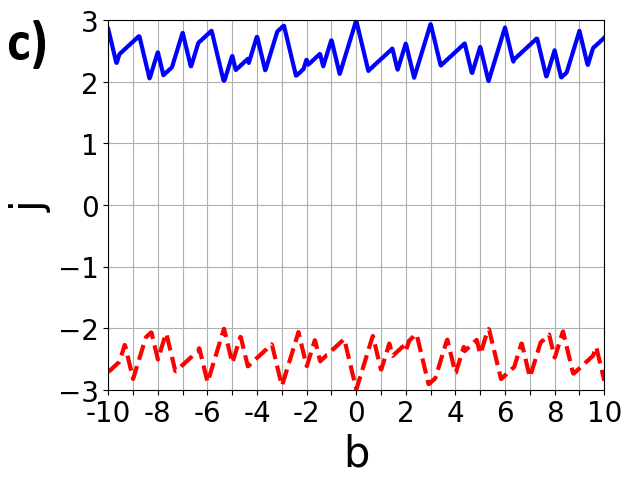}
    }
    \hfill
    \subfigure{%
        \includegraphics[width=0.24\linewidth]{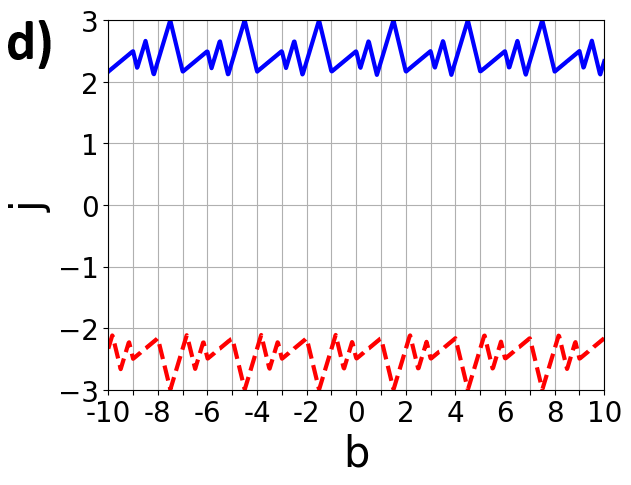}
    }
    \caption{Critical current versus magnetic field for a MW-SQUID based on three identical nanowires, with disordered positions. (a) The coordinates of the wires are 0, 1/3, 1, so the size of the first cell is 1/3 and the size of the second cell is 2/3. Thus, the ratio of their dimensions is $k/m=2$. Thus, $m=1$ and $k=2$. The $I_c(b)\pm$ curves for this device have a period of $m+k=3$. (b) The coordinates of the wires are 0, 0.4, and  1. The second cell is 3/2 times larger than the first cell. Thus $k/m=3/2$. Thus, this device $I_c(b)\pm$ curves have a period of $m+k=5$. (c) The wires' positions are 0, $\pi/10$, 1. The relative sizing of the 2 cells, that is $[1 - \pi/10]/[\pi/10]$, is irrational. Thus, this device's $I_c(b)\pm$ curves are not periodic. (d) Fully disordered system of 3 nanowires. The positions are [0, 1/3, 1]. The critical phases are $[2\pi, \pi, 3\pi]$. The critical currents are $[1, 0.5, 1.5]$. The period of the critical current is 3, although the peaks have shifted due to the applied phase disorder. }
    \label{fig:3_wire_position_disorder_icb}
\end{figure*}

In other words, the period of Little-Parks oscillations would be such that by introducing an integer number of vortices in each loop, we can compensate the Meissner phase shifts produced by the applied magnetic fields. Some complication occurs because the Meissner phase shifts are different in the first cell and the second cell because the cells now have different dimensions. Note that the Meissner current is proportional to $b$ and the Meissner phase shift is proportional to the magnetic field and the distance between the wires, which is what we refer to as the size of the cell formed the the wires. Then the Meissner phase shift in cell 2 is greater than that of cell 1 by exactly the factor $k/m$. To achieve total Meissner current suppression, one must apply a flux of magnitude $k+m$ to the entire device. In this case, the phase shift of the first cell would be proportional to $m/(k+m)$, and the phase shift in the second cell is proportional to $k/(k+m)$. Thus, the ratio is $k/m$. A complete compensation of the Meissner phase shits can be achieved by introducing $m$ vortices in the first cell and $k$ vortices in the second cell. In other words, a vorticity state $[m,k]$ thereby completely reduces the Meissner phase shifts and restores the condition that all wires have the same phase bias. Thus, if the field is $b=m+k$ and the vorticity is $[m,k]$, then the device's physical state is exactly equivalent to its state at zero field and zero vorticity everywhere. This is the essence of the generalized Little-Parks effect. This means that the new period of the device is $\Delta b=(m + k)$. Examples are shown in Fig.\ref{fig:3_wire_position_disorder_icb}a.

If the ratio $(1-x_2)/x_2$ can be presented as a ratio of two integer numbers, i.e., $(1-x_2)/x_2=k/m$, where $k$ and $m$ are the smallest integers that can represent this ratio, then $\Delta b = k+m$ is the smallest period of the pattern of VSR and the critical current versus $b$ function. An example of rational cell sizes is shown in Fig.\ref{fig:3_wire_position_disorder_icb}b. 
It follows that if $(1-x_2)/x_2$ is irrational, then the device is not periodic, as shown in Fig.\ref{fig:3_wire_position_disorder_icb}c. This lack of periodicity is the consequence of the single-valuedness of the wave function in a single superconducting loop, which, in other words, represents the fact that a superconducting loop can only hold an integer number of vortices. As always, it is assumed that the nanowires forming the SQUID are very thin so that a vortex core cannot be stable inside the nanowire itself.

\begin{figure*}[t]
    \centering
    \subfigure{%
        \includegraphics[width=0.24\linewidth]{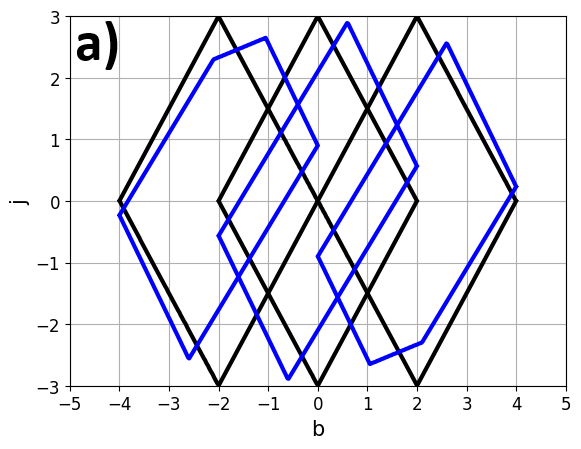}
    }
    \hfill
    \subfigure{%
        \includegraphics[width=0.24\linewidth]{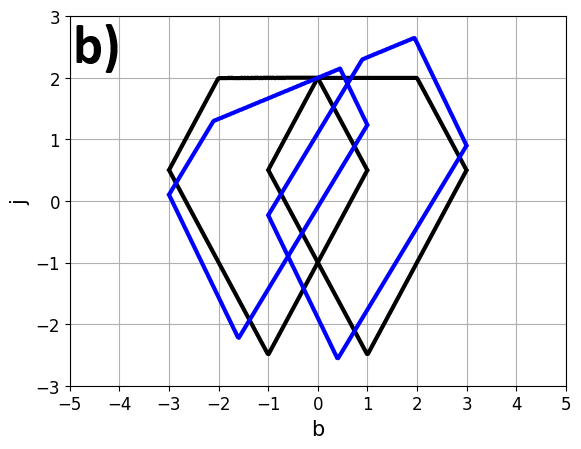}
    }
    \hfill
    \subfigure{%
        \includegraphics[width=0.24\linewidth]{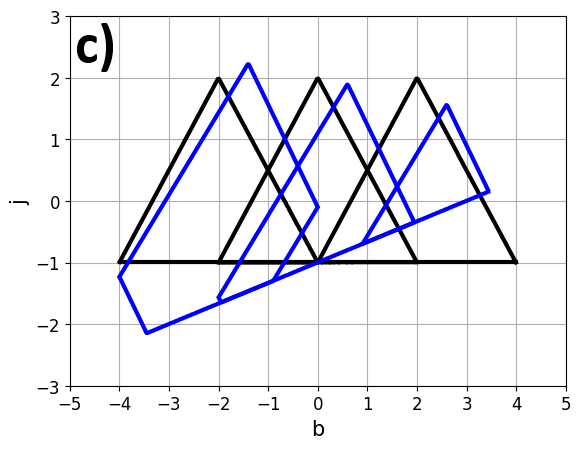}
    }
    \hfill
    \subfigure{%
        \includegraphics[width=0.24\linewidth]{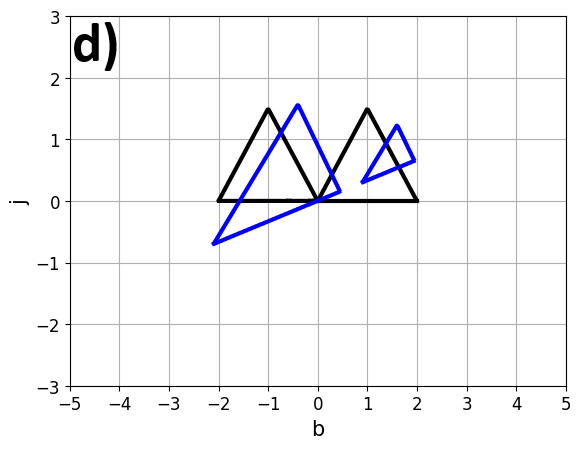}
    }
    \caption{Complete disorder in a 3-SQUID. A symmetric device (black curve) corresponds to a wire geometry of $[0, 0.5, 1]$, critical phases of $[2\pi, 2\pi, 2\pi]$, and critical currents of $[1, 1, 1]$. As an example, we choose the following parameters for the disordered device. The wire locations are $[0, 1/3, 1]$, the critical phases are $[\pi, 2\pi, 3\pi]$, and the critical currents are $[0.5, 1, 1.5]$. We apply the complete disorder analysis to the following stability region shapes: 
        (a) Diamond-shaped regions with vortex configurations $[-1, -1][0,0][1, 1]$
        (b) Flat-top diamond regions with vortex configurations $[-1,0], [0,1]$
        (c) Big triangular regions with vortex configurations $[0,-2], [1,-1], [2,0]$
        (d) Small triangular regions with vortex configurations $[1, -2][2, -1]$
    }
    \label{fig:total_disorder}
\end{figure*}

Lastly, we applied a complete disorder comprising of the phase, current, and position disorder with the same values as in the above individual disorder analysis sections. We analyzed how the complete disorder affected diamond VSRs (Fig.\ref{fig:total_disorder}a), flat-top diamond VSRs (Fig.\ref{fig:total_disorder}b), triangular VSRs (Fig.\ref{fig:total_disorder}c), and small triangular VSRs (Fig.\ref{fig:total_disorder}d). The result of such complete disorder is a linear combination of all the effects from the individual disorder effects discussed above. The critical current curve of completely disordered systems of nanowires is still periodic, with the new period being $m + k$ (see the definitions above). Only the positions of the nanowires change the period of the critical current curves, as shown in Fig.\ref{fig:3_wire_position_disorder_icb}d.

\section{Four-Nanowire SQUID}

First, we discuss the ideal case of a SQUID containing four identical parallel nanowires (4-SQUID), the coordinates being $x_i=[0, 1/3, 2/3, 1]$. Note that here and below, $i$ is the wire number, from left to right. It is found that the diamond-shaped VSR is still a dominant feature, but other types of VSR shapes also occur, depending on the distribution of vortices in the superconducting loops of the device. A system of four identical, equidistant superconducting nanowires produces VSRs of shapes that are qualitatively different and more complex than those of the 3 identical, equidistant nanowire system. For four identical nanowires, each VSR can be uniquely defined by a two-element array $\Delta v_ = [v_{2,3} - v_{1,2}, v_{3,4} - v_{1,2}]$. 

\begin{figure*}[htbp]
    \centering
    \subfigure{%
        \includegraphics[width=0.32\linewidth]{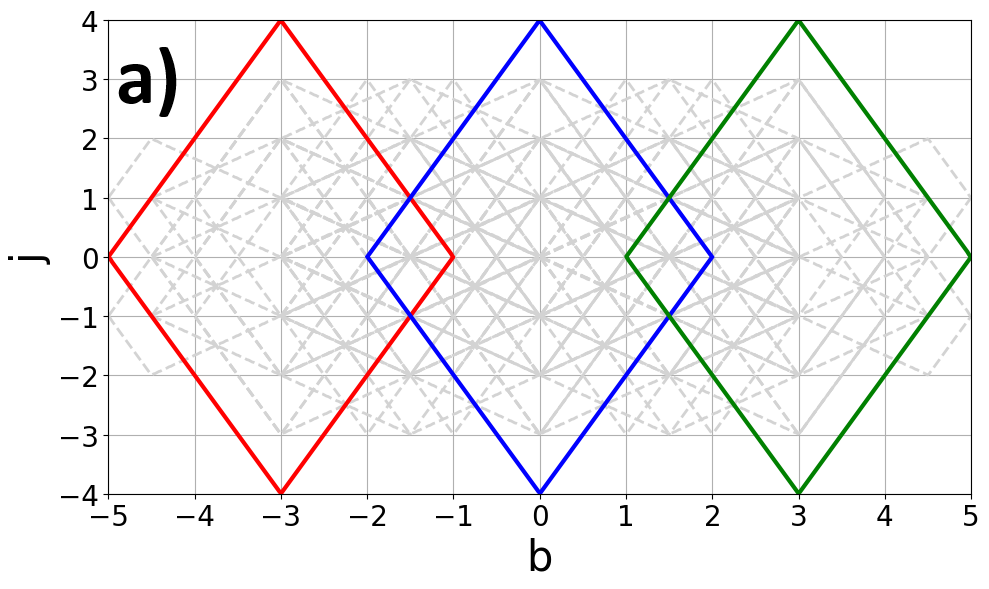}
    }
    \hfill
    \subfigure{%
        \includegraphics[width=0.32\linewidth]{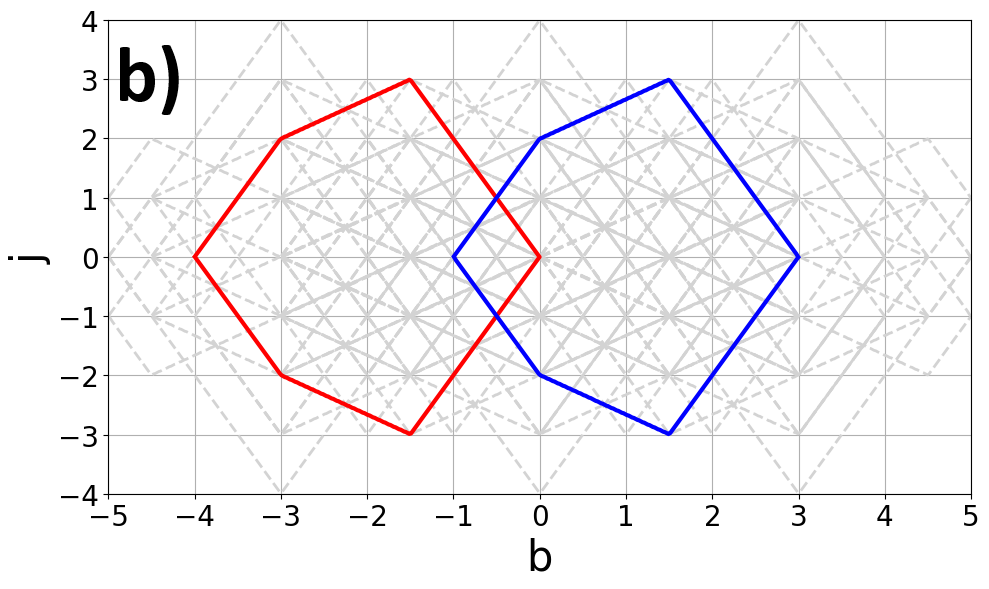}
    }
    \hfill
    \subfigure{%
        \includegraphics[width=0.32\linewidth]{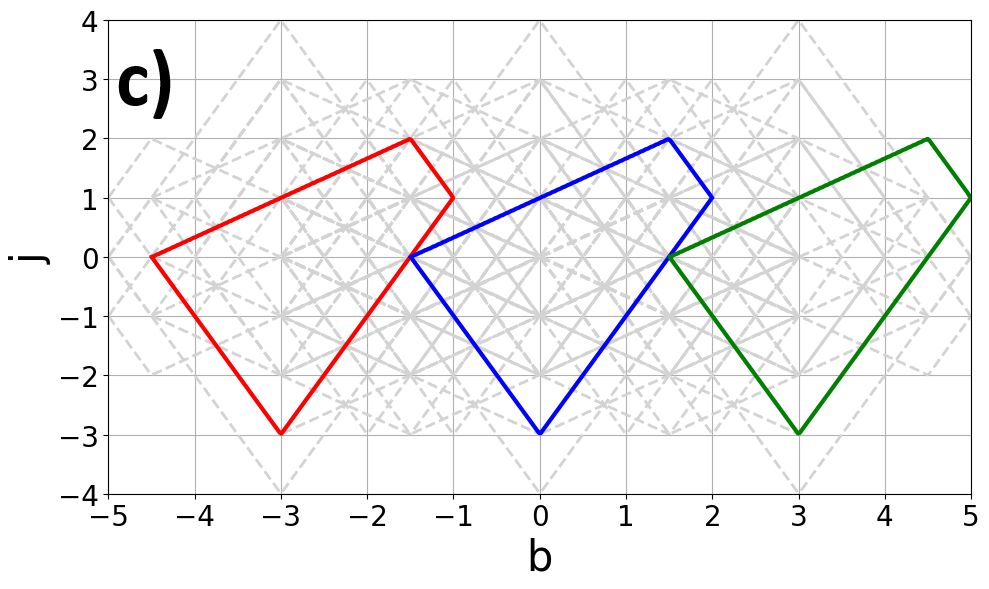}
    }
    \hfill
    \subfigure{%
        \includegraphics[width=0.32\linewidth]{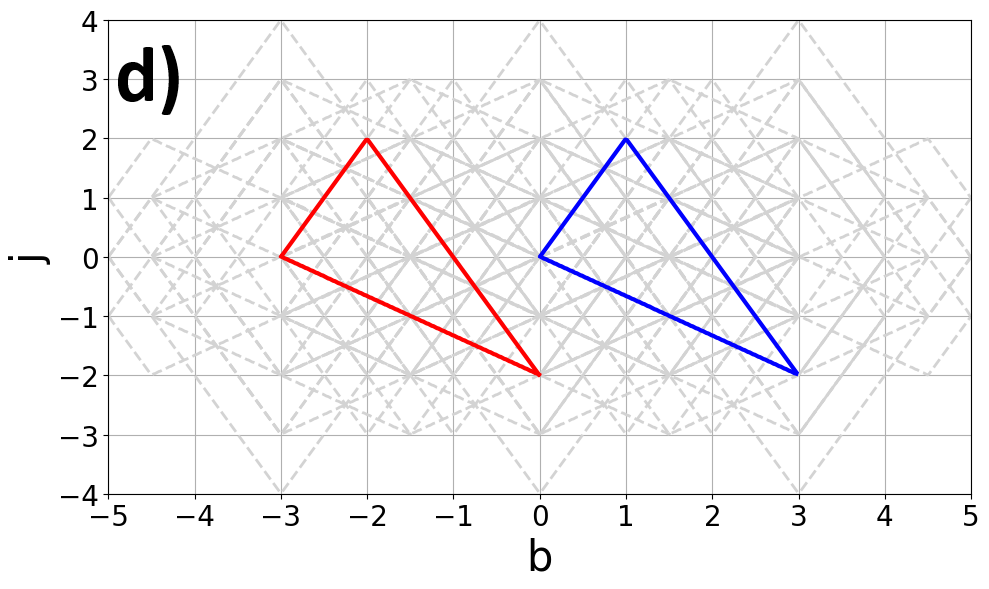}
    }
    \hfill
    \subfigure{%
        \includegraphics[width=0.32\linewidth]{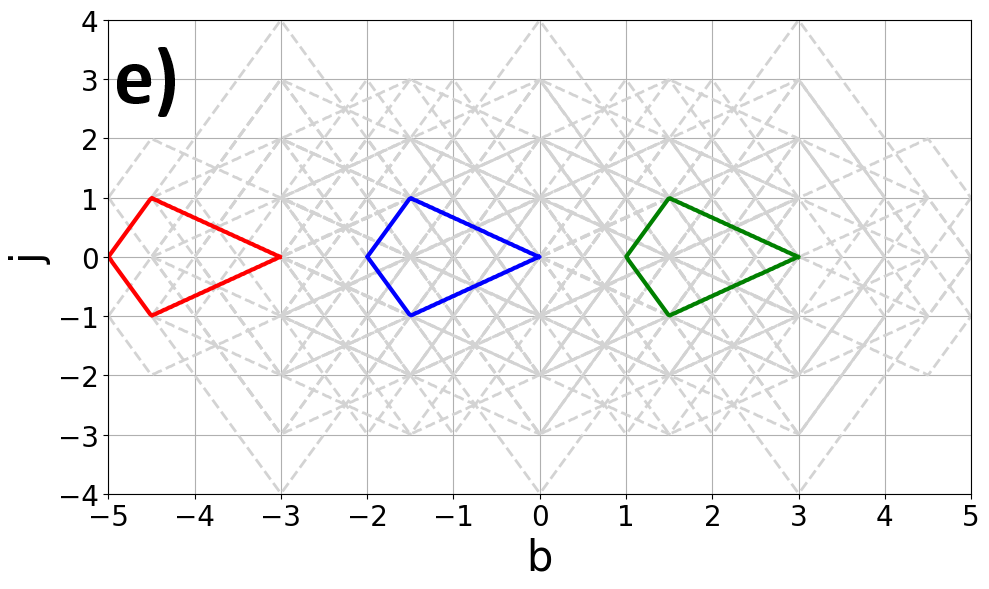}
    }
    \hfill
    \subfigure{%
        \includegraphics[width=0.32\linewidth]{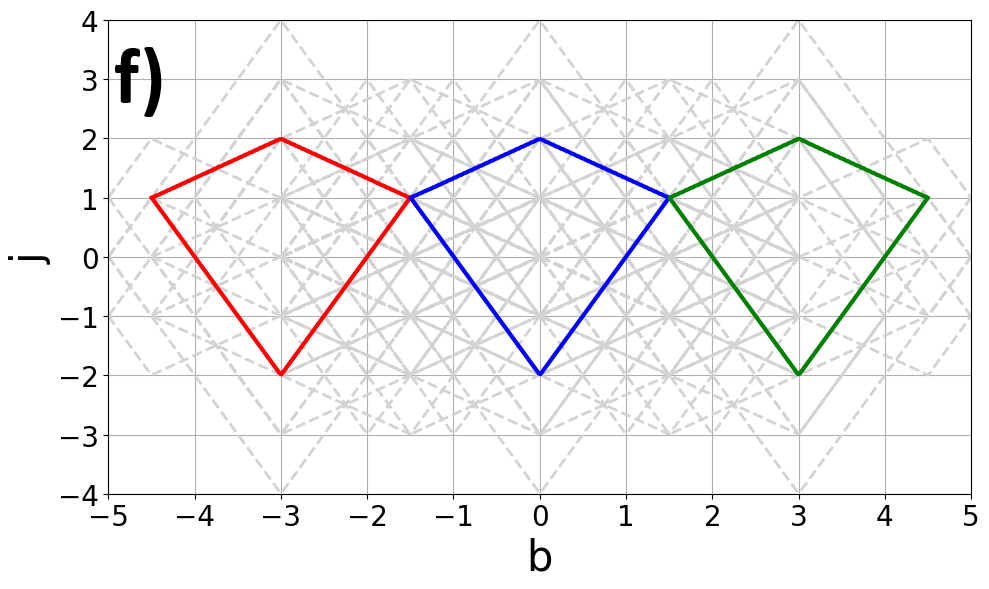}
    }
    \caption{Six example vorticity stability regions (VSR) for a device containing four identical, equidistant nanowires with all critical phases set to $2\pi$ and the critical currents set to 1. The coordinates are 0, 1/3, 2/3, and 1. The x-axis is the normalized magnetic field, and the y-axis is the normalized total bias current flowing through the device. Note that the gray dashed lines in all figures indicate the boundaries of various possible stability regions. (a) Diamond VSR. Such correspond to the following vortex configurations: $[-1,-1,-1]$, $[0,0,0]$, and $[1,1,1]$ from left to right. (b) Glider-shaped VSR. Such occurs with the following vorticity configurations $[-1, 0, -1]$, $[0,1,0]$ corresponding to the stability regions, shown from the left to the right. (c) Trapezoidal VSR. Such correspond to the following vortex configurations: $[-1,1,0]$, $[0,2,1]$, and $[1,3,2]$ from the left to the right. (d) Tilted triangular VSR. Such correspond to the following vortex configurations: $[0,0,-2]$ and $[1,1,-1]$, appearing from left to right. (e) Flipped kite VSR. Such correspond to the following vortex configurations: $[0,-3,0]$, $[1,-2,1]$, and $[2,-1,2]$ from left to right. (f) Kite VSR. Such correspond to the following vortex configurations: $[-2, -1, 0]$, $[-1,0,1]$, and $[0,1,2]$ from left to right.}
  \label{fig:4-wire-stability-regions}
\end{figure*}

We numerically generated some representative VSRs characterizing 4-SQUIDs, assuming that the nanowires are equidistant and have $\phi_{c,i} = 2\pi$. Six VSR shapes, as examples, are displayed in Fig.\ref{fig:4-wire-stability-regions}(a-f). Diamond VSRs (Fig.\ref{fig:4-wire-stability-regions}a) are formed when $\Delta v = [0,0]$, or, in other words, when each cell has the same number of fluxons. 

We also observe more exotic shapes. That is, glider-shaped VSRs (Fig.\ref{fig:4-wire-stability-regions}b) are formed when $\Delta v = [1, -1]$. This means that the 2nd cell has 1 more fluxoid (vortex) than the 1st cell, and the 3rd cell has one less fluxoid than the 2nd cell. These VSRs have 6 vertices, which is one more compared to the flat-top diamond VSR found for symmetrical 3-SQUIDs. Trapezoidal shapes VSRs (Fig.\ref{fig:4-wire-stability-regions}c) are formed when $\Delta v = [2, -1]$. Tilted Triangular VSRs (Fig.\ref{fig:4-wire-stability-regions}d) are formed when $\Delta v = [0, -2]$. Horizontal kite VSRs (Fig.\ref{fig:4-wire-stability-regions}e) are formed when $\Delta v = [-3, 3]$ and they are symmetric with respect to the j-axis. Vertical kite VSRs (Fig.\ref{fig:4-wire-stability-regions}f) are formed when $\Delta v = [1, 1]$ and they are symmetric with respect to the b-axis. For example, if the vorticity array is $[-1,0,1]$, then changing the magnetic field from $b$ to $-b$ leaves the VSR unchanged. Note that Fig.\ref{fig:4-wire-stability-regions} has a multitude of dashed lines. They show the boundaries of all possible VSRs. The upper and lower envelopes of this pattern represent the critical currents of the device, $I_{c,+}(b)$ and $I_{c,-}(b)$, respectively.

If the critical phase of the wires is sufficiently low, the VSR pattern can become topologically disjoint. Consider, for example, the VSR sequence in Fig.\ref{fig:disjoint-VSR-equidistant}b. As usual, the upper envelope of the pattern represents the positive critical current $j_{c,+}(b)$ and the lower envelope of the VSR sequence represents $j_{c,-}(b)$. We note that there are $b$ values at which $j_{c,+}=j_{c,-}=0$ (Fig.\ref{fig:disjoint-VSR-equidistant}b). This is a border-line case, in the sense that if the critical phase would be slightly lower than the VSR diamonds would not teach each other and, therefore, a non-superconducting phase would exist between the diamonds. As it should be clear by now, superconducting regime is only possible inside VSRs.

 To better understand how the Little-Parks effect is realized in MW-SQUIDs, consider an example, suppose that there are no vortices and $b=0$. The total supercurrent then reaches the maximum possible value $j=4$ if the phase bias of each wire is equal to its critical phase. This current represents the principal maximum of the critical current $j_c(b)$ located at $b=0$.
This point is the top vertex of the diamond VSR. To evaluate the period, one needs to find the lowest non-zero field at which the total current can reach the level $j=4$, provided that the phase bias of the first wire is properly adjusted, namely if $\phi_1=\phi_c$. Here, $\phi_c$ is the critical phase, which is assumed to be the same for all wires. For the maximum to occur, all wires have to contribute the maximum possible current, which is the case if the phase bias on each wire equals the same value, namely the critical phase. The lowest non-zero magnetic field at which such "equal-phase-bias" condition is achieved is $b=3$. 

At $b=3$ the phase bias of the i-th wire is $\phi_i=\phi_1+2\pi b x_i-2\pi v_{1,i} \rightarrow \phi_1+6\pi x_i-2\pi v_{1,j}$. The coordinate of the i-th wire is $x_i=(i-1)/3$, therefore $\phi_i=\phi_1+6\pi (i-1)/3-2\pi v_{1,i} = \phi_1+2\pi (i-1)-2\pi v_{1,i}$. Next, the vorticity needs to be adjusted to compensate for the phase shits generated by the Meissner currents. The equal-phase-bias state ($\phi_i=\phi_1$) is achieved if $v_{1,i}=(i-1)$, in which case $\phi_i=\phi_1+2\pi (i-1)-2\pi (i-1)=\phi_1$, i.e., the phase bias is the same for all wires. In such a case, if the phase bias of the first wire equals its critical phase, then the phase biases of all other wires equal to their critical phases also. Thus, each wire contributes the maximum possible current of 1, and the total current is $j=4$. Thus, we conclude that there is a principal critical current maximum at $b=3$. 

This is in agreement with the conclusion that the period of the VSR sequence is $\Delta b =3$. Note that if the field is lower, then the Meissner phase shift between the wires is less than $2\pi$ and therefore it cannot be exactly compensated for by introducing a vortex (fluxoid) between the wires. The vorticity distribution used to achieve this maximum is $v_{1,i}=(i-1)$, which means that each cell has one vortex, and the total number of vortices in the device considered is three. Thus, again we see that $\Delta b=3$.

Another way to understand the same result is as follows. If $b=3$, then each nanowire's phase bias is augmented with respect to its left neighbor by $2\pi$, due to the Meissner current in the electrodes. At this point, if each cell accepts one vortex, then the vortex will compensate the Meissner phase shift, and the resulting net phase shift between the wires becomes zero, as it is at $b=0$. Therefore, the system will exhibit a principal maximum of the critical current at $b=3$, because all wires will be able to achieve their critical phase states simultaneously.

A remarkable fact is that the 4-wire-SQUID exhibits secondary critical current maxima between the principal maxima. In Fig.\ref{fig:disjoint-VSR-equidistant}b, the first secondary maximum occurs at $b=1$, corresponding to the vorticity $[0,0,1]$. The second secondary maximum (blue) occurs at $b=1.5$ and corresponds to the vorticity $[0,1,0]$. The third secondary maximum (yellow) corresponds to $b=2$ and corresponds to the vorticity $[1,1,0]$. In total, there are three secondary maxima between each pair of neighboring principal maxima. In this respect the system is somewhat similar to an optical diffraction grating with four equidistant identical slits. The analogy occurs because in the case of wave interference, a phase shift of a wave by $2\pi$ is insignificant since the wave transforms into itself. In the case of a nanowire SQUID a phase change of $2\pi$ is insignificant since if the phase shifts by $2\pi$ between two neighbor wires then one can introduce a vortex between the two wires and in this way eliminate the phase shift, since each vortex produced a phase shift of $-2\pi$ between the wires. Yet, the 4-slit diffraction grating exhibits only two secondary interference maxima between each pair of neighboring principal maxima, while our device exhibits three secondary maxima. This happens because the current-phase relation is assumed linear for each nanowire. The agreement between the optical diffraction grating and the SQUID is much closer if the weak links have a sinusoidal CPR\cite{pekker-2005}.

If the critical phase of the wires is less than $(3/4)\pi$, then the sequence of the VSR regions becomes topologically disjoint. The example shown in Fig.\ref{fig:disjoint-VSR-equidistant}b represents a critical situation of $\phi_{c,i}=3/4\pi$, in which case the diamond VSR and the other VSR are only touching at one point. Thus, if the critical phase was slightly lower, then the VSR sequence would be disjoint. In such a case, there are regions of magnetic field in which the system cannot be superconducting with any combination of vortices in the cells. Therefore, if $\phi_{c,i}<3/4\pi$, then the system exhibits a sequence of quantum transitions between normal and superconducting states if the magnetic field is swept up at zero temperature. 

A 4-nanowire SQUID responds to disorder in a similar fashion to a 3-nanowire SQUID. A phase-disordered system of four nanowires (with equidistant nanowires with identical critical currents) will have shifted VSR top vertices. The position of the top vertex is defined by the condition that two nanowires switch simultaneously. Therefore, if the critical phase difference between these two nanowires is changed, then the position of the top vertex along the b-axis changes as well. Additionally, the VSRs are still periodic with a period of $\Delta b = 3$. This is based on the fact that neither Meissner phase shifts nor the phase shifts generated by trapped vortices depend on the critical phase of the nanowires. Therefore, adding one fluxoid to each cell will reproduce the same VSR shape but shifted to the right by $\Delta b = 3$.  Because the VSRs are still periodic with $\Delta b = 3$, it follows that the magnitude of shifting is the same for every VSR shape. 

A critical current disordered system of four equidistant nanowires with identical critical phases can alter the geometry of any VSR, except for the position of the top and bottom VSR vertex (if there is one). This can be explained because top vertices are formed when 2 superconducting nanowires switch simultaneously. Since the critical phase criterion does not depend on the critical current parameter, it follows that a critical current disordered system has no influence on the top vertex of each VSR. The VSRs of the same shape remain periodic with the same period of $\Delta b = 3$ even if the critical current disorder is present. 

For a position-disordered system, the period of the critical current curves will change. This means that to recreate the same VSR shape but shifted along the $b$ axis, one cannot just add one fluxon to each cell but rather several fluxons that is proportional to the relative size of the cell. We derive this new period. Consider a position-disordered system of 4 superconducting nanowires of positions $[0, x_2, x_3, 1]$ where $0 < x_2 < x_3 < 1$ and $x_2,x_3$ are real numbers. Then the size of the superconducting cells is $[x_2, x_3-x_2,1-x_3]$. Then, to derive the new period, simply multiply $x_2, x_3-x_2, 1-x_3$ by the smallest number $c$ such that $[cx_2, c(x_3-x_2), c(1-x_3)]$ are integers. Moreover, to horizontally shift a VSR of a vorticity state $v_i$ while preserving its shape, one must add a fluxon distribution of $[cx_2, c(x_3-x_2), c(1-x_3)]$ to $v_i$. Since $cx_2, c(x_3-x_2), c(1-x_3)$ are all integers, this is always possible. Then we can augment the magnetic field by $\Delta b = c(x_2 + (x_3-x_2) + (1-x_3)) = c$, in which case the net phase shift between the nanowires would be the same as it was before the vortices were added. For example, if before the vortices were added the wires were ''in phase'' and produced a maximum possible current then after the vortices are added and the magnetic field applied as calculated above then the wire will also be in phase and give rise to the maximum of the critical current. As a first simple example, we calculate the period of 4 identical, equidistant nanowires. In this case, the size of the superconducting cells is $[1/3, 1/3, 1/3]$. Clearly $c=3$, which implies that the period is $\Delta b = 3$. Now, we calculate the period of a device where the 1st cell is twice as large as the 2nd and 3rd cells. The cell size array is $[1/2, 1/4, 1/4]$. Clearly $c=4$, and therefore, the period is $\Delta b = 4$. Additionally, if the sizes of any of the superconducting cells are incommensurate, then the device is not periodic.
In the completely disordered system of 4 superconducting nanowires, the distortion of the VSR is a combination of the effects generated by its individual phase, current, and position disorder, due to the linearity of the model considered.  

\section{Generalized properties of symmetrical MW-SQUIDs with many wires}

In this section we discuss symmetrical MW-SQUIDs, i.e., such that all the wires are identical and equidistant. Such SQUIDs are symmetrical with respect to 180-degree rotations around x-axis, or around y-axis, or around z-axis. As will be discussed in detail below, these geometrical symmetries of the device dictate certain symmetries of the corresponding VSRs. 

The devices considered have $n$ equidistant identical nanowires and therefore have $n-1$ identical superconducting cells. So, the period of the critical current function and of the VSR pattern is $\Delta b=n-1$. This is the essence of the generalized Little-Parks effect for a periodic symmetrical arrays of nanowires. The physical meaning is as follows. Suppose we apply a magnetic field such that the Meissner phase shift between each pair of neighbor wires is $2\pi k$, where $k$ is an integer. Then it is possible to introduce vortices in the cells and compensate exactly the effect of the magnetic field. More specifically, if $k$ vortices is introduced then the compensation is exact, assuming that the magnetic field is increased by $\Delta b=k(n-1)$. Thus, all the currents in the device are the same as in its original state. Thus, the principal maxima of the critical current will occur at $b=(n-1)k$, where $k$ is an integer. 

\begin{figure}[b]
    \centering
    \includegraphics[width=0.8\linewidth]{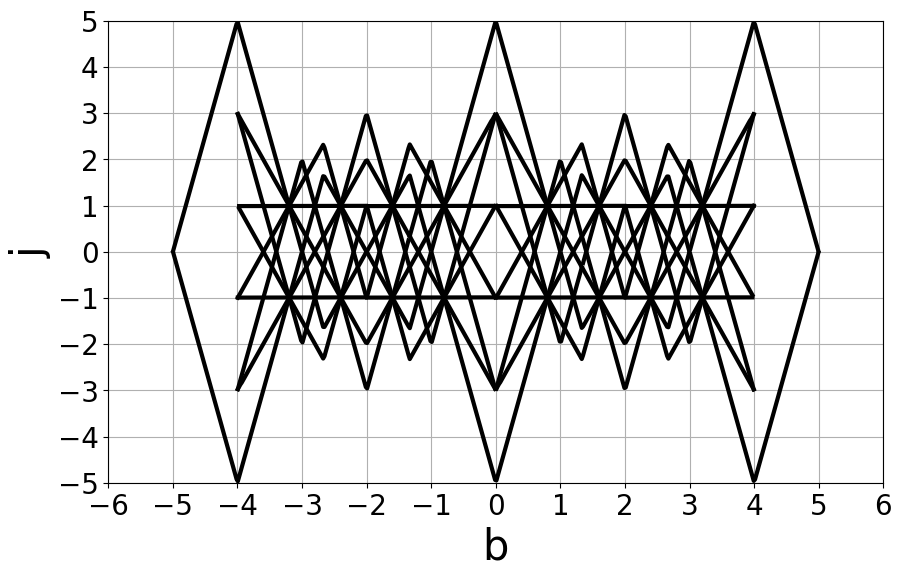}
    \caption{Examples of VSR for a symmetrical five-wire device. The wires are equidistant. The critical phase is $\pi$ for each wire. The large diamonds (black) correspond to the vorticity distributions of [-1,-1,-1,-1], [0,0,0,0], and [1,1,1,1].  
        }
    \label{fig:5-wire-VSR}
\end{figure}

In general, a symmetrical $n$ nanowire SQUID produces critical current curves somewhat similar to the optical diffraction patterns. It exhibits principal maxima and and also secondary maxima. Yet, the number of the secondary maxima is defined by different possible vorticity combinations and can differ from the basic optical diffraction grating. General illustrations to this phenomenon can be seen in Fig.\ref{fig:disjoint-VSR-equidistant} and Fig.\ref{fig:5-wire-VSR}.

\begin{figure}[t]
    \centering
    \includegraphics[width=\linewidth]{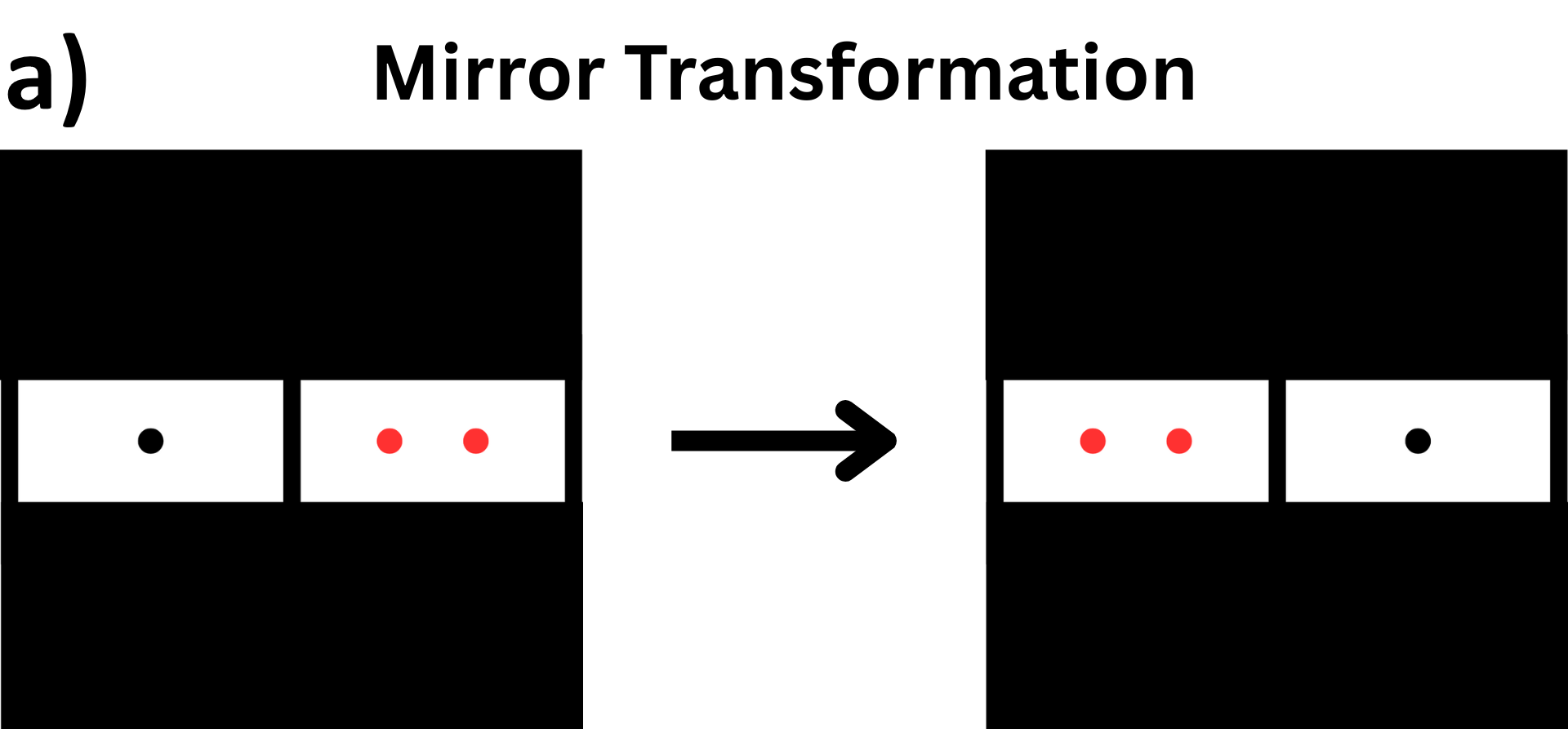}
    \includegraphics[width=\linewidth]{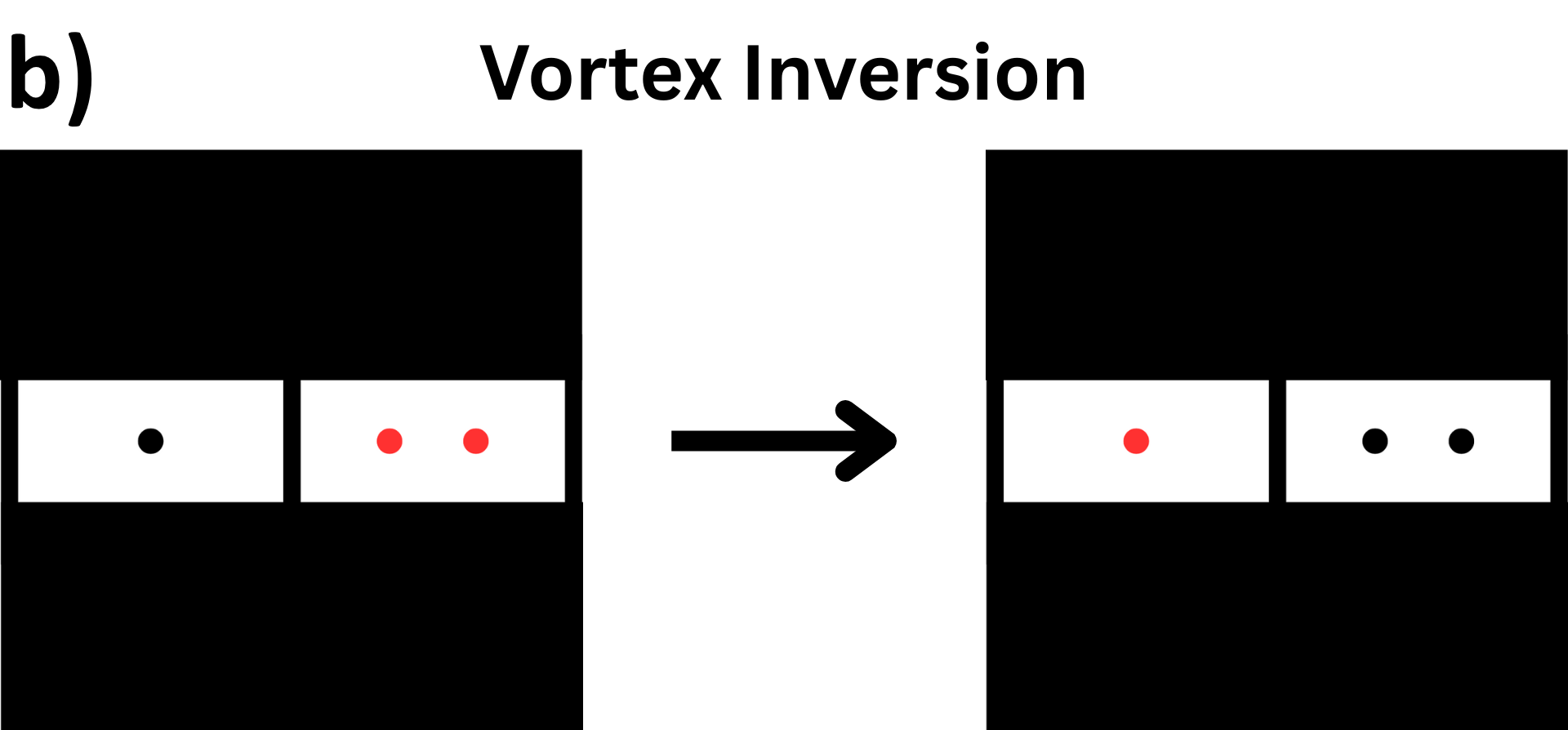}
    \caption{Example transformations done on vortices for three nanowires. Note that initially, the device has 1 vortex (black) in the left cell and two anti-vortices (red) in the right cell. (a) A vorticity parity transformation, $P_v$, is done on the ensemble of vortices. It is equivalent to a mirror reflections of their locations, while their polarity is not transformed. (b) A vortex inversion (V-inversion) is done on the ensemble of vortices.}
    \label{fig:vortex-transformations}
\end{figure}

We now discuss general symmetries of VSRs produced by an $n$-nanowire symmetrical MW-SQUID. We define the following two transformations: (1) The vorticity parity transformation ($P_v$-transformation) reflects the locations of the vortices with respect to the device center, while preserving the their polarity. In other words, in the $P_v$-transformation, the vortices on the left are shifted to the right, while those on the right move to the left. For example, if there is a vortex in the second cell from the left side of the device, it will end up in the second cell from the right side of the device, etc. (2) The vortex inversion (V-inversion) reverses the polarity of the vortices. Both of these transformations are illustrated in Fig.\ref{fig:vortex-transformations} for the case of three superconducting nanowires. 

\begin{figure*}[htbp]
    \centering
    \includegraphics[width=0.32\linewidth]{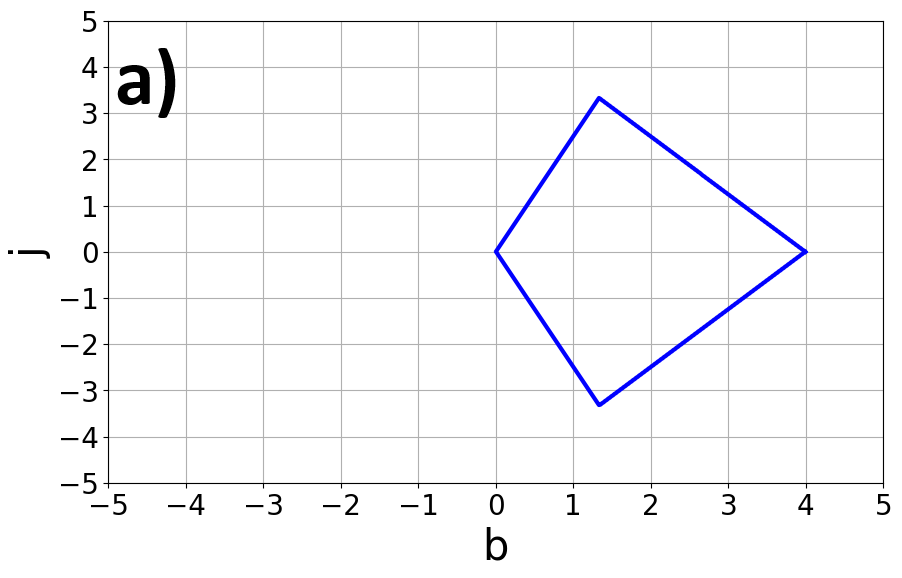}
    \includegraphics[width=0.32\linewidth]{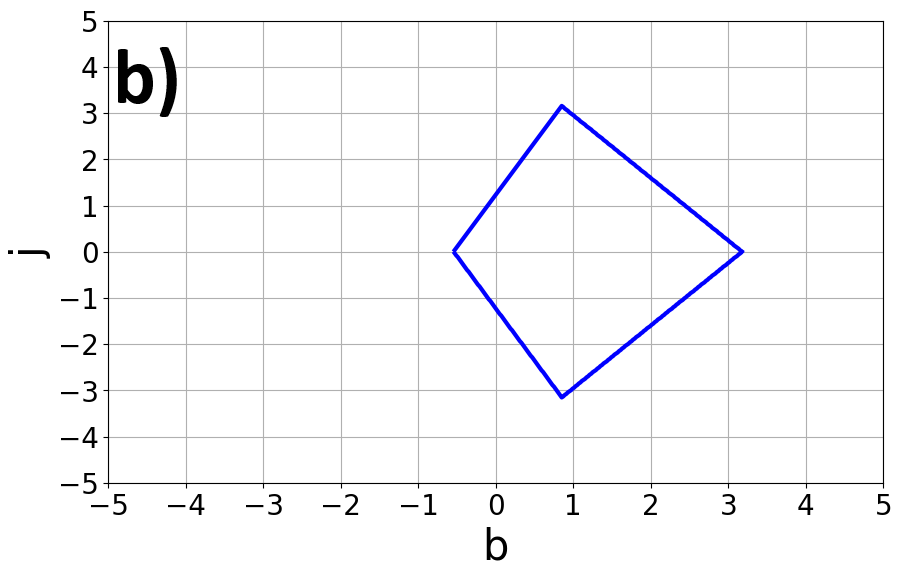}
    \includegraphics[width=0.32\linewidth]{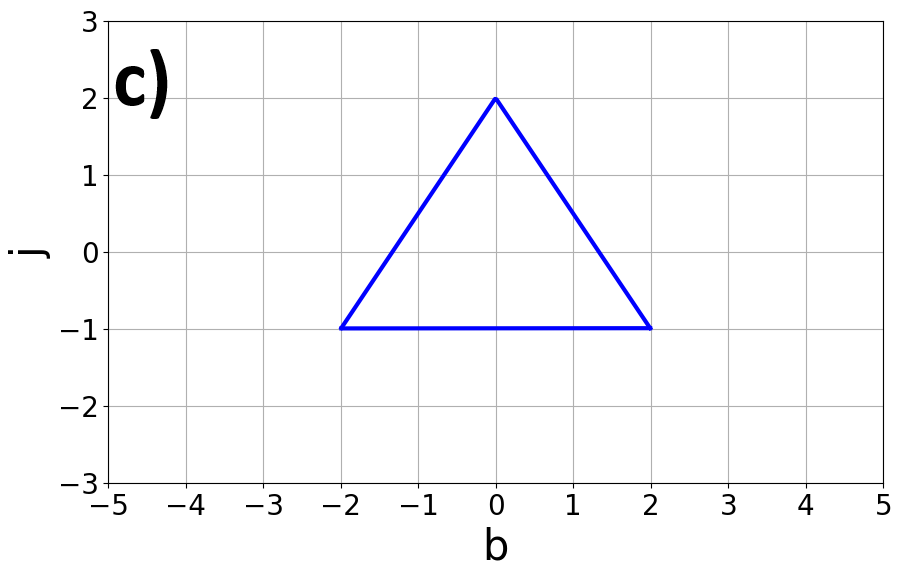}
    \includegraphics[width=0.32\linewidth]{images/Vortex_Symmetries/1001_mirror_symmetric_device_edited.png}
    \includegraphics[width=0.32\linewidth]{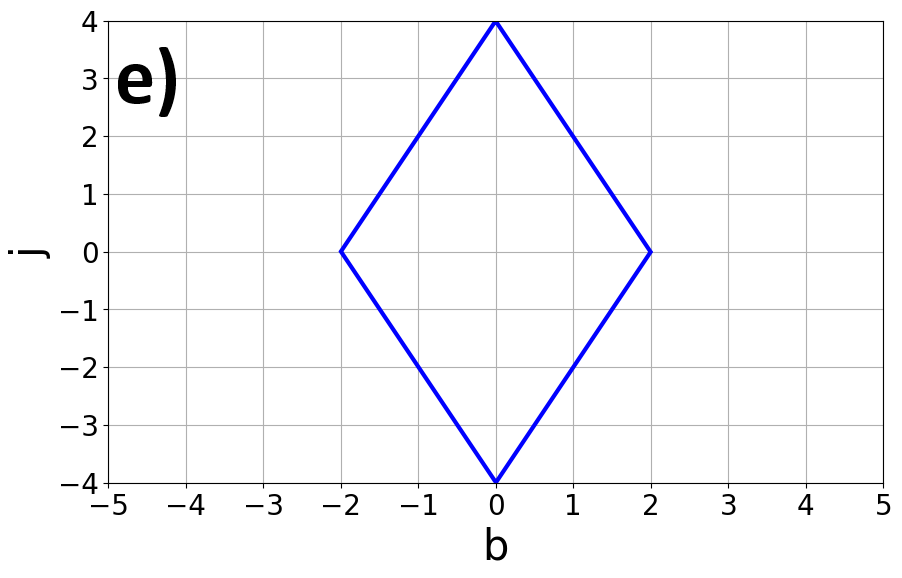}
    \includegraphics[width=0.32\linewidth]{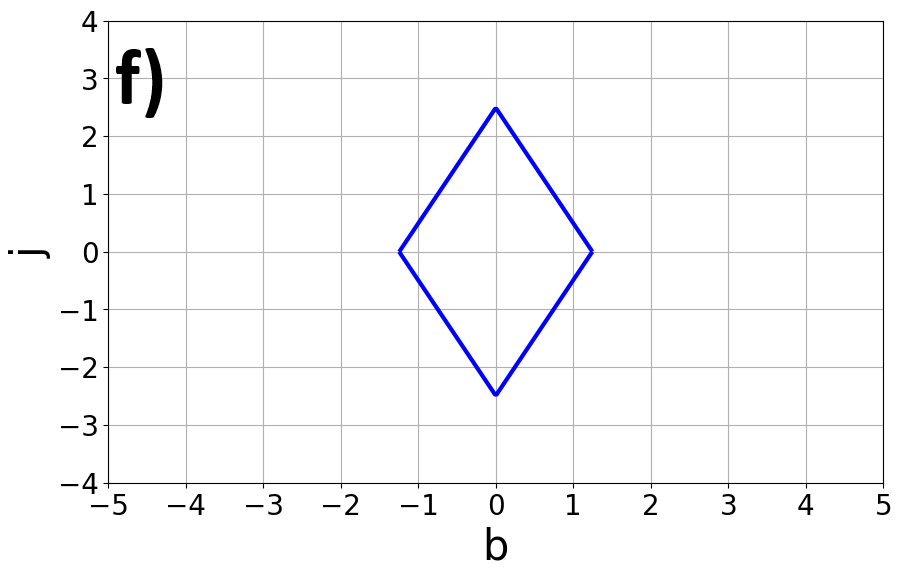}
    \caption{Vortex symmetries for various MW-SQUID configurations. (a) The device consists of five identical, equidistant nanowires with $\phi_{c,i} = 2\pi$. The VSR corresponds to the vorticity state $[1,0,0,1]$ which is $P_v$-inversion symmetric. Therefore this VSR is $I$-inversion symmetric, that is, reflecting the VSR across the $b$ (or $B$) axis produces the same VSR. (b) The device consists of five nanowires with positions $[0, 0.2, 0.5, 0.8, 1]$, critical phases of $[8,6,7,6,8]$ (rad), critical currents of $[1.2, 1, 0.6, 1, 1.2]$. The device is mirror-symmetric. Introducing a vorticity state of $[1,0,0,1]$ generates a VSR that is $j$-inversion symmetric. (c) The device consists of three identical, equidistant nanowires with $\phi_{c,i} = 2\pi$. The VSR corresponding to the vorticity state $[1, -1]$ is plotted. This VSR is $b$-inversion (same as $B$-inversion) symmetric, that is, reflecting the VSR across the $j$ (or $I$) axis produces the same VSR. (d) The device consists of three equidistant nanowires with critical phases of $[8,6,8]$ (rad) and critical currents of $[1.2, 0.6, 1.2]$. Introducing a vorticity state of $[1,-1]$ generates a VSR that obeys the $b$-inversion symmetry. (e) The device consists of four identical, equidistant nanowires. The vorticity state $[0,0,0]$ produces a VSR that is simultaneously both $I$ and $B$ symmetric. (f) The device consists of four nanowires with positions $[0, 0.1, 0.9, 1]$, critical phases of $[2\pi, \pi, \pi, 2\pi]$, and critical currents of $[1.5, 0.5, 0.5, 1.5]$. This device is mirror-symmetric. Introducing a vorticity state of $[0,0,0]$ produces a VSR that is both $I$ and $B$ symmetric. }
    \label{fig:vortex-symmetries}
\end{figure*}

Consider a vorticity state that is symmetric with respect to the $P_v$-transformation. Remember also that in this section we consider devices which are, in particular, mirror symmetric with respect to the y-axis. So, both the vortex distribution in the cells and the arrays of the nanowires are mirror-symmetric with respect to the y-axis. Then, we find that the corresponding VSR is invariant with respect to the $j$-inversion. That is, the VSR is the same when mirror-reflecting it across the $b$ axis. An example mirror-symmetrical vorticity state would be $[1,0,0,1]$, for five equidistant, identical superconducting nanowires. Clearly, the corresponding VSR is mirror-symmetric with respect to the b-axis, as shown in Fig.\ref{fig:vortex-symmetries}(a). If the critical phase is not all the same for all the nanowires, but the mirror symmetry with respect to $y$-axis is present then the $j$-inversion symmetry of the VSR is still present also (Fig.\ref{fig:vortex-symmetries}(b)). 

To understand this conclusion, we first put a reminder that the devices considered here are mirror-symmetric with respect to the x-axis of the device and also with respect to its y-axis. Therefore, a rotation by 180 degrees around the z-axis will keep the device unchanged. So, if we rotate the device 180 degrees around the z-axis, the physical device will transform into itself, while the current in the device will change from $j$ to $-j$. Now, if the vortex distribution is $P_v$ symmetric, then the vortex ensemble will not change due to the considered rotation either. (The assumption is that the vortices follow the device frame as the device is rotated around the z-axis.) So, since the device and the vortices remain the same and the current changes the sign (while the magnetic field with respect to the device remains unchanged), one concludes that the same critical current would be measured, meaning that the VSR is symmetric with respect to the $j$-inversion, or equivalently, $I$-inversion symmetric.

Now, consider a vorticity state that is symmetric with respect to the combination of the $P_v$-transformation and the V-inversion. Such combined transformation is denoted as $P_vV$ inversion. We argue below that if the vortex distribution satisfies the $P_vV$ inversion symmetry and the device is symmetric with respect to its 180-degree rotation around the $y$-axis then the corresponding VSR is invariant with respect to the $b$ inversion. That is, the VSR is the same when reflecting it across the $j$ axis. To illustrate this type of symmetry consider a device with three equidistant identical nanowires. Such 3-SQUID is symmetric with respect to its rotation around its $y$-axis (which coincides with the middle wire). Consider a vorticity state  $[1,-1]$, i.e., the left cell has one vortex and the right cell has one anti-vortex. Such state satisfies the $P_vV$ symmetry. The corresponding VSR is mirror-symmetric with respect to the j-axis (vertical axis), which is equivalent to the $b$-inversion symmetry (see the examples in Fig.\ref{fig:vortex-symmetries}(c-d)). Note that although these VSRs are symmetrical with respect to the $b$-inversion, they are not symmetrical with respect to the $j$-inversion. Thus, the device with the considered vortex distribution acts as a superconducting diode. Note again that for this $b$-inversion symmetry to work it is not necessary that the nanowires are identical. An example with different nanowires is shown in Fig.\ref{fig:vortex-symmetries}(d)).

The general rule given above follows from the simple analysis. Suppose we rotate our device 180 degrees around the y-axis (flip it upside-down). And assume that the vorticity is attached to the device during this rotation. Then, due to the device's mirror symmetry about the y-axis, the physical device will transform into itself. Now, if the vortex distribution is symmetric with respect to the combination of the $P_v$-transformation and the V-inversion, then the vortex ensemble will remain invariant under this rotation. Therefore, rotating the device with the vortices about the y-axis is a symmetry. Therefore, the critical current would not change. However, a rotation about the y-axis in the lab frame is equivalent to a $b$ inversion in the device frame. The critical current would be the same in the device frame as in the lab frame. It follows that the VSR must be symmetric with respect to the $b$ inversion, or, equivalently, $B$-inversion symmetric.

We can now argue that a zero vorticity state must be I-symmetric and also B-symmetric. We can also prove that only a zero vorticity state can be both I-symmetric and B-symmetric. To prove this, remember that a VSR is both $I$ symmetric and $B$ symmetric if its vorticity state is symmetric with respect to the $P_v$-transformation and a combination of the $P_v$-transformation and the V-inversion, or equivalently, is symmetric with respect to just the V-inversion. The only vorticity state that is V-inversion invariant is the zero vorticity state, as shown in Fig.\ref{fig:vortex-symmetries}(e-f). This proves that only the zero vorticity state is both $I$ symmetric and $B$ symmetric. Such VSR has the diamond shape. Note that this conclusion is applicable only to mirror-symmetrical MW-SQUIDs.

A diamond (rhombic) VSR is always produced when each superconducting cell has the same number of fluxons, assuming identical and equidistant nanowires. Let us analize the VSR of the zero vorticity state first. We first derive the location of the top and bottom vertices for a diamond with a zero vorticity state. The Meissner phase shift formula, which relates the phase bias of the i-th and k-th nanowires, is $\phi_i = \phi_k + 2\pi b(x_i - x_k) - 2\pi v_{k,i}$. Then, for simplicity, let all vorticity values be equal to 0, then $\phi_i = \phi_k + 2\pi b(x_i - x_k)$. Now, suppose wires $i$ and $j$ reach their critical state simultaneously, that is $\phi_i  = \phi_{c,i}$ and $\phi_k = \phi_{c,k}$. Then, the Meissner phase shift formula becomes $\phi_{c,1} = \phi_{c,2} + 2\pi b_{top} (x_i - x_k)$. So, the top vertex of the VSR occurs at $b_{top} = (\phi_{c,1} - \phi_{c,2})/[2\pi(x_i - x_k)]$. If all the wires are identical, i.e., $\phi_{c,i}=\phi_c$, then the top vertex of the VSR is $b_{top}=0$. Setting $b=0$, the total normalized supercurrent is $j(0) = \sum_{k=1}^n \phi_k/\phi_c$. At $b=0$, the maximum supercurrent occurs if $\phi_k = \phi_c$. Then $j_{c,+}(0) = \sum_{k=1}^n 1 = n$. Similarly, if $\phi_i = \phi_k = -\phi_c$, then $b_{bottom} = 0$ which is the location of the bottom vertex. The corresponding critical current is $j_{c,-}=-n$. There is only one top and one bottom vertex. This is because all the nanowires have the same critical phases, which means that, at $b=0$, all of them switch to the normal state simultaneously.

We now derive the location of the right and left vertices for the zero-vorticity state ($v_{i,k}=0$). Recall the Meissner phase shift formula $\phi_i = \phi_k + 2\pi b(x_i - x_k)$ or $\phi_i - 2\pi b(x_i - x_k) = \phi_k$. Here and below, we assume $i>k$. Now, given the critical phase criterion for the k-th wire, $-\phi_c \leq \phi_k \leq \phi_c$, we plug it into the formula for $\phi_k$ to obtain $-\phi_c \leq \phi_i - 2\pi b(x_i - x_k) \leq \phi_c \Rightarrow -\phi_c + 2\pi b(x_i - x_k) \leq \phi_i \leq \phi_c + 2\pi b(x_i - x_k)$. We now analyze the left inequality, according to which the phase difference across the i-th wire must satisfy $-\phi_c + 2\pi b(x_i - x_k) \leq \phi_i$. Then, if $b=b_{right,k,i}$, where $b_{right,k,i}$ is defined by $-\phi_c + 2\pi b_{right,k,i}(x_i - x_k) = \phi_c$ then we get $\phi_c \leq \phi_i$. Moreover, from the critical phase criterion, we have that $\phi_i \leq \phi_c$. Then the interval of the phase bias values allowed for the i-th wire shrinks to just one point, which is the critical value, since we get $\phi_c \leq \phi_i \leq \phi_c$. That means that the phase difference across the i-th wire must be $\phi_c$. In other words, at $b=b_{right,k,i}$ the difference between the negative and the positive critical currents shrinks to zero. Therefore, $b_{right,k,i}$ is the critical $b$ value, such that the wires $k$ and $i$ can be subcritical only if $b<b_{right,k,i}$. The entire SQUID can remain superconducting if all wires are superconducting. Therefore,  the magnetic field $b_{right} = \min_{k,i}(b_{right,k,i})$  corresponds to the location of the right vertex because this is the maximum field at which all loops, $j,k$, can remain superconducting. Because $b_{right} > 0$  (recall our convention that $i>k$), this is the location of the right vertex of the zero-vorticity VSR. Thus, to minimize $b_{right,k,i}$, we set $x_i - x_k = 1$. Therefore $b_{right} = \phi_c/\pi$. This formula is verified by our numerical model.

Similarly, in the right inequality, if $\phi_c + 2\pi b_{left,k,i}(x_i - x_k) = -\phi_c$ or $b_{left,k,i} = -\phi_c/[\pi(x_i - x_k)]$, then the phase difference across the i-th wire is bounded between $-\phi_c \leq \phi_i \leq -\phi_c$. , i.e. $\phi_i=-\phi_c$.
So the wire must be in the critical state, and therefore the device has to be normal. The smallest $|b_{left,k,i}|$ defines the location of the left vertex. The minimum possible $|b_{left,i}|$, among all pairs of wires, is achieved if $x_i - x_k = 1$, i.e., if $x_k=0$ and $x_i=1$. Note, $b_{left} = -b_{right}$ as expected because the zero vorticity state (for equidistant, identical nanowires) must be symmetric with respect to the $b$ inversion. 

The zero vorticity state VSR must also be symmetric with respect to $j$ inversion (see the discussion above). Therefore, it follows that the supercurrent at the left and right vertices must be zero. 

Generally, the top and bottom vertices for the zero vorticity state in a symmetric MW-SQUID are defined when the phase difference across two superconducting nanowires is critical and of the same sign and magnitude. The right and left vertices, on the other hand, take place when the phase differences across two superconducting nanowires equals to their critical phase, but has opposite signs, so the total current is zero (assuming no critical current disorder). 

Now, as a simple example, to show that the VSR analyzed above has the diamond shape for symmetrical SQUIDS with $n$ identical, equidistant nanowires, we we present a simple calculation that demonstrates that the boundary of the zero-vorticity VSR, which is the dependence of the corresponding critical current on the magnetic field, is a straight line. Recall that $j = \sum_{i=1}^n\phi_i/\phi_c$. Using the formula $\phi_i = \phi_k + 2\pi b(x_i - x_k)$, we obtain that $j(b,\phi_k) = n\phi_k/\phi_c + 2\pi b (\left(\sum_{i=1}^n x_i\right) - nx_k)/\phi_c$.  Next, we derive the slope of the critical current in the interval between $b=b_{top} = 0$ and $b=b_{right}$. As the bias current is ramped up, there will be a wire that reaches its critical phase first. Let us say $k$ is the number of that wire. Then, the critical current is defined by the switching of the k-th wire, and therefore  $j_{(c,+),k}(b) = \max_{\phi_k}[j(b,\phi_k)]$. If $\phi_k=\phi_c$ then the current equals the critical current, so we have $j_{(c,+),k}(b) = n\phi_c/\phi_c + 2\pi b (\sum_{i=1}^n x_i - nx_k)/\phi_c$.  Because the wires are equidistant, it follows that $x_i = (i-1)/(n-1)$, then using Gaussian summation formula, we obtain $\sum_{i=1}^n x_i = n(n-1)/[2(n-1)] = n/2$. The meaning of this is that the mean value of $x_i$ for a symmetrical device equals 1/2. Then the formula for the critical current is simplified to $j_{(c,+),k}(b) = n + (2\pi bn/\phi_c) (1/2 - x_k)$. Notice, there are a total of $n$ possible critical currents indexed by $k$. The smallest critical current would be the critical current of the entire device. For sufficiently small but positive $b$ values, the formula shows that the lowest critical current corresponds to the switching of the wire with the largest $x_k$, i.e., the $k=n$ wire. Therefore $j_{c,+}(b) = n - (\pi bn/\phi_c) $. The slope of the VSR boundary is $-\pi n/\phi_c $. Note that the sensitivity of the SQUID to the magnetic field increases with $n$ since the slope of the critical current is proportional to $n$.  The right vertex of the VSR can be found by requiring $j_{c,+}(b_{right})=0$. Therefore, $b_{right}=\phi_c/\pi$, in agreement with the previous derivation.
Using the fact that the zero vorticity state is both $I$ symmetric and $B$ symmetric, we can construct a diamond shape. This proves our claim that the zero vorticity state for all MW-SQUIDs with identical equidistant nanowires must be rhombic.  

\section{Generalized properties of asymmetrical MW-SQUIDs with many wires}

In this section we analyze asymmetrical $n$-wire SQUIDs, in which the wires are not equidistant or they are different or both. In general, multiple-wire SQUIDs respond to disorder in a similar fashion as the $2,3, \& 4$ nanowire SQUIDs. A critical phase disordered $n$ nanowire (equidistant nanowires with identical critical currents) SQUID will have horizontally shifted top and bottom VSR vertices (if they exist) because the nanowires will not switch at the same time at the original $b_{top}$, but rather a pair of nanowires switch at a new $b_{top}'$ value. Moreover, because the position of the wires has not been changed, the period of this device is still $\Delta b = n-1$ where $n$ is the number of nanowires.

Based on the results obtained on nanowire SQUIDs with less than six nanowires, we expect that the VSRs in an $n$-nanowire symmetrical SQUID will become topologically disjoint if $\max_i(\phi_{c,i}) < \pi (n-1)/n$. That is, the maximum critical phase out of all the wires is less than $\pi(n-1)/n$. Our computer model of multiple-nanowire SQUIDs confirms this formula. If the VSRs are disjoint, superconducting-normal quantum transitions should occur when sweeping the magnetic field at zero temperature. Moreover, a 100\% supercurrent modulation is achieved in this case.

A random critical current disordered $n$ nanowire SQUID (equidistant nanowires with identical critical phases), in a similar fashion to the $4$ nanowire SQUID, will alter the geometry of each point in the VSR except for the top and bottom VSR vertices, if there is one. An important consideration here is that VSR is analyzed in the $j$ and $b$ plane, where $j$ is the total current divided by the mean critical current. Therefore the top and the bottom vertices of the VSR co not shift in the vertical direction. The top and the bottom vertices do not shift horizontally since the critical current parameter has no effect on the critical phase criterion.

A position-disordered $n$ nanowire SQUID will have an altered period, which can be viewed as a generalized Little-Parks oscillation effect. Consider a position disordered $n$ nanowire SQUID with nanowires located at $[0, x_2, x_3, \dots, x_{n-1}, 1]$. Then, its cell sizes are $[x_2, x_3 - x_2, \dots, 1 - x_{n-1}]$. To derive the new period of this device, we need to find the smallest integer number $c$ such that if all the dimensions of the cells are multiplied by $c$ then all of them become integers. Suppose that such number $c$ does exist. If now the normalized magnetic field is increased by $c$ (i.e., $\Delta b=c$), then the Meissner phase shift on each cell will increase by an integer multiple of $2\pi$ and therefore can be canceled by adding the corresponding number of vortices in the cells. Therefore, to horizontally shift, without any distortion, a VSR of an initial vorticity state $v_i$, one must add a fluxon distribution of $c[x_2, x_3 - x_2, \dots, 1 - x_{n-1}]$ to $v_i$.  To calculate the new period, simply add $cx_2 + c(x_3 - x_2) + c(x_4 - x_3) + \dots + c(1 - x_{n-1}) = c$. Thus, the new period is $c$. 

So far we have analyzed the effects of different types of disorder. In a completely disordered MW-SQUID, the distortion of the VSR is simply a linear combination of the effects generated by the critical phase, critical current, and the position disorder. 

Let us prove that any given VSR is $IBV$ symmetric, where $IBV$ represents an inversion of the magnetic field, the bias current and the directions of the persistent currents of the vortices. We consider the phase difference across the $k$-th wire as our independent variable. Then the phase differences across all other nanowires are given by the Meissner phase formula: $\phi_i = \phi_{k} + 2\pi b(x_i - x_k) - 2\pi v_{k,i}$. Consider now the supercurrent formula $j_i = j_{c,i} \phi_i/\phi_{c,i}$. We perform an $I$ (or $j$) inversion which transforms $j_i \rightarrow -j_i$, a $B$ (or $b$) inversion which transforms $b \rightarrow -b$, and a vortex inversion which transforms $v_{k,i} \rightarrow -v_{k,i}$. (To clarify, the current inversion can be achieved by changing the polarity of the ammeter measuring the current through the device.) Then $j_i \rightarrow j_i' = -(j_{c,i}/\phi_{c,i}) (\phi_k - 2\pi b(x_i - x_k) + 2\pi v_{k,i}) \iff j_i'=  (j_{c,i}/\phi_{c,i} )(-\phi_k + 2\pi b(x_i - x_k) - 2\pi v_{k,i})$. Since the positive critical current at a fixed magnetic field is defined as the maximum of the total supercurrent with respect to the phase difference across the k-th nanowire, or $I_{c,+} = \max_{\phi_k}\left(\sum_i j_i\right)$ where $\phi_k \in [-\phi_{c,k}, \phi_{c,k}]$ is an independent variable, $\phi_k$ is interchangeable with $-\phi_k$ due to the symmetry of the range $[-\phi_{c,k}, \phi_{c,k}]$. A similar argument can be made for $I_{c,-} = \min_{\phi_i}\left(\sum_i j_i\right)$. Thus, we obtain the exact same critical current formula, proving that a VSR is always $IBV$ symmetric, even in MW-SQUIDS with non-periodic positions, different critical phases and different critical currents of the composing nanowires.

\begin{figure*}[t]
    \centering
    \includegraphics[width=0.32\linewidth]{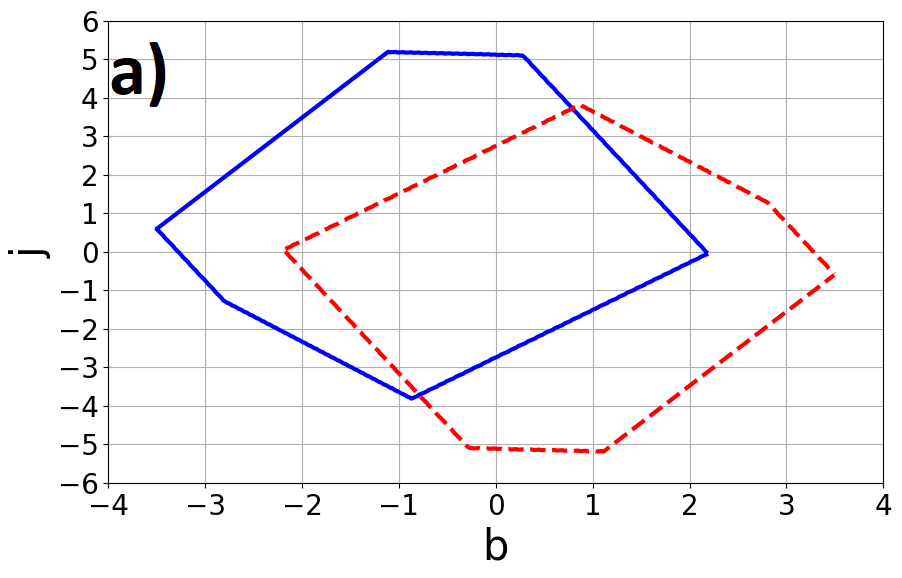}
    \includegraphics[width=0.32\linewidth]{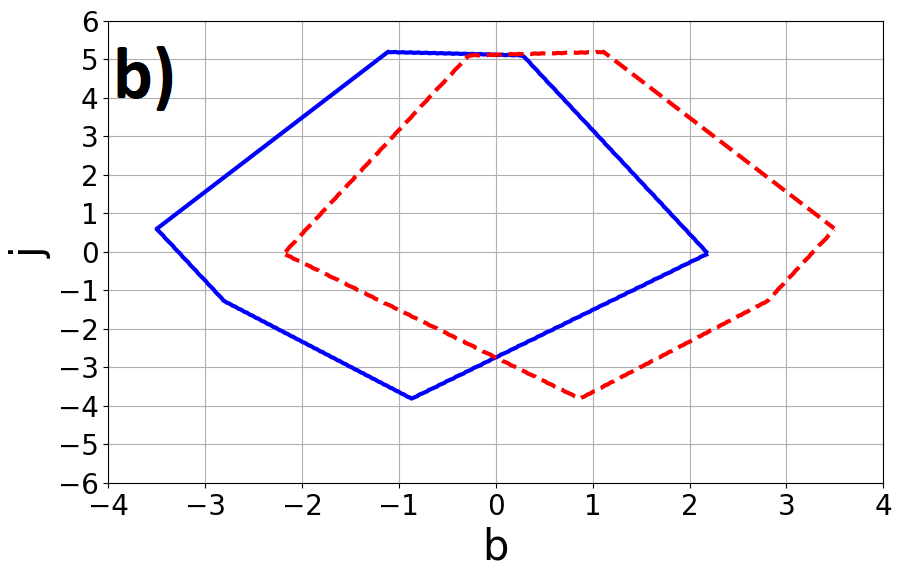}
    \includegraphics[width=0.32\linewidth]{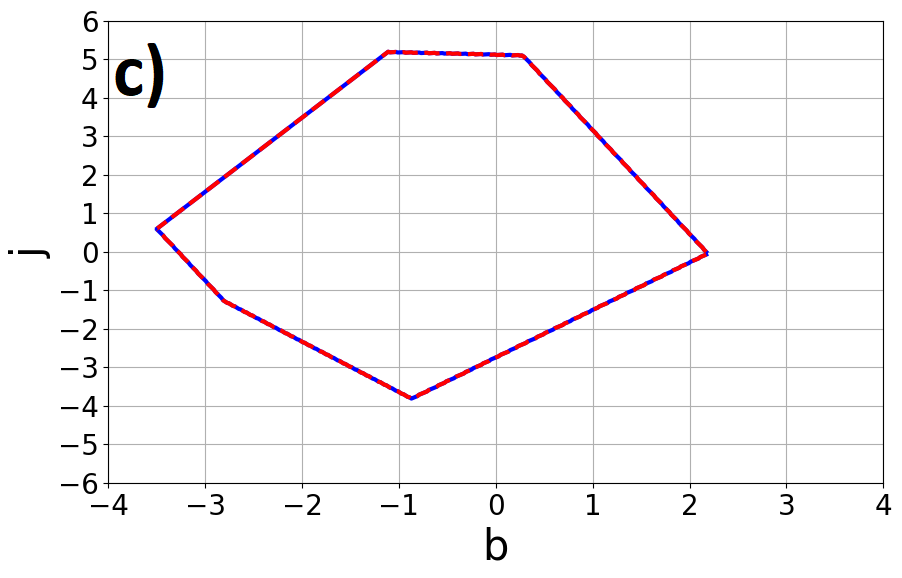}
    \caption{$IBV$ symmetry for a disordered device of eight superconducting nanowires. The wire positions are $[0, 0.1, 0.15, 0.3, 0.43, 0.7, 0.8, 1]$, the critical phases are $[12, 12.5, 8, 15, 9, 11, 13, 10]$ (rad), and the critical currents are $[1.2, 0.9, 1.1, 1.3, 0.7, 1.1, 0.8, 1]$. The y-axis is the normalized bias current, and the x-axis is the normalized magnetic field. (a) The state $v_i = [0,-1,1,0,-1,1,0]$ (solid line) and $v_i' = [0,1,-1,0,1,-1,0]$ (dashed line)  are shown. Note, $v_i = -v_i'$. (b) $v_i = [0,-1,1,0,-1,1,0]$ (solid line) is same as in (a). The VSR corresponding to $v_i' = [0,1,-1,0,1,-1,0]$ (dashed line) has been inverted with respect to the $b$ axis. (c) $v_i = [0,-1,1,0,-1,1,0]$ (solid line) is plotted as in (a). The VSR corresponding to $v_i' = [0,1,-1,0,1,-1,0]$ (dashed line) has been inverted with respect to the $j$ and $b$ axes. Both VSRs match. }
    \label{fig:8_wire_vortex_symmetries}
\end{figure*}

An example of this $IBV$ symmetry is shown in Fig.\ref{fig:8_wire_vortex_symmetries} for a disordered device of 8 nanowires. Two random vorticity states $v_i$ and $v_i'$ were plotted on the $I-B$ plane, where $v_i = -v_i'$. Inverting the VSR corresponding to the vorticity state $v_i'$ with respect to the $j$ axis and the $b$ axis maps this VSR onto the VSR corresponding to the vorticity state $v_i$. This demonstrates $IBV$ symmetry. 

\begin{figure*}[t]
    \centering
    \subfigure{%
        \includegraphics[width=0.32\linewidth]{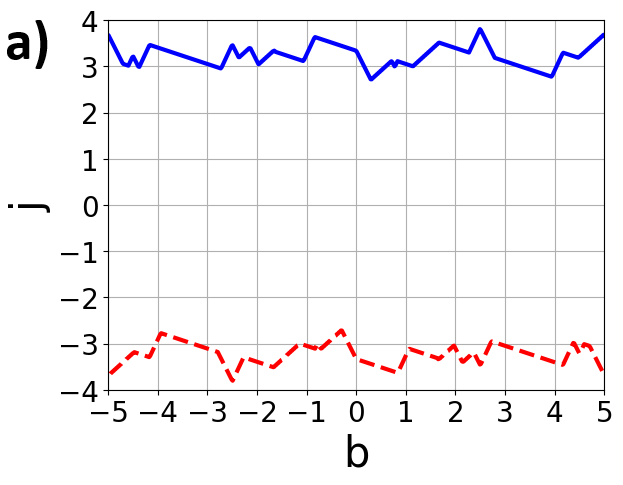}
    }
    \hfill
    \subfigure{%
        \includegraphics[width=0.32\linewidth]{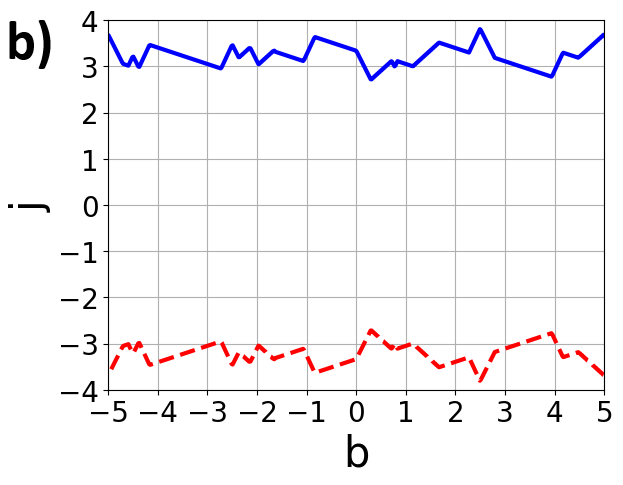}
    }
    \hfill
    \subfigure{%
        \includegraphics[width=0.32\linewidth]{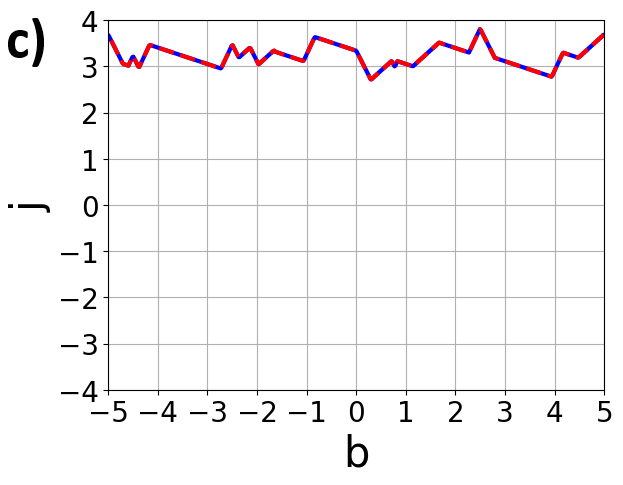}
    }
    \caption{$IB$ symmetry in Ic(b) curve for four nanowires of positions $[0, 0.3, 0.6, 1]$, critical phases of $[4\pi, 2\pi, \pi, 3\pi]$, and critical currents of $[1.2, 0.7, 1.3, 0.8]$. (a) $I_{c,+}$ is shown as the blue solid curve, and $I_{c,-}$ is shown as the red dotted curve. (b) $I_{c,+}$ is unchanged, while $B$ is inverted in the $I_{c,-}$ curve. (c) $I_{c,+}$ is shown unchanged, while the dashed curve is multiplied by -1 to realize the current inversion. Thus obtained new dashed curve represent the function $-I_{c,-}(-B)$ and it matches exactly the function $I_{c,+}$. This illustrates the $IB$ symmetry of the critical current.}
    \label{fig:4_wire_icb_symmetry}
\end{figure*}

The maximum critical current curve for any disordered MW-SQUID is always $IB$ symmetric, as will be shown below. Let us first explain the meaning of this symmetry. First, we perform a $B$ inversion on the function $I_{c,+}(B)$ which turns into $I'_{c,+}(B) = I_{c,+}(-B)$, which is equivalent to the inversion of this curve with respect to the $I$-axis. Secondly, we perform an $I$ inversion on $I_{c,+}'(B)$ and get $I''_{c,+}(B) =-I_{c,+}(-B)$. Then the new function $I''_{c,+}(B)$ is equivalent to the negative critical current function, i.e.,  $I''_{c,+}(B)=I_{c,-}(B)$. The same rule applies to $I_{c,-}(B)$, namely, if $I $is inverted and $B$ is inverted, then the new curve coincides with the original positive critical current curve. Thus, $IB$ inversion is a symmetry of the two curves, $I_{c,+}(B)$ and $I_{c,-}(B)$, taken together.  An example of $IB$ symmetry is shown in Fig.\ref{fig:4_wire_icb_symmetry} for a disordered system of 4 nanowires. 

To prove that the maximum critical current curve is $IB$ symmetric, we can use the same expression for total current as above. Namely, if a B (or $b$) inversion and an $I$ (or $j$) inversion is performed, then the total current is $j_i'= j_{c,i}/\phi_{c,i} (-\phi_k + 2\pi b(x_i - x_k) + 2\pi v_{k,i})$. To obtain the maximum critical current curve, we have to maximize this supercurrent with respect to both $\phi_k$ and $v_i$. Because both variables are independent, the resulting critical current will be the same as before the $IB$ inversion is performed.

\section{\label{sec:particle-physics-symmetries}Analogies between symmetries of multiple nanowire SQUIDS and symmetries of fundamental physical laws}

So far, we have discussed the symmetries that the $I_c(B)$ curve exhibits, as well as each individual vorticity stability region (VSR). We have defined this invariance to be the product of several transformations, namely, the current inversion denoted as $I$-inversion, a magnetic field inversion denoted as $B$-inversion, and a vortex-to-antivortex transformation, which we call $V$-inversion. In this section, we draw parallels between this superconducting device symmetry and the fundamental symmetries occurring in physics in general, such as $CPT$ symmetry, which denotes charge, parity, and time symmetry. 
The CPT symmetry can be summarized as follows. The empty space-time obeys the PT symmetry, which means that if all four coordinates of the space-time are reversed, including x,y,z, and t, the space-time remains geometrically unchanged, i.e., all the distances and the angles remains unchanged. If particles are present, then, for the symmetry to hold, one needs to invert all four coordinates of the space-time (i.e., perform the $PT$ inversion) and also apply a particle-to-antiparticle inversion (C-inversion or the charge inversion). Then the laws of physics remain unchanged.

Our goal here is to understand if there is some analogy between the general CPT symmetry and the $IBV$ symmetry that holds for the SQUIDs considered here. The first simple step is to realize that the C-inversion is analogous to the V-inversion, in which vortices are inverted to become anti-vortices, i.e., the persistent currents forming vortices change sign. The next step is to compare the symmetry of the empty spacetime, which is the PT symmetry, and the symmetry of our device without any vortices. Since the properties we consider are static (like the critical current function), therefore, physical time is not included in the analysis. Therefore, there are only three coordinates that can be inverted, namely x, y, and z. If we perform z-inversion and y-inversion, this is equivalent to rotating the device around the x-axis by 180 degrees. Such a rotation changes the sign of $B$ and the sign of $I$ for the device. Therefore, performing $zy$-inversion is equivalent to the $IB$ inversion. But, since the device does not change, one concludes that $zy$-symmetry and $IB$-symmetry are valid for the electrical transport properties of empty (no vortices) MW-SQUIDs.

To proceed with these parallels, our goal is to prove that $IBV$ symmetry is isomorphic to the CPT symmetry. Isomorphic means that one can put into one-to-one correspondence the elements of both groups, and then the products of two elements would also be in the same one-to-one correspondence. 

All observations indicate that $CPT$ is a perfect symmetry in nature \cite{Peskin:1995ev}. Therefore, it is logical that $CPT = e$ where $e$ is the identity element. Moreover, if the Charge, Parity, or Time inversion is applied twice to a system, then it is natural to expect that the system will not change. In other words, $C^2 = P^2 = T^2 = e$. Namely, the CPT symmetry group has four distinct elements: $\{e, C, P, T\}$ where $e$ is the identity. One can also argue that if $CPT=e$ then $CPTT=T$ and then $CP=T$. Same way $CT=P$ and $PT=C$. Thus, it follows that the group of these transformations has four distinct elements. Therefore, the CPT group of transformations is isomorphic to the Klein four-group. As the last logical step, we just need to prove that $IBV$ is also isomorphic to the Klein four-group. Proving this would show that $IBV$ and $CPT$ (by transitivity) are isomorphic, which permits us to construct a parallel between $IBV$ and $CPT$.

Firstly, we remind that the $I$-inversion is defined by $j \rightarrow -j$, which is the current inversion. Next, the $B$-inversion is defined by $b \rightarrow -b$, which is the magnetic field inversion. Finally, the $V$-inversion is defined by $v_i \rightarrow -v_i$, which is the vortex inversion. We now define a set $S = \{e, I, B, V\}$ where $e$ is the identity element. On this set, we define a binary operation on $S$ which behaves like the application of the inversions listed above, one after another. For example, $IB$ would mean that we first perform $B$ inversion and then $I$ inversion. 

From our definitions of $I$, $B$, $V$, it is clear that $I^2 = B^2 = V^2 = e$. Therefore, $I = I^{-1}$, $B = B^{-1}$, and $V = V^{-1}$. 
Additionally, we impose the physical constraint $IBV  = e$. This is because for any VSR, if we invert the bias current, the magnetic field, and the vorticity state, we would get back the same VSR ($IBV = e$). From this relation, we can derive that $IBVV = eV$ is equavalent to $IB = V$. Similarly for $BV = I$ and $IV = B$. Thus, it follows that $S$ is a group with four distinct transformations. Therefore, it can be concluded that $IBV$ is isomorphic to the Four-Klein group. In other words, $IBV$ exhibits the same group structure and behavior as $CPT$.

As a result of this isomorphism, one can construct a one-to-one correspondence map from the $IBV$ symmetry group to the $CPT$ symmetry group. In this analogy, we associate the $C$-inversion with the $V$-inversion, the $B$-inversion with the $P$-inversion, and the $T$-inversion with our $I$-inversion.

It is of interest to identify situations in which one would expect the $IB$ symmetry of the maximum critical current to be broken. This symmetry originates from the ability of the system to adjust vorticity values of the superconducting loops such that the maximum critical current is achieved. This may be the case if thermal fluctuations are sufficient. A similar situation occurs if the critical phase of the individual wires is sufficiently near the $\pi/2$ value, in which case there would be only one allowed vorticity state for any given magnetic field, like in ordinary SQUIDs. Also, the vorticity changes, usually randomly, if the current is pushed above the critical current and the devices switches to the normal state. Yet, each individual vorticity does not always obey the $IB$ symmetry. Examples of such VSRs, which do not obey $IB$ symmetry, are shown in Fig.\ref{fig:4-wire-stability-regions}(c-d). So if a system has topological defects or vortices which produce such asymmetric VSR, and if those vortices are rigidly trapped and do not escape even if the critical current is exceeded, then the $IB$ symmetry of the critical current may be broken. Such situation might occur if a vortex is trapped in the electrodes. Thus the presence or absence of the $IB$ symmetry can be used as a detector of trapped vortices in the superconducting electrodes. Moreover, if a vortex is present in one of the electrodes and it is not subjected to the $V$ inversion then even the $IBV$ symmetry of individual vorticity stability regions may be broken also, depending on the position of the trapped vortex.

\section{\label{sec:nonparallel_wires}The case of non-parallel wires}

It is interesting to discuss whether $IBV$ symmetry still holds even in the case of nanowires that are not parallel to the $y$-axis. In such a case, the device is not symmetrical with respect to the mirror reflection with respect to the x-axis. To analyze this, we first derive the phase correlation between two nanowires in a device of only two superconducting nanowires. Then we generalize the Meissner phase correlation equation to SQUIDs containing $n$ nanowires. 

First, assume a device with two superconducting nanowires that are not parallel to the $y$-axis. Let the first nanowire be described with two coordinates: $X_{1,bottom}$ and $X_{1,top}$ where $X_{1,bottom}$ is the connection point of the nanowire with the 'bottom' electrode (without loss of generality) and $X_{1,top}$ is the connection point of the nanowire with the 'top' electrode. Similarly, the second nanowire can be described with $X_{2,bottom}$ and $X_{2,top}$. Assume $X_{2,top} > X_{1,top}$ and $X_{2,bottom} > X_{1,bottom}$. Next, we take a contour integral of the phase gradient $(\oint \vec{\nabla} \varphi \cdot d\vec{l})$ through these two superconducting nanowires. The closed trajectory starts at the bottom left-hand corner ($X_{1,bottom}$) of this device and proceeds clockwise to the second wire and then returs to the first fire. From this closed contour integral, we obtain $\phi_1 + \nabla \varphi (X_{2,top} - X_{1,top}) - \phi_2 + \nabla \varphi (X_{2,bottom} - X_{1,bottom}) = 2\pi n$ where $n$ is the winding number, representing the number of vortices trapped between these two wires. Here, $\nabla \varphi$ is the phase gradient on the 'top' electrode. Similarly, the phase gradients on the 'bottom' electrode would be $-\nabla \varphi$ since the Meissner currents flow in the opposite direction relative to the 'top' electrode. Here and everywhere, the gauge for the vector potential is chosen the same way as \cite{hopkins_SQUID}. Thus, our phase correlation simplifies down to $\phi_1- \phi_2 + \nabla \varphi (X_{2,top} - X_{1,top} + X_{2,bottom} - X_{1,bottom}) = 2\pi n$. 

Analogously, it is easy to show that in general, the phase correlation between two superconducting nanowires in an $n$-wire interference device is $\phi_i- \phi_j + \nabla \varphi (X_{j,top} - X_{i,top} + X_{j,bottom} - X_{i,bottom}) = 2\pi v_{i,j}$. This equation implies that the Meissner phase correlation between the wires depends only on their center position. The center position is calculated as $X_{j, center}=(X_{j,top} + X_{j,bottom})/2$. With this notation the Meissner phase correlation formula becomes $\phi_i- \phi_j + 2\nabla \varphi (X_{j,center} - X_{i,center}) = 2\pi v_{i,j}$. The rotation of the device around the x-axis by 180 degrees does not change the position of the center of each nanowire, and therefore, the equation for the Meissner phase is invariant under such rotation of the device. 

We now study whether $IBV$ symmetry still holds. Given this generalized Meissner phase formula, rotating the device about the $x$-axis by 180 degrees (which is equivalent to the $IBV$ inversion) bidirectionally maps $X_{i,bottom} \leftrightarrow X_{i,top}$ and $X_{j,bottom} \leftrightarrow X_{j,top}$. Therefore, the center locations of the wires do not change, and the Meissner phase equations remain the same. In general, as was discussed above, the shape of each VSR and the critical current curves are derived from the Meissner phase equation augmented by the critical phase conditions. Since the Meissner phase correlation equation does not change under the 180-degree rotation around the $x$-axis, it follows that the shape of the VSRs and the critical current curves will be unchanged.  So, the $IBV$ symmetry applies to the VSRs of the devices where the wires are not parallel to the $y$-axis. Therefore, the $IB$ symmetry is valid for the maximum and minimum critical current curves of such devices. Moreover the ensemble of all the VSRs is $IB$ symmetric also.

\section{\label{sec:discussion}Discussion}

An example VSR of a device with many wires (5 in this case) is shown in Fig.\ref{fig:5-wire-VSR}. The MW-SQUID considered is a symmetrical device in which the wires are identical and equidistant. The dominant feature remains the sequence of large diamonds corresponding to the vorticity of the [k,k,k,k] type, where k is an integer. As the number of nanowires in the device increases, the height of the diamonds also increases. In fact, if the critical phase of the wires is the same, then at zero magnetic field, all of them can reach the critical current simultaneously, so the total critical current is the sum of the critical currents of all wires, even if the critical currents would be different. On the other hand, if the wires would have a different critical phase, then they cannot reach their corresponding critical current simultaneously at zero magnetic field, and therefore the total critical current at $B=0$ is less than the sum of all critical currents of individual wires, and the maximum of $I_c(b)$ would be shifted away from $b=0$ field (not shown).  In this respect, the MW-SQUID is qualitatively different from the traditional SQUID made of JJ tunnel junctions, since in such junctions the critical phase is always $\pi/2$ and so the maximum critical current always takes place at $b=0$.

Although the height of the central diamond is proportional to the total number of wires, the width does not change as the number of wires increases. Therefore the derivative of the critical current with respect to the magnetic field gets large with the number of wires. Thus such nanowire arrays can be used as more sensitive magnetic field sensors. For example, the critical current of the state $[0,0,0,0]$ goes to zero at $b=1$ because at this field the phase shift between wire-1 and wire-5 is $2\pi$ and the critical phase in the example considered (Fig.\ref{fig:5-wire-VSR}) is $2\pi$. Therefore, no matter which phase we assign to wire-1, the absolute value of the phase bias on wire-5 will be equal to or larger than its critical phase, i.e., the device cannot be superconducting if $b=1$ in this example. 

The pattern is periodic, and the period can be understood as follows. Suppose, for example, $k=1$, i.e., there is one fluxoid in each cell. Then the phase shift between each pair of neighbor wires is $2\pi$. It is possible to reduce this phase shift by applying a suitable magnetic field. That is, if the magnetic field is $b=4$, then the Meissner phase shift between the first wire and the last wire is $8\pi$ and the phase shift between neighbor wires is exactly $2\pi$. Thus, the phase shift due to the trapped vortices is exactly compensated by the Meissner phase shift, and therefore the wires can again have the same phase bias and can be pushed to the critical phase simultaneously by increasing the bias current up to $5I_{c,1}$.

The VSR of other vorticity states, not equal to $[k,k,k,k]$, produces much lower critical currents because for such states, there is no magnetic field at which all wires can have the same phase. Therefore, it is impossible to achieve a situation in which all wires contribute their maximum possible supercurrent. In general, the critical current of the device resembles the interference pattern of an optical diffraction grating.  It is a sequence of sharp principal maxima and some lower magnitude secondary maxima in between. In the example of Fig.\ref{fig:5-wire-VSR}, one can distinguish five secondary maxima in $I_c(b)$ between each pair of neighboring principle diamonds, $[k,k,k,k]$ and $[k+1,k+1,k+1,k+1]$. Thus, the principal maxima of the critical current correspond to the principal diamond VSRs having vorticity $[k,k,k,k]$. Other vorticity states produce much smaller maxima of the critical current.  

In an MW-SQUID with equidistant nanowires, the threshold at which the VSRs become topologically disconnected or disjoint is $\max_i(\phi_{c,i}) < \frac{n-1}{n}\pi$ where $n$ is the number of wires. This threshold also defines the condition when the minima of the critical current drop to zero. When the VSRs are not disjoint, this provides an avenue for programmable superconducting memory. Namely, at the locations of VSR overlap, supercurrent-controlled memory read-out techniques may be used \cite{eduard_memory}. This is because the critical current is a function of the vorticity of the device.
In particular, for devices with three or more nanowires, there may be more than two states that overlap at fixed magnetic fields. In which case, more information may be stored, thus creating complex logic devices.

A superconducting memory device that leverages superconducting nanowires provides significant advantages, namely the high degree of customization in the positions of the nanowires, critical phases, critical currents, as well as the vorticity states, which may be analyzed by our model. 

Specific VSRs provide a perfect diode effect. It is prodced by the triangular VSR. It occurs, for example, if the critical phases of all the nanowires to $2\pi$. Namely, the triangular VSRs of such a device is such that $I_{c,-} = 0$ and $I_{c,+} > 0$. Moreover, an extremely strong diode effect, with 3x the difference between the $I_{c,+}^{v_i}$ and $I_{c,-}^{v_i}$ is exhibited in the flat top diamond stability region for 3 equidistant, identical superconducting nanowires with $\phi_{c,i} = 2\pi$.

Previously, it has been shown that superconducting diodes can be used as rectifiers \cite{Song2023}. Generally speaking. this perfect diode effect might allow the development of logic gates constructed based on diode logic. In our case, the logic operations would be realized based on the fact that the critical current in one polarity is zero, while in the other polarity, it is distinct from zero. 

\section{\label{sec:conclusion}Conclusions}

We analyze a linear model that generates vorticity stability regions (VSR) and the critical current as a function of perpendicular magnetic field for planar superconducting quantum interference devices (SQUIDs) made with many parallel nanowires. The nanowires are assumed to have linear current-phase relationships, which are limited by critical currents and the corresponding critical phase value. We investigated in detail the VSRs of two, three, and four-nanowire devices. For two nanowires with all critical phases being the same, all VSR regions have rhombic (diamond) shapes. For devices containing three wires (3-SQUIDs) and, correspondingly, two loops, a greater variety of VSR shapes is predicted. The list includes regular diamond shapes, flat-top/flat-bottom diamonds of various sizes (depending on the magnitude of the critical phases and the vorticity differences between the loops), and triangular shapes of various sizes (depending on the magnitude of the critical phases and the vorticity differences between the loops). For example, when the critical phase is $4\pi$, if the vorticity difference is 1 or 2, then the generated VSR shape is a flat-top or flat-bottom diamond shape. If the vorticity difference is greater than 2, then the VSR shape is a triangular shape. The tendency is that as the magnitude of the vorticity difference increases, the area of the VSR decreases.
There are many more distinct VSR shapes for 4-SQUID than 3-SQUID. For devices containing four wires (4-SQUID) with all critical phases being the same, some of the generated VSRs include the diamond shape, glider shape, trapezoidal shape, tilted triangular shape, kite shape, and flipped kite shape.

We have argued that the maximum critical current versus magnetic field curve, $I_c(B)$, is invariant under a combination of $I$-inversion and $B$-inversion, even if there are many parallel nanowires, they are all different and the spacing between them is all different. Thus, one can say that the critical current curve obeys $IB$ symmetry. This symmetry would be broken only if there are randomly positioned vortices in the electrodes, which are permanently trapped and cannot be removed even if the device is biased above its critical current. Also, it was found that the VSR of any arbitrary nanowire device is invariant if, in addition to $I$ and $B$, a vortex-to-anti-vortex transformation (or $V$ inversion) is performed. Thus, we show that the VSR obeys the $IBV$ symmetry, again, assuming that there are no vortices trapped in the electrodes of the SQUID. Moreover, the VSRs of a device with nanowires that are not parallel to the $y$ axis also obey the $IBV$ symmetry, assuming the electrodes can be maintained vortex-free.  

We predict the existence of perfect diodes. For example, for 3 identical, equidistant superconducting nanowires (3-SQUID) with all critical phases set to $2\pi$, if one sets the vorticity of the device to be $[-1,-4]$, or $[0,-3]$, etc., then one observes a perfect diode effect. That is, the critical current in the negative direction is exactly zero, while the critical current in the positive direction is above zero and it is in fact relatively large. In other words, such a device can support supercurrents of only one polarity. The shape of this VSR is a triangular, as shown in Fig.\ref{fig:3-wire-stability-regions}e.

Generally speaking, superconducting diode effects are predicted to occur in system with broken inversion symmetry. Simply speaking, if the left side of the device is not equivalent to the right side then a superconducting diode effect takes places. Or, if the vortex numbers or polarities are different in the left cells of the device and the right cells of the device then again the corresponding positive and negative critical currents will be different in magnitude.

We distinguish two different transformations that can be applied to the ensemble of vortices in the context of mirror-symmetric MW-SQUIDs: vorticity-mirror ($P_v$) transformation and vortex-to-antivortex transformation (V-inversion). For vorticity states that are $P_v$ symmetric, the corresponding VSR is symmetric with respect to the $j$ inversion, that is, $I_{c,+}(B)=-I_{c,-}(B)$. If a vorticity state is symmetric with respect to the combined $P_vV$ transformations, then the corresponding VSR is symmetric with respect to the $b$ inversion, i.e., $I_c(B)=I_c(-B)$. We use this result to predict that a vorticity state that is symmetric with respect to the combined $P_vVP_v$ transformation (which is equivalent to the $V$ inversion) must be $j$ symmetric and also $b$ symmetric. The only vorticity state that is $V$ symmetric and therefore separately $j$ symmetric and $b$ symmetric is the zero vorticity state. 

For systems of $n$ equidistant nanowires, the sequence of the VSRs regions is disjoint if the maximum critical phase across all wires is less than $\frac{n-1}{n}\pi$. This means that one can observe S-N quantum transitions when sweeping the magnetic field at zero temperature. Moreover, the minima of the critical current envelope curve hit zero, which means that the modulation of the critical current is 100\%. Such widely adjustable devices could find applications in systems in which tuning of the kinetic inductance is essential.

We also investigated the effects of disorder of the critical phases, critical currents, positions, and all of the above. When applying a critical phase disorder, the top and bottom vertices of VSRs can shift horizontally and vertically, while the period of the device remains unchanged. When applying a critical current disorder (without critical phase disorder), all parts of a VSR can be altered except for the top and bottom vertices, if viewed in the normalized coordinates. Moreover, the period of the device remains the same. When applying a position disorder, the period of the device changes. The new periodicity of the device, with nanowires positioned aperiodically, constitutes a generalized Little-Parks effect. In essence, if the number of vortices in each loop of the device is increased by an integer number, \textit{which is proportional to the size of the loop}, then it becomes possible to change the magnetic field in such a way that it perfectly compensates the effect of the additional vortices in all loops. As a result, the critical current and the device energy return to their original values prior to the introduction of the additional vortices. Such increase of the magnetic field represents the period of the critical current function. The period in the normalized magnetic field is equal to the common denominator of all the cells dimensions, expressed in normalized coordinates.

\section*{Acknowledgments}

The work was supported in part by the NSF DMR-2104757 and by the NSF OMA 2016136 Quantum Leap Institute for Hybrid Quantum Architectures and Networks (HQAN).

\section*{Code}

The referenced MW-SQUID model can be found at \url{https://github.com/cliffsun/Nanowire-Memory-Model}.

\section*{References}

\vspace{-10pt}

\bibliography{references}

\end{document}